\documentclass[sigconf,
              nonacm,
]{acmart}

\AtBeginDocument{%
  \providecommand\BibTeX{{%
    Bib\TeX}}}

\setcopyright{acmcopyright}
\copyrightyear{2022}
\acmYear{2022}
\acmDOI{XXXXXXX.XXXXXXX}

\acmConference[PPDP'22]{PPDP'22}{September 20--22,2022}{Tbilisi, Georgia}
\acmPrice{}%15.00}
\acmISBN{}%978-1-4503-XXXX-X/18/06}

\usepackage{todonotes}
\usepackage{xcolor}
\usepackage{hyperref}
\usepackage{verbatim}
\usepackage{amsmath,mathtools}
\usepackage{subcaption}
\usepackage{amsthm}
\usepackage{tikz}
\usepackage{MnSymbol}
\usepackage{paralist}
\usepackage{enumitem}
\usepackage[capitalise]{cleveref}
\usepackage[all]{xy}

\begin{document}
\sloppy
\title[A Probabilistic Call-by-Need Lambda-Calculus -- Extended Version]{\texorpdfstring{A Probabilistic Call-by-Need Lambda-Calculus\\Extended Version}{A Probabilistic Call-by-Need Lambda-Calculus -- Extended Version}}

\author{David Sabel}
\email{david.sabel@lmu.de}
\affiliation{%
  \institution{LMU Munich}
  \city{Munich}
  \country{Germany}
}
\orcid{0000-0002-5109-3273}
\author{Manfred Schmidt-Schau{\ss}}
\email{schauss@ki.cs.uni-frankfurt.de}
\affiliation{%
  \institution{Goethe-University Frankfurt}
  \city{Frankurt}
  \country{Germany}
}
\orcid{0000-0001-8809-7385}

\author{Luca Maio}
\email{l.maio@campus.lmu.de}
\affiliation{%
  \institution{LMU Munich}
  \city{Munich}
  \country{Germany}
}

% \renewcommand{\shortauthors}{Sabel, D., Schmidt-Schau{\ss}, M., Maio, L.}

\newcommand{\longversion}[1]{}
\newcommand{\shortversion}[1]{#1}

\newcommand{\commuting}[4]{
\xymatrix{
\cdot\ar[d]_{#1}   & \cdot\ar[l]_{#2}\ar@{-->}[d]^{#4} 
\\
\cdot   & \cdot\ar@{-->}[l]^{#3}
 }
}
\newcommand{\commutingtr}[4]{
\xymatrix@R=4mm@C=4mm{
\cdot\ar[dd]_{#1}  && \cdot  \ar@{-->}[dl]^{#4}\ar[ll]_{#2}
\\
&\cdot \ar@{-->}[dl]^{#3} 
\\
\cdot && 
 }
}
\newcommand{\commutingtriangle}[3]{
\xymatrix{
\cdot\ar[d]_{#1}  & \cdot\ar[l]_{#2}\ar@{-->}[dl]^{#3}
\\
\cdot
 }
}
\newcommand{\PS}{\mathtt{PS}}

\newcommand{\problab}{\ensuremath{\mathit{prob}}}
\newcommand{\problabr}{\ensuremath{\mathit{probr}}}
\newcommand{\problabl}{\ensuremath{\mathit{probl}}}
\newcommand{\probid}{\ensuremath{\mathit{probid}}}
\newcommand{\probassoc}{\ensuremath{\mathit{probassoc}}}
\newcommand{\probprob}{\ensuremath{\mathit{probdistr}}}
%%%   Manfred: das ist eher eine Distributivitaet
\newcommand{\probreorder}{\ensuremath{\mathit{probreorder}}}
\newcommand{\probcomm}{\ensuremath{\mathit{probcomm}}}
\newcommand{\sr}{\ensuremath{\mathit{sr}}}
\newcommand{\llet}{\ensuremath{\mathit{llet}}}
\newcommand{\lletin}{\ensuremath{\mathit{llet\textit{-}in}}}
\newcommand{\llete}{\ensuremath{\mathit{llet\textit{-}e}}}
\newcommand{\lllrule}{\ensuremath{\mathit{lll}}}
\newcommand{\lacs}{\ensuremath{\mathit{lacs}}}
\newcommand{\gcone}{\ensuremath{\mathit{gc\text{-}1}}}
\newcommand{\gctwo}{\ensuremath{\mathit{gc\text{-}2}}}
\newcommand{\gc}{\ensuremath{\mathit{gc}}}
\newcommand{\ucpone}{\ensuremath{\mathit{ucp\text{-}1}}}
\newcommand{\ucptwo}{\ensuremath{\mathit{ucp\text{-}2}}}
\newcommand{\ucpthree}{\ensuremath{\mathit{ucp\text{-}3}}}
\newcommand{\ucp}{\ensuremath{\mathit{ucp}}}
\newcommand{\absrule}{\ensuremath{\mathit{abs}}}

\newcommand{\xch}{\ensuremath{\mathit{xch}}}
\newcommand{\cp}{\ensuremath{\mathit{cp}}}
\newcommand{\cpx}{\ensuremath{\mathit{cpx}}}
\newcommand{\cpxin}{\ensuremath{\mathit{cpx\textit{-}in}}}
\newcommand{\cpxe}{\ensuremath{\mathit{cpx\textit{-}e}}}
\newcommand{\cpin}{\ensuremath{\mathit{cp\textit{-}in}}}
\newcommand{\cpe}{\ensuremath{\mathit{cp\textit{-}e}}}
\newcommand{\abs}{\ensuremath{\mathit{abs}}}
\newcommand{\cpcx}{\ensuremath{\mathit{cpcx}}}
\newcommand{\cpcxin}{\ensuremath{\mathit{cpcx\textit{-}in}}}
\newcommand{\cpcxe}{\ensuremath{\mathit{cpcx\textit{-}e}}}
\newcommand{\caserule}{\ensuremath{\mathit{case}}}
\newcommand{\casec}{\ensuremath{\mathit{case\textit{-}c}}}
\newcommand{\casein}{\ensuremath{\mathit{case\textit{-}in}}}
\newcommand{\casee}{\ensuremath{\mathit{case\textit{-}e}}}
\newcommand{\seq}{\ensuremath{\mathit{seq}}}
\newcommand{\seqc}{\ensuremath{\mathit{seq\textit{-}c}}}
\newcommand{\seqin}{\ensuremath{\mathit{seq\textit{-}in}}}
\newcommand{\seqe}{\ensuremath{\mathit{seq\textit{-}e}}}
\newcommand{\lbeta}{\ensuremath{\mathit{lbeta}}}
\newcommand{\lapp}{\ensuremath{\mathit{lapp}}}
\newcommand{\lseq}{\ensuremath{\mathit{lseq}}}
\newcommand{\lcase}{\ensuremath{\mathit{lcase}}}
\newcommand{\cpd}{\ensuremath{\mathit{cpd}}}
\newcommand{\cpS}{\ensuremath{\mathit{cpS}}}
\newcommand{\cps}{\ensuremath{\mathit{cpS}}}

\newcommand{\SR}{\ensuremath{\text{SR}}}
\newcommand{\SRlll}{\ensuremath{\text{SRlll}}}
\newcommand{\SRlbeta}{\ensuremath{\text{SRlbeta}}}
\newcommand{\Sug}{\ensuremath{\text{Sug}}}
\newcommand{\Scp}{\ensuremath{\text{Scp}}}
\newcommand{\Scpd}{\ensuremath{\text{Scpd}}}
\newcommand{\ScpS}{\ensuremath{\text{ScpS}}}
\newcommand{\Slll}{\ensuremath{\text{Slll}}}
\newcommand{\Slapp}{\ensuremath{\text{Slapp}}}
\newcommand{\Scpx}{\ensuremath{\text{Scpx}}}
\newcommand{\Sllet}{\ensuremath{\text{Sllet}}}

\newcommand{\letbind}{{=}}
\newcommand{\LNEED}{L_\mathit{need}}
\newcommand{\LNEEDPROB}{L_{\mathit{need,}\oplus}}
\newcommand{\LRPROB}{L_{\mathit{need,}\oplus}^{case,seq}}
\newcommand{\maycon}{{\downarrow}}
\newcommand{\SVAR}{\mathit{Var}_{\!\!\mathit{sym}}}
\newcommand{\CVAR}{\mathit{CV}_{\!\!\mathit{sym}}}
\newcommand{\AVAR}{\mathit{SV}_{\!\!\mathit{sym}}}
\newcommand{\ENV}{\mathcal{E}}
\newcommand{\NCCDVC}{\mathit{NCC}_{\!\mathit{dvc}}}
\newcommand{\Dom}{\mathtt{Dom}}
\newcommand{\Cod}{\mathtt{Cod}}
\newcommand{\simlet}{\sim_{\mathit{let}}}
\newcommand{\simsubst}{\sim_{\mathit{subst}}}
\newcommand{\alphaRename}{\ensuremath{\mathit{AR}}}
\newcommand{\siftDown}{\ensuremath{\mathit{sift}}}
\newcommand{\siftDownWithScope}{\ensuremath{\mathit{sift'}}}
\newcommand{\splitNCC}{{\mathit{split}_{\!\mathit{ncc}}}}
\newcommand{\splitC}{\ensuremath{\mathit{split}}}
\newcommand{\CFO}{\ensuremath{\mathit{CFO}}}
\newcommand{\CAO}{\ensuremath{\mathit{CAO}}}
\newcommand{\CCO}{\ensuremath{\mathit{CCO}}}
\newcommand{\davidins}[1]{{\color{red}#1\color{black}}}
\newcommand{\delete}[1]{{\color{black!50!white}#1\color{black}}}
\newcommand{\ignore}[1]{}
\newcommand{\ROT}[1]{\color{red}\underline{#1}\color{black}}
\newcommand{\BLAU}[1]{\color{blue}{#1}\color{black}}
\newcommand{\prob}[2]{\ensuremath{#1 \oplus #2}}
\newcommand{\lucacomment}[1]{\fbox{\begin{minipage}{.9\columnwidth}\color{green!30!black}LM: #1\end{minipage}}}
\newcommand{\lucainsert}[1]{{\color{green!30!black}#1}}

\newcommand{\manfredcomment}[1]{\fbox{\begin{minipage}{.9\columnwidth}\color{blue}MS: #1\end{minipage}}}
\newcommand{\manfredinsert}[1]{\BLAU{#1}}
\newcommand{\manfredins}[1]{{\color{blue}#1\color{black}}}
\newcommand{\davidcomment}[1]{\fbox{\begin{minipage}{.9\columnwidth}\color{red}DS: #1\end{minipage}}}

\newtheorem{maintheorem}[theorem]{Main Theorem}
  \theoremstyle{acmdefinition}
\newtheorem{remark}[theorem]{Remark}
\newtheorem{algorithm}[theorem]{Algorithm}

\newcommand{\HEAD}[1]{%
\begin{tikzpicture}%
\node at (0,0) [rectangle]{\rule{0mm}{1ex}\rule{3mm}{0mm}};%
\node at (0,0) {\rotatebox{90}{\parbox{4mm}{$\!\!\!\mbox{\small #1}\!\!\!$}}};\end{tikzpicture}}
\newcommand{\BIGOR}{\mathop{\,{\scalebox{2}[1.3]{\ensuremath{|}}}}}
\newcommand{\BIGBIGOR}{\mathop{\,{\scalebox{2.7}[2.3]{\ensuremath{|}}}}}
\newcommand{\BIGBIGBIGOR}{\mathop{\,{\scalebox{2.7}[3.3]{\ensuremath{|}}}}}
\newcommand{\BIGBIGBIGBIGOR}{\mathop{\,{\scalebox{2.7}[5]{\ensuremath{|}}}}}
\newcommand{\ari}{\mathit{ar}}
\newcommand{\calA}{\ensuremath{\mathcal{A}}}
\newcommand{\calK}{\ensuremath{\mathcal{K}}}
\newcommand{\calT}{\ensuremath{\mathcal{T}}}
\newcommand{\calC}{\ensuremath{\mathcal{C}}}
\newcommand{\LM}{\mathrm{LM}}
\newcommand{\LMP}{\mathrm{LMP}}

\newcommand{\tletrx}[2]{({\tletr~#1~\tin~#2})}
\newcommand{\tletr}{{\tt let}}
\newcommand{\tletrec}{{\tt let}}
\newcommand{\tin}{{\tt in}}
\newcommand{\tif}{{\tt if}}
\newcommand{\tthen}{{\tt then}}
\newcommand{\telse}{{\tt else}}
\newcommand{\ttrue}{{\tt True}}
\newcommand{\tfalse}{{\tt False}}
\newcommand{\tList}{{\tt List}}
\newcommand{\tnil}{{\tt Nil}}
\newcommand{\tCons}{{\tt Cons}}
\newcommand{\tcase}{{\tt case}}
\newcommand{\tof}{{\tt of}}
\newcommand{\tcons}{{\tt Cons}}
\newcommand{\tBool}{{\tt Bool}}
\newcommand{\tseq}{{\tt seq}}
\newcommand{\Env}{{\mathit{Env}}}
\newcommand{\env}{{\mathit{env}}}
\newcommand{\ialts}{{\mathit{alts}}}
\newcommand{\ialt}{{\mathit{alt}}}
\newcommand{\LR}{\ensuremath{\mathrm{LR}}}
\newcommand{\TRANS}{\mathsf{T}}
\newcommand{\LRP}{\mathrm{LRP}}
\newcommand{\FV}{{\mathit{FV}}}
\newcommand{\BV}{{\mathit{BV}}}
\newcommand{\Var}{{\mathit{Var}}}
\newcommand{\LV}{{\mathit{LV}}}
\newcommand{\wrt}{{w.r.t.}}
\newcommand{\ie}{{{i.e.}}}
\newcommand{\eg}{{{e.g.}}}
\newcommand{\FIGURE}{{{Fig.}}}
\newcommand{\SECTION}{{{Sect.}}}
\newcommand{\tletrxnk}[2]{\tletr~#1 ~{\tt in}~#2}
\newcommand{\letwork}{{\ensuremath{\mathtt{letw}}}}
\newcommand{\dotcup}{\ensuremath{\mathaccent\cdot\cup}}
\newcommand{\MBA}{\mathbb{A}}
\newcommand{\sort}{\mathit{cl}}
\newcommand{\SPOS}{\mathit{sp}}
\newcommand{\CCSv}{\mathit{CCSv}}
\newcommand{\sval}{{\tt sval}}
\newcommand{\ilabel}{\mathit{label}}
\newcommand{\sem}{\mathit{sem}}
\newcommand{\tid}{{\tt id}}
\newcommand{\tbot}{{\tt Bot}}
\newcommand{\vect}[1]{{\overrightarrow{#1}}}
\newcommand{\EMPTYCONTEXTCLASS}{\ensuremath{\mathit{Triv}}}
\newcommand{\oarituple}[1]{\langle #1 \rangle}
\newcommand{\RULEFONT}[1]{\normalfont\scriptsize\sffamily(#1)\!\!}
\newcommand{\Sol}{\mathit{Sol}}
\newcommand{\CC}{{{\mathit{Ch}}}}  %\color{blue}
\newcommand{\CCK}{\CC} % {{{\mathit{Ch}}^{\mathcal{K}}}}
\newcommand{\CCKI}[1]{\CC_{#1}} %{{{\mathit{Ch}}^{\mathcal{K}}_{#1}}}
\newcommand{\CCKII}[1]{\CC_{#1}}%{{{\mathit{Ch}}^{\mathcal{K}_{#1}}_{#1}}}
\newcommand{\CCKIII}[2]{\CC_{#2}}%{{{\mathit{Ch}}^{\mathcal{K}_{1}}_{2}}}
\newcommand{\CV}{\mathit{CV}}
\newcommand{\CVM}{\mathit{CV\!\!}_M}
\newcommand{\VarM}{\mathit{Var\!}_M}
\newcommand{\CVA}{\mathit{CV\!\!}_A}
\newcommand{\VarA}{\mathit{Var\!}_A}
\newcommand{\FAIL}{\ensuremath{\textit{FAIL}}}
\newcommand{\Variable}{{{\textbf{\normalfont\bfseries Var}}}}   %\color{blue}
\newcommand{\Expression}{\textbf{\normalfont\bfseries Exp}} 
\newcommand{\SymVar}{\mathit{Sym\!Var}}
\newcommand{\HExpression}{\textbf{\normalfont\bfseries HExp}} 
\newcommand{\Binding}{\textbf{\normalfont\bfseries Bind}} 
\newcommand{\Environment}{\textbf{\normalfont\bfseries Env}}

\newcommand{\order}{\mathit{order}} 
\newcommand{\oari}{\mathit{oar}} 
\newcommand{\tvarlift}{{\ensuremath{\mathtt{var}}}}  %%\color{red}
\newcommand{\LRS}{{\texttt{\normalfont\tt LRS}}}
\newcommand{\LRSX}{{\texttt{\normalfont\tt LRSX}}}
\newcommand{\LRSXALPHA}{{\texttt{\normalfont\tt LRSX$\alpha$}}}
\newcommand {\UNIFLRS}{{\ensuremath{\mathsf{\textsc{UnifLRS}}}}}
\newcommand {\MATCHLRS}{{\ensuremath{\mathsf{\textsc{MatchLRS}}}}}
\newcommand{\MBK}{\mathbb{K}}
\newcommand{\UV}{{\mathit{MV}}}
\newcommand{\UVM}{\BLAU{\mathit{MV\!}_I}} 
\newcommand{\UVF}{\ROT{\mathit{MV\!}_F}} 
\newcommand{\n}[1]{\mathsf{#1}} %% ``nor
\newcommand{\ANS}{\mathsf{Ans}} %% ``nor
\newcommand{\CALSR}{\mathsf{SR}}
\newcommand{\ICtxts}{\textit{ICtxt}}
\newcommand{\Typ}{\textit{Typ}}
\newcommand{\CHFI}{\ensuremath{\mathit{CHFI}}}
\newcommand{\CHF}{\ensuremath{\mathit{CHF}}}
\newcommand{\CHFP}{\ensuremath{\mathit{PF}}}
\newcommand{\CHFL}{\mathit{CHFL}}
\newcommand{\slashslash}{\,/\!/\,} 
\newcommand{\tauop}{\xi} 
\newcommand{\Fcand}{F_{\mathit{cand}}}

\newcommand{\NSR}{\ensuremath{\mathit{NSR}}}
\newcommand{\infsr}{\ensuremath{{\mathit{infSR}}}}
\newcommand{\tBot}{{\tt Bot}}
\newcommand{\ITT}{{\mathit{IT}}}
\newcommand{\italt}{{\mathit{alt}}}
\newcommand{\iRED}{{\mathit{Red}}}
\newcommand{\RCtxt}{\mathbb{R}}
\newcommand{\SCtxt}{\mathbb{S}}
\newcommand{\itrace}{{\mathit{trace}}}
\newcommand{\nullproc}{{\mathbf{0}}}
\newcommand{\PAR}{\ensuremath{\,\texttt{\tt |}\,}}
\newcommand{\OR}{\mid}
\newcommand{\MVAR}[2]{{#1}\,{\boldsymbol{\mathsf{m}}}\,{#2}}
\newcommand{\EMPTYMVAR}[1]{\MVAR{{#1}}{-}}
\newcommand{\Proc}{\textit{Proc}}
\newcommand{\IProc}{\textit{IProc}}
\newcommand{\Expr}{\textit{Expr}}
\newcommand{\Eval}{\textit{Eval}}
\newcommand{\evaluation}{\ensuremath{{{\downrsquigarrow}}}}
\newcommand{\MExpr}{\textit{MExpr}}
\newcommand{\IMExpr}{\textit{IMExpr}}
\newcommand{\Id}{I}
\newcommand{\NEW}{\nu}
\newcommand{\THREAD}[2]{{#1}\,{\Leftarrow}\,{#2}}
\newcommand{\LAZYTHREAD}[2]{{#1} \xLongleftarrow{\!\text{\normalfont\itshape lazy}\!}{#2}}

\newcommand{\SHARE}[2]{{#1}={#2}}
\newcommand{\arity}{\mathrm{ar}}
\newcommand{\ttcommand}[1]{{\ensuremath{\mathtt{#1}}}}
\newcommand{\tTrue}{\ttcommand{True}}
\newcommand{\tFalse}{\ttcommand{False}}
\newcommand{\treturn}{\ttcommand{return}}
\newcommand{\tbind}{\ensuremath{\,\texttt{>}\!\texttt{>}\!\texttt{=}\,}}
\newcommand{\tfuture}{\ttcommand{future}}
\newcommand{\ttakeMVar}{\ttcommand{takeMVar}}
\newcommand{\tunsafeInterleaveIO}{\ttcommand{unsafeInterleaveIO}}
\newcommand{\tunsafePerformIO}{\ttcommand{unsafePerformIO}}
\newcommand{\tMVar}{\ttcommand{MVar}}
\newcommand{\tnewMVar}{\ttcommand{newMVar}}
\newcommand{\tputMVar}{\ttcommand{putMVar}}
\newcommand{\MTHREAD}[2]{{#1} \xLongleftarrow{\!\text{\normalfont \sffamily main}\!}{#2}}
\newcommand{\PCtxt}{\mathbb{D}}
\newcommand{\PCtxts}{\textit{PCtxt}}
\newcommand{\IPCtxts}{\textit{IPCtxt}}
\newcommand{\Ctxt}{\mathbb{C}}
\newcommand{\Ctxts}{\textit{Ctxt}}
\newcommand{\MultiCtxt}{\widetilde{\mathbb{C}}}
\newcommand{\MCtxt}{\mathbb{M}}
\newcommand{\MCtxts}{\textit{MCtxt}}
\newcommand{\IMCtxts}{\textit{IMCtxt}}
\newcommand{\ECtxt}{\mathbb{E}}
\newcommand{\ECtxts}{\textit{ECtxt}}
\newcommand{\IECtxts}{\textit{IECtxt}}
\newcommand{\FCtxt}{\mathbb{F}}
\newcommand{\FCtxts}{\textit{FCtxt}}
\newcommand{\IFCtxts}{\textit{IFCtxt}}
\newcommand{\LCtxt}{\mathbb{L}}
\newcommand{\LCtxts}{\textit{LCtxt}}
\newcommand{\IRedCtxts}{\textit{IRCtxts}}
\newcommand{\IRedCtxt}{\mathbb{R}}
\newcommand{\ACtxts}{\textit{ACtxts}}
\newcommand{\LHatCtxts}{\textit{$\widehat{LCtxt}$}}
\newcommand{\ACtxt}{\mathbb{A}}
\newcommand{\KCtxt}{\mathbb{K}}
\newcommand{\reduce}[1][]{\xrightarrow{\textit{\tiny CHF}{#1}}}
\newcommand{\reducename}[1][]{\xrightarrow{\textit{\tiny CHF}{#1}}}
\newcommand{\reducechf}[1][]{\xrightarrow{\textit{\tiny CHF}{#1}}}
\newcommand{\reducechfi}[1][]{\xrightarrow{\textit{\tiny CHFI}{#1}}}
\newcommand{\reducepfi}[1][]{\xrightarrow{\textit{\tiny PFI}{#1}}}
\newcommand{\reducepfmi}[1][]{\xrightarrow{\textit{\tiny PFMI}{#1}}}
\newcommand{\reducepf}[1][]{\xrightarrow{\textit{\tiny PF}{#1}}}
\newcommand{\GCtxt}{\mathbb{G}}
\newcommand{\GCtxts}{\textit{GCtxt}}
\newcommand{\mustcon}{{\Downarrow}}
\newcommand{\mustdiv}{{\Uparrow}}
\newcommand{\maydiv}{{\uparrow}}
\newcommand{\tundefined}{\text{{\normalfont\ttfamily undefined}}}
\newcommand{\iRed}{\mathit{Red}}
\newcommand{\TC}{\mathit{TC}}
\newcommand{\cand}{{{\ {\le_{\mathit{cand}}}\ }}}
\newcommand{\candc}{{(\le_{\mathit{cand}})^c}}
\newcommand{\closed}[1]{(#1)^c}
\newcommand{\open}[1]{(#1)^o}
\newcommand{\PureInfF}{\mathit{PFI}}   
\newcommand{\PureInfFM}{\mathit{PFMI}}

\newcommand{\EC}{\mathit{ExCv}}

\begin{abstract}
To support the understanding of declarative probabilistic programming languages, we introduce a lambda-calculus with a fair binary probabilistic choice that chooses between its arguments with equal probability. The reduction strategy of the calculus is a call-by-need strategy that performs lazy evaluation and implements sharing by recursive let-expressions.
Expected convergence of expressions is the limit of the sum of all successful reduction outputs weighted by their probability. 
We use contextual equivalence as program semantics: two expressions are contextually equivalent if and only if the expected convergence of the expressions plugged into any program context is always the same. 
We develop and illustrate techniques to prove equivalences including a context lemma, two derived criteria to show equivalences and a syntactic diagram-based method.
This finally enables us to 
show correctness of a large set of program transformations with respect to the
contextual equivalence.

\end{abstract}



% %   \includegraphics[width=\textwidth]{sampleteaser}

\maketitle

\bibliographystyle{ACM-Reference-Format}
\section{Introduction}
Probabilistic programming aims at expressing probabilistic problems and models using programming language techniques.
Purely functional programming languages like Haskell allow expressing programs in a formal, declarative high-level manner. 
Equational reasoning and techniques for program transformations are available for those languages.
Lazy evaluation combined with sharing results in call-by-need evaluation
(see e.g.~\cite{ariola:95,ariola:97} for call-by-need lambda-calculi).
It enables an efficient implementation of lazy functional languages. 
In this paper, we combine both worlds and thus investigate call-by-need functional languages, extended with a probabilistic operator $\prob{}{}$, such that $\prob{s}{t}$ performs a fair choice between programs $s$ and $t$. 
Our hypothesis is, that such a language supports the declarative construction of probabilistic programs and models, provides a large set of correct program transformations, and allows to apply techniques from program transformations and program equivalence to them. 
%The goal of this paper is thus to develop and investigate a semantics (in form of a contextual semantics) for a call-by-need functional language, enriched with probabilistic evaluation.
 
%We are interested in the  
We present and develop an operational as well as a contextual semantics of call-by-need evaluation in  combination with a probability operator. 
The goal of the formal development is to enable us proving correctness of a rich set of semantic equivalences, which are reminiscent of denotational semantics
%This requires first a formal development of adequate notions, then to introduce proof techniques and lemmas to work with these notions, and finally to apply them to prove semantic equivalences.
% We begin with a 
We present a minimal probabilistic call-by-need lambda-calculus to develop the notions,  explore 
%\ignore{try
the techniques, and figure out the core properties of such a calculus. Thus we enrich the untyped lambda-calculus with recursive $\tletrec$-bindings, to implement sharing, and with a fair binary probabilistic choice $\oplus$. 

% \ignore{
% At the very end of the paper we discuss extensions to make the language more realistic as a core language of a probabilistic functional language and we sketch that our results still hold for the extended language. Further studies, like considering a typed language or built-in numbers, are left for future work. 
% Since our correctness proofs make use of automated techniques from unification and rewriting it is not too hard to extend the result to more sophisticated languages.
% 
% \lucacomment{Wird hier nicht teilweise das Selbe gesagt wie am Ende der Einleitung, bzgl. Erweiterung zu core-sprache mit seq etc. am Ende des Papers?}
% DS: Ja, man wiederholt sich manchmal;-)
% }

As semantic equivalence, we define a Morris' style contextual equivalence \cite{morris:68}, adapted to the setting of probability and expectation, i.e.~two programs $s,t$ are contextually equivalent if, and only if, the programs $C[s]$
and $C[t]$ behave the same, where $C$ is any program context, i.e.~any surrounding program. Since arbitrary contexts are permitted as a test, in a deterministic setting, observing whether $C[s]$ and $C[t]$ terminate, usually suffices to discriminate all obviously different programs while identifying as many programs as it makes sense; (for investigations in  non-deterministic functional languages see \cite{moran-sands-carlsson:99,moran-sands-carlsson:2003,kutzner-schmidt-schauss:98,sabel-schmidt-schauss-MSCS:08}).
In the probabilistic setting, termination of programs depends on the concrete random execution. In our approach 
we replace observing termination with observing the expectation of termination (i.e.~the limit of the sum of the probabilities of all successful evaluations), where contextual equivalence holds if this expected termination is the same for $C[s]$ and $C[t]$, again for any surrounding program context $C$.

Our operational semantics performs (randomized) execution of an expression resulting in an expression (a weak head normal form (WHNF) in the case of success). Thus, it does not evaluate an expression to a multi-distribution representing all possible outputs.
However, for the semantic equivalence, we collect all successful executions such that, in principle, a multi-distribution could be reconstructed. However, the core test in the notion of contextual equivalence checks whether a WHNF is the result or not (and thus in the probabilistic case it observes the expectation of a Bernoulli-experiment). 
\ignore{Manfred: Different WHNFs are discriminated by surrounding contexts since contextual equivalence tests expected convergence in all contexts. In summary, this means that} 
Contextual equivalence does not need
the information about the whole multi-distribution. 
Since the operational semantics stops if a WHNF is reached, i.e. abstractions, %%% (which represent functions),
the corresponding multi-distribution would be based on WHNFs,
% is a multi-distribution of WHNFs, 
and thus in general, it is not possible to construct a (non-multi!) distribution from it, since this would require to identify all contextually equivalent WHNFs, which is undecidable.
%in the multi-distribution, but this problem is undecidable in general.

For our program equivalence, our next goal is to show concrete laws for program transformations, that usually occur as local compiler optimizations like garbage collection, partial evaluation, inlining, and also to prove algebraic 
laws of the $\prob{}{}$-operator. We prove these equivalences to emphasize that the notions are defined in the right way. 
However, due to the quantification over all contexts, establishing contextual equivalences is usually hard and requires techniques. Thus, as a first step, we prove a novel context lemma (see \cite{Mason-Smith-Talcott:96,schmidt-schauss-sabel-gencontext:10} for some work on context lemmas in other calculi) in \cref{thm:context-lemma-ext}, which shows that it
is sufficient to take into account the more specific class of reduction contexts $R$ to conclude contextual equivalence. To prove the context lemma, we had to restrict the formulation: equal expected convergence of $R[s]$ and $R[t]$ must also hold if the number of probabilistic evaluation steps is bounded by any fixed number. However, to make the context lemma applicable to expressions that have a different number of probabilistic evaluation steps, the formulation allows for a (fix) difference between the bound for $R[s]$ and the bound for $R[t]$.

Based on the context lemma, we prove the correctness of two criteria to establish contextual equivalences: one criterion ( \cref{prop-surface-almost-same-prob-seq}) establishes contextual equivalence by allowing to pre-evaluate the expressions, such that same reduction successors, on different (probabilistic) evaluation paths, can be combined before they are compared, while the other criterion (
\cref{prop-surface-same-prob-seq}) requires the same probabilistic choices during the evaluation for the compared expressions. The former allows proving the correctness of algebraic laws in \cref{cor:problaws}. The latter is a preparation for further correctness proofs of program transformations, that are used as local optimizations in compilers: here we apply a syntactic method to show correctness. This so-called diagram method \cite{schmidt-schauss-schuetz-sabel:08,sabel-schmidt-schauss-MSCS:08,rau-sabel-schmidtschauss:12,sabel:wpte:2019} computes all overlaps between transformation steps and reductions of the operational semantics, and joins these overlaps, such that the results are complete sets of diagrams. The diagrams are then used to inductively show the (expected) convergence equivalence. The diagram computation and the inductive construction are semi-automated by techniques and tools, from unification and term rewriting, that were developed in previous work \cite{schmidtschauss-sabel-PPDP:2016,rau-sabel-schmidtschauss:12,sabel:wpte:2019}
for deterministic and non-deterministic calculi. We show that they are transferable and still very useful for the probabilistic setting. In \cref{theo:all-correct} our results on correct program transformations are summarized.
At the very end of the paper, we show that our language can be extended to a more realistic core language, by adding data constructors, case-expressions, and a seq-operator for strict evaluation. We sketch that all proofs are transferable to the extended calculus and that further program transformations can be shown to be correct by our automated technique.

\paragraph*{Related Work}
The core concept of probabilistic lambda-calculi has been researched since the late seventies, with \cite{saheb-djahromi-prob-LCF:78} introducing a probabilistic variant of LCF, 
%a version of typed lambda-calculus,
%introducing a probabilistic choice on a term-level. 
and not much later \cite{jones-plotkin-prob-powerdomain:89} investigated a primitive lambda-calculus-based language with a probabilistic choice operator modeling a fair coin.
The study of pure probabilistic lambda-calculus in regards to semantics, program and contextual equivalence as well as confluence and standardization, however, has happened much more recently,
with work mostly utilizing call-by-name and call-by-value evaluation strategies. 

Operational semantics for a probabilistic lambda-calculus 
were defined in \cite{dal-lago-prob-op-sem:2012},
covering both call-by-value and call-by-name.
The probabilistic operator is treated as a non-deterministic extension, yielding a probability distribution as the result of the reduction. Several call-by-name and call-by-value small- and big-step semantics are defined and proven equivalent. 
Furthermore, an equivalence of the given call-by-value and call-by-name semantics is proven by simulating one with the other.
%DS nicht so wichtig: which is of interest since in standard lambda-calculus this does not hold.
%Crubillé and Dal Lago went on to study a typed version of the call-by-value probabilistic lambda-calculus with regards to probabilistic applicative bisimulation in \cite{crubille-dal-lago-bisimulation:2014}. 
% Kommt unten, was das gleiche Paper...

The relation between contextual equivalence and bisimilarity in probabilistic lambda-calculi was analyzed in 
\cite{dal-lago-et-al-coinductive-equivalence:2014} for a call-by-name setting and in \cite{crubille-dal-lago-bisimulation:2014}
for a call-by-value setting. Contextual equivalence is defined analgous to our definition by observing the expected convergence 
(also called the probability of convergence).
While also giving a general overview of probabilistic lambda-calculi, Dal Lago discusses in \cite{dal-lago-on-probabilistic:2020}  operational semantics, contextual equivalence,  expressive power and termination of a typed call-by-value calculus, expanding on Plotkin's PCF. Dal Lago distinguishes between randomized lambda-calculi and Bayesian lambda-calculi. Our calculus is a randomized lambda-calculus since it performs random evaluation of the choice-operator.

Regarding a polymorphically typed call-by-value higher-order language with, probabilistic extensions (among others), in
\cite{bizjak-birkedal-logical-relations:2015}. 
a logical relation, CIU-equivalence, and contextual equivalence are shown to coincide. Their notion of contextual equivalence is analogous to ours since it compares the probabilities of termination, the coincidence of CIU-equivalence is similar to our context lemma, while adapted to the call-by-value setting.

Even more recent studies include \cite{faggian2019lambda} and \cite{dal-lago-et-al-decomposing:2020}, which both mainly focus on confluence and standardisation, but with different approaches.
In \cite{faggian2019lambda}, probability distributions and surface contexts are used to achieve confluence in their calculus.
% Zu speziell
%Expanding upon this, \cite{faggian2022strategies} proposes a technique to prove normalization in an asymptotic setting, using the call-by-value and call-by-name probabilistic lambda-calculi as concrete examples, 
%since in probabilistic settings, it is common to encounter terms that will terminate with a probability of 1, but not in a finite number of steps. 

In \cite{dal-lago-et-al-decomposing:2020} different probabilistic choices can be shared or not-shared by using labels to make them a common or separate events -- hence a probabilistic event lambda-calculus is introduced. With this decomposition of the probabilistic choice operator, they achieve confluence, with call-by-name and call-by-value variants. For our call-by-need calculus, choices are shared by default, but they can be duplicated if they occur below abstractions (since abstractions are not evaluated by the operational semantics).

We do not focus on confluence and related notions, since
contextual equivalence does not require confluence (like other equivalence notions, like convertibility). However, our diagram-based
 proof technique is related to local confluence, where, however, the rewrite relations are mixed of the operational semantics and transformation steps.
%However, \cite{dal-lago-et-al-decomposing:2020} provide a different version of probabilistic lambda-calculus altogether, the probabilistic event lambda-calculus, where the operator of probabilistic choice additionally carries a label,
%which can then be bound in a generator, a second syntactical addition to standard lambda-calculus, thus decomposing the typical probabilistic operator. Through this, they achieve confluence, with call-by-name and call-by-value variants
%differing in regards to the interpretation of this decomposition.

%\davidcomment{Das 
%\texttt{
%https://www.cambridge.org/core/books/foundations-of-probabilistic-programming/on-probabilistic-calculi/A09F32F3D5014D55178703729DB1C4BA
%}
%müsste eigentlich ein guter Pointer sein?
%Da wird z.B. ein altes Plotkin-Papier zitiert (Jones \& Plotkin)}

%\davidcomment{Ich habe ein paar TeX-Fehler behoben (unicode-Zeichen).
%Und: Besser die veröffentlichten Papiere zitieren statt der ArXiv-Versionen (bei dblp.org findet man meistens die passenden bibtex-eintraege}

\paragraph*{Outline}
 
In \cref{sec:calculus} we introduce the syntax, operational semantics, and contextual equivalence of the probabilistic lambda-calculus $\LNEEDPROB$. In \cref{sec:context-lemma} we prove the context lemma. In \cref{sec:program-transf} we introduce a set of program transformations and prove two criteria to show the correctness of transformations. Most of the correctness proofs are obtained in \cref{sec:diagram} using the diagram method. In \cref{sec:extensions} we discuss extensions of the calculus. We conclude in \cref{sec:conclusion}.
Due to space constraints, details are given in the appendix. Outputs of our automated tools and automated termination proofs can be found via \url{https://p9471.gitlab.io/prob-lneed/}.

\section{\texorpdfstring{The Calculus $\LNEEDPROB$}{The Calculus LNEEDPROB}\label{sec:calculus}}
\subsection{Syntax and Operational Semantics}
We define the syntax of the call-by-need lambda-calculus with a binary, probabilistic operator $\prob{}{}$.
\begin{definition}[Syntax of Expressions and Environments]
Let $\Var$ be an infinite, countable set of variables. We use $x,y,z,x_i,y_i,z_i$ for variables of $\Var$. The syntax of \emph{expressions} $s,t,r\in\Expr$ and \emph{environments} $\env \in \Env$ of the probabilistic call-by-need letrec-calculus $\LNEEDPROB$ 
is given by the following grammar:
\[
\begin{array}{rcl}
s,t,r \in \Expr &::=& x~|~\lambda x.s ~|~ (s~t)~|~(\prob{s}{t})~|~\tletr~\env~\tin~s
\\
\env\in\Env &::=& x_1 = s_1, \ldots, x_n = s_n
\end{array}
\]
\end{definition}
Thus, the syntax consists of the lambda-calculus (variables $x$, abstractions $\lambda x.s$, applications $(s~t)$) extended by recursive $\tletrec$ (in $\tletrec~x_1=s_1,\ldots,x_n=s_n~\tin~t$
the scope of $x_i$ is $s_1,\ldots,s_n,t$) and the probabilistic operator $\prob{s}{t}$ which allows to randomly choose between $s$ and $t$ (both with the same probability of $\frac{1}{2}$). We
abbreviate $\lambda x_1.\lambda x_2.\ldots\lambda x_n.s$ with $\lambda x_1,x_2,\ldots,x_n.s$.
As a side condition, all variables on left-hand sides of the same environment must be pairwise different. The bindings in an environment are treated like a multiset, and thus their order is irrelevant.
The binders $\lambda x$ and bindings $x = s$ in environments induce (as usual) notions of scope, free and bound variables, 
%free and bound occurrences,
$\alpha$-renaming 
and $\alpha$-equivalence $=_\alpha$. We use the distinct variable convention and assume that variables at binders are pairwise disjoint and that free variables are disjoint from bound variables. If needed, the convention can always be kept by %performing 
$\alpha$-renaming. %with fresh variable names. 
For expressions $s,t$ and a variable $x$ we denote with $s[t/x]$ the capture free substitution of all free occurrences of $x$ in $s$ with (freshly $\alpha$-renamed copies of) expression $t$.

At this point, we do not include data like constants or constructors and case-distinctions into the syntax, since this can be simulated and it only would distract from the main focus, which is the treatment of random events in programming. However, we discuss extensions of the calculus in \cref{sec:extensions}.
%% \davidins{}

%\manfredcomment{noch mehr sagen}
%\davidcomment{Ja hier sehe ich es noch als Baustelle, da man Erweiterungen noch später hinzufügen kann.}

We abbreviate some prominent expressions: $\mathit{K}=\lambda x,y.x$, $\mathit{K2}=\lambda x,y.y$, 
$id = \lambda x.x$,
$\omega = \lambda x.(x~x)$,
$\Omega = (\omega~\omega)$,
$\mathit{Bot}=\tletrec~x=x~\tin~x$.

% \subsection{Operational Semantics}
For defining the evaluation and the semantics of expressions, 
we use contexts which are expressions that have exactly one hole
instead of a subexpression. The hole is denoted with $[\cdot]$.
In addition, we introduce surface contexts which are the subset of all contexts that do not have their hole below a $\lambda$-binder.
\begin{definition} \emph{Contexts} $C\in\Ctxt$
and \emph{surface contexts} $S \in \SCtxt$ 
are defined by the following grammars:
$$
\begin{array}{r@{~}c@{~}l}
C \in \Ctxt &::= &[\cdot] ~|~ \lambda x.C ~|~ (C~t) ~|~ (t~C) ~|~ (\prob{C}{t})~|~(\prob{t}{C})\\
&&~|~\tletr~\env~\tin~C~|~\tletrec~\env,y=C~\tin~t
\\
S\in\SCtxt           &::= &[\cdot] ~|~ (S~t) ~|~ (t~S) ~|~ (\prob{S}{t})~|~(\prob{t}{S})\\
&&~|~\tletr~\env~\tin~S~|~\tletrec~\env,y=S~\tin~t
\end{array}
$$
Substituting the hole of $C$ by expression $s$ is written as $C[s]$.
\end{definition}

For defining the evaluation strategy, we use application contexts $A\in\ACtxt$ and reduction contexts $R\in\RCtxt$. The former have the context hole in the function position of applications (and only inside applications). Reduction contexts  extend application contexts
such that 
the application also can be in the $\tin$-expression of a $\tletrec$-expression, or
 in the right-hand-side of a $\tletrec$-binding. In the latter case, the binding must be \emph{needed},~\ie~there is a chain of bindings 
$x_1 = A_1[x_{2}],\ldots,x_n=A_n[x_{n+1}], x_{n+1} = A_{n+1}$ and the start variable of the chain $x_1$ is in the focus of the $\tin$-expression of the $\tletrec$ ( \ie~the \tin-expression is $A[x_1]$).
To ease notation, we abbreviate chains of bindings $x_j = s_{f(j)}, x_{j+1}=s_{f(j+1)},\ldots,x_m=s_{f(m)}$ as $\{x_i = s_{f(i)}\}_{i=j}^m$, where this means an empty environment if $m < j$.
The above chain $x_1 = A_1[x_{2}],\ldots,x_n=A_n[x_{n+1}]$ 
is thus abbreviated by
$\{x_i = A_i[x_{i+1}]\}_{i=1}^n$.
This results in the following definition:
\begin{definition}\label{def:contexts-A-und-S-und-R}
 \emph{Application contexts} $A \in \ACtxt$ and reduction contexts $R \in \RCtxt$ are defined by the following grammars:
$$
\begin{array}{r@{~}c@{~}l}
A\in\ACtxt           &::= & [\cdot] ~|~ (A~s) 
\\
R\in\RCtxt          &::=& A ~|~ \tletr~\env~\tin~A\\
&&~|~ \tletr~\env,
\{x_i = A_i[x_{i+1}]\}_{i=1}^n, x_{n+1} = A_{n+1}~\tin~A[x_1]
\end{array}
$$
\end{definition}
Note that the inclusions $\ACtxt \subseteq \RCtxt \subseteq \SCtxt \subseteq \Ctxt$ hold.

In the following we define reduction, evaluation, and finally contextual equivalence also for open expressions, since this is more general than the restriction to closed expressions (see \cite{schmidt-schauss-schuetz-sabel:08,schmidt-schauss-sabel-gencontext:10}).

The standard (call-by-need) reduction of the calculus $\LNEEDPROB$  defines the operational semantics where pure lambda expressions are evaluated using lazy evaluation and sharing (with $\tletrec$-bindings). The probabilistic operator is evaluated non-deterministically by choosing the left or the right expression.
\begin{definition}
The \emph{standard reduction} $\xrightarrow{\sr}$ of $\LNEEDPROB$ is defined as the union of the steps $\xrightarrow{\sr,\lbeta}$, $\xrightarrow{\sr,\lapp}$, $\xrightarrow{\sr,\cp}$ (which is the union of $\xrightarrow{\sr,\cpin}$ and $\xrightarrow{\sr,\cpe}$),
$\xrightarrow{\sr,llet}$  (which is the union of $\xrightarrow{\sr,\lletin}$ and $\xrightarrow{\sr,\llete}$), and 
$\xrightarrow{\sr,\problab}$ (which is the union of $\xrightarrow{\sr,\problabl}$ and $\xrightarrow{\sr,\problabr}$), where the rules $\xrightarrow{\sr,a}$ are defined in \cref{fig-sr-rules} with label $(\sr,a)$.

The transitive closure of $\xrightarrow{\sr}$ is denoted with $\xrightarrow{\sr,+}$ and the reflexive-transitive closure is denoted with $\xrightarrow{\sr,*}$. With $\xrightarrow{\sr,a \vee b}$ we denote the union of $\xrightarrow{\sr,a}$ and $\xrightarrow{\sr,b}$.
\end{definition}
\begin{figure*}[tpb]
$\begin{array}{@{}l@{\qquad\quad}l@{}}
\begin{array}{@{}l@{~}l}
(sr,\lbeta) &R[((\lambda x.s)~t)] \\&\to R[\tletr~x=t~\tin~s]
\\                 
(sr,\problabl)&R[\prob{s}{t}] \to R[s]
\\                 
(sr,\problabr)&R[\prob{s}{t}] \to R[t]
\\                 
(sr,\lapp) 
&R[((\tletr~\env~\tin~s)~t)] 
\\&\to R[\tletr~\env~\tin~(s~t)]
\\
(sr,\lletin)&
\tletr~\env_1~\tin~\tletr~\env_2~\tin~s
\\&   \to
   \tletr~\env_1,\env_2~\tin~s
\\
\end{array}&\begin{array}{@{}l@{~}l@{}}
(sr,\cpin)&
\tletr~\{x_i{=}x_{i+1}\}_{i=1}^{n-1},x_{n} =\lambda y.s, \env~\tin~A[x_1]
\\
&\to  
                     \tletr~\{x_i = x_{i+1}\}_{i=1}^{n-1},x_n=\lambda y.s, \env~\tin~A[\lambda y.s]
\\                 
(sr,\cpe)&\begin{array}[t]{@{}l@{}}
\tletr~\{x_i=A_i[x_{i+1}]\}_{i=1}^{n-1},x_n = A_n[y_1], \{y_j =y_{j+1}\}_{j=1}^{m-1}, y_m = \lambda z.s,\env~\tin~A[x_1]
\end{array}
\\
&
\to  \begin{array}[t]{@{}l@{}}
\tletr~\{x_i=A_i[x_{i+1}]\}_{i=1}^{n-1},x_n = A_n[\lambda z.s], \{y_j =y_{j+1}\}_{j=1}^{m-1}, y_m = \lambda z.s,\env~\tin~A[x_1]
     \end{array}
\\& \text{ where $A_n \not= [\cdot],n {\geq} 1, m {\geq} 1$}
\\                 
(sr,\llete)&
\tletr~\{x_i = A_i[x_{i+1}]\}_{i=1}^{n-1},x_n = (\tletr~\env_1~\tin~s), \env_2~\tin~A[x_1]
\\&\to
\tletr~\{x_i = A_i[x_{i+1}]\}_{i=1}^{n-1},x_n = s,\env_1,\env_2~\tin~A[x_1]
\\
\end{array}
\end{array}
$
\caption{Standard Reduction-Rules\label{fig-sr-rules}}
\end{figure*}

% \davidins{..} 
Reduction rule $(sr,\lbeta)$ is the sharing-variant of $\beta$-reduction where the argument is shared by a new binding.
% \longversion{It evaluates the application of an abstraction to an argument by sharing the argument with a new $\tletrec$-binding instead of substituting the formal parameter by copies of the it.}
Rules $(sr,\cpin)$ and $(sr,\cpe)$ inline a binding,
% the delayes
%such a 
if it is needed 
% if the expression that is bound by a $\tletrec$-binding
% is 
%i) 
% needed 
%(the variable occurs in needed position), 
and already evaluated (and hence is an abstraction).
%ii) 
% the expression is already evaluated and thus it is an abstraction.
Rules $(sr,\lapp)$, $(sr,\lletin)$, $(sr,\llete)$ rearrange $\tletrec$-enviroments w.r.t.~applications and nesting of $\tletrec$-expressions. Rules $(\sr,\problabl)$ and $(\sr,\problabr)$ evaluate $\prob{}{}$-expression by choosing either the left or the right argument. We call these two rules also \emph{prob-reductions} and all other reductions are called \emph{non-prob-reductions}.

As usual in lazy functional programming languages, successfully evaluated expressions are identified with 
%standard reductions will evaluate expressions to 
weak head normal forms. 
%Thus they form the set of successfully evaluated expressions. 
\begin{definition}[Weak Head Normal Form, Evaluation]
Let $s$ be an expression. Then $s$ is a \emph{weak head normal form} (WHNF) if it is an abstraction, or of the form $(\tletr~\env~\tin~\lambda x.s')$.
%%  a $\tletrec$-expression where the $\tin$-expression is an abstraction. 
A sequence of reductions $s\xrightarrow{\sr,*}t$ where $t$ is a WHNF is called an \emph{evaluation} of $s$. We write $\Eval(s)$ for the set of all evaluations of expression $s$.
\end{definition}
\begin{example}\label{ex:red}
% \davidcomment{Beispiele kürzen und/oder in den Anhang schieben}
Let $a,b,c,d$ be different abstractions (for instance, $a {=} s_1, b{=}s_2, c{=}s_3, d{=}s_4$ where $s_i{=}\lambda x_1,x_2,x_3,x_4.x_i$).  There are two evaluations of the expression  
    $
    \tletrec\,z{=}\prob{\mathit{K}}{\mathit{K2}}\,\tin\,(z\,(z\,a\,b)\,(z\,c\,d)).
$
Due to sharing only $a$ and $d$ are possible results (plus some additional environment which is garbage), i.e. ~the evaluations end with
$\tletrec\,z{=}\mathit{K},\,x{=}x_1,y_1{=}b,x_1{=}a, y{=}(z\,c\,d)\,\tin\,a$
and $\tletrec\,z{=}\mathit{K2},\,x{=}(z\,a\,b), y{=}y_1,y_1{=}d,x_1{=}c\,\tin\,d$
resp. (the complete reduction sequences can be found in the appendix, \cref{ex:eval:app}).

However, abstractions are not shared, but copied and thus if we shift the $\oplus$-operator under the $\lambda$, we get the expression 
 $$\tletrec\,z{=}\lambda x.\lambda y.\prob{x}{y}\,\tin\,z\,(z\,a\,b)\,(z\,c\,d)$$
which now has four evaluations (see  in the appendix, \cref{ex:eval:app})
ending with all four possibilities:
$$\begin{array}{@{\bullet~}l}
\tletrec\,z{=}\lambda x.\lambda y.\prob{x}{y},x_1{=}a,y_1{=}b,x{=}x_1,y{=}(z\,c\,d)\,\tin\,a
\\
\tletrec\,z{=}\lambda x.\lambda y.\prob{x}{y},x_1{=}a,y_1{=}b,x{=}y_1,y{=}(z\,c\,d)\,\tin\,b
\\
\tletrec\,z{=}\lambda x.\lambda y.\prob{x}{y},x{=}(z\,a\,b),x_1{=}c,y_1{=}d,y{=}x_1\,\tin\,c
\\
\tletrec\,z{=}\lambda x.\lambda y.\prob{x}{y},x{=}(z\,a\,b),x_1{=}c,y_1{=}d,y{=}y_1\,\tin\,d
 \end{array}
$$

\end{example}

%\davidins{..}  gecheckt.
By inspecting the definition of reduction contexts and the standard reduction rules the following lemma can be verified:
\begin{lemma}
For every expression $s$, there is either no standard reduction applicable (if $s$ is a WHNF, or $s$ is of the form $R[x]$ where the shown occurrence of $x$ is free), or there is exactly one standard reduction applicable (which is not a prob-reduction), or
a $\xrightarrow{\sr,\problabl}$ and a $\xrightarrow{\sr,\problabr}$-reduction are applicable,
where the placed $\prob{}{}$-expression is the same for both reductions.
Thus the redex of the standard reduction is unique and standard reduction is deterministic up to $\problab$-reductions. 
\end{lemma}

\begin{corollary}
Each evaluation $s \xrightarrow{\sr,a_1} \cdots \xrightarrow{\sr,a_n} s'$ of an expression $s$ is uniquely determined by $s$ and the subsequence of $a_1,\ldots,a_n$ where all labels that are not $\problabl$ nor $\problabr$ are removed.
\end{corollary}

\begin{corollary}
For every expression $s$, every evaluation of $s$ is a finite sequence, and the set of all evaluations of $s$ is countable. 
\end{corollary}

\begin{definition}
For a reduction sequence $s \xrightarrow{\sr,a_1}\cdots \xrightarrow{\sr,a_n} t$, with $\PS(s \xrightarrow{\sr,a_1}\cdots \xrightarrow{\sr,a_n} t)$ we denote the subsequence of labels $a_i\ldots,a_k$ which is derived from the sequence $a_1,\ldots,a_n$ after removing all $a_j$ with $a_j \not\in\{\problabl,\problabr\}$.
We call the sequence the \emph{prob-sequence} of reduction sequence.

To identify a single evaluation $s \xrightarrow{\sr,*} t$ in the set $\Eval(s)$, we write $s \evaluation_L t \in \Eval(s)$, where $L = \PS(s \xrightarrow{\sr,*} t)$.
 \end{definition}
 %%}

\subsection{Contextual Equivalence}

The defined operational semantics does not track the probability of different events where an event is a single evaluation together with the WHNF at the end of the evaluation. We now define a weighted reduction
which keeps track of the probability.
\begin{definition}[Weighted Expressions and Reduction]
A \emph{weighted expression} is a pair $(p,s)$ where $p \in (0,1]$ is a rational
number and $s$ is an $\LNEEDPROB$-expression.
% \todo[inline]{It ``calculates'' part of the multidistribution that is represented by the expression (we will define this later).}
% 
Let $(p,s)$ be a weighted expression. A \emph{weighted standard reduction step} $\xrightarrow{wsr}$ (or $\xrightarrow{wsr,a}$ to make the rule explicit) on $(p,s)$ is defined as
follows:
\begin{itemize}
 \item $(p,s) \xrightarrow{wsr,a} (p,t)$ iff $s \xrightarrow{\sr,a} t$ and $a \not\in \{\problabl,\problabr\}$
 \item $(p,s) \xrightarrow{wsr,a} \left(\frac{p}{2},t\right)$ iff $s \xrightarrow{\sr,a} t$ and $a \in \{\problabl,\problabr\}$
\end{itemize}
Again we use $\xrightarrow{wsr,*}$ and $\xrightarrow{wsr,+}$ for the reflexive-transitive, or transitive closure of $\xrightarrow{wsr}$. 
\end{definition}
An evaluation of a weighted expression $(p,s)$ is a sequence $(p,s)\xrightarrow{wsr,*}(q,t)$ where $t$ is a WHNF.
%We will use these notations for evaluations of weighted expressions $(p,e)$ and for weighted standard reduction. 
Clearly, the evaluations of a weighted expression are  countable. Again the sequence of labels $a$ for each prob-reduction $\xrightarrow{wsr,a}$ together with the weighted expression $(p,s)$ uniquely identifies an evaluation of $(p,s)$).
\begin{definition}[Expected Convergence]
For a weighted expression $(p,s)$ we denote with $\Eval(p,s)$ the set of evaluations of $(p,s)$.
A single evaluation in this set is notated as $(p,s)\evaluation_L(q,t)$,
i.e. if $(p,s)\evaluation_L(q,t)\in\Eval(p,s)$ then there is an evaluation $(p,s)\xrightarrow{wsr,*} (q,t)$
where $(q,t)$ is the resulting weighted expression, $t$ is a WHNF,
and $L = \PS(s\xrightarrow{\sr,*} t)$.
The \emph{expected convergence} $\EC(p,s)$ of a weighted expression $(p,s)$ 
%% is the \davidins{the real number that is the} limit of the series $\sum\limits_{(p,e)\evaluation(p',e')\in\Eval(p,e)} p'$, 
is the 
(perhaps infinite) sum $\EC(p,s)
=\sum_{(p,s)\evaluation_L(q,t)\in\Eval(p,s)} q$, and the \emph{expected convergence $\EC(s)$ of an expression} $s$ is $\EC(s)=\EC(1,s)$.
For $s$ with $\EC(s) = q$, we also write $s \downarrow_q$.
%If the expected convergence of expression $e$ is $q \in [0,1]$, then we write $e \downarrow_q$.
\end{definition}

%\lucacomment{Muss die Summe unendlich sein? Im Fall von bspw. (1,\prob{A}{B}) ist die Summe doch tatsächlich über 2 Elemente oder?}
%\davidcomment{Ja stimmt, nicht immer}

%% \manfredinsert Eine andere Variante.. \\
\begin{proposition}\label{prop:weighted-exp-formula}
 Expected convergence 
 %of weighted expressions $(p,s)$
 is well-defined, i.e.~for $(p,s)$,~the limit $\sum_{(p,s)\evaluation_L(q,t)\in\Eval(p,s)} q$ always exists and is unique. In particular, its value is independent
 of the enumeration of the countable set $\Eval(p,s)$.
 In addition, $\EC(p,s) = p\cdot\EC(s)$.
 \end{proposition}
 \begin{proof}
 This holds, since $\Eval(p,s)$ is enumerable, all $q$ with $(p,s)\evaluation_L(q,t)\in \Eval(p,s)$ are non-negative, and since in case of an infinite set 
any partial sum is bounded by 1.
In fact, a stronger property holds:~for all finite $Q \subseteq \Eval(p,s)$: $\sum_{(p,s)\evaluation_L(q,t)\in Q} q \leq 1$.
Thus the series converges absolutely, and  the limit always exists and is unique.
The equation $\EC(p,s) = p{\cdot}\EC(s)$ follows, since $p$ can be multiplied into the summands of the (infinite) sum. 
 \end{proof}

% Note that the limit always exists \manfredinsert{and it is unique, i.e. independent from the sequence of summation}: 
%if $\Eval(1,e)$ is finite, then this is clear. 
%Otherwise, all elements of the series are positive
%and any partial sum is bounded by 1
%(i.e.~for all finite $Q \subseteq \Eval(p,e)$: 
%$\sum\limits_{\evaluation(p',e')\in Q} p' \leq 1$).
%Thus the monotone convergence criterion  for series shows that the limit always exists.
%\manfredcomment{converges absolutely?}
 
%\manfredinsert{Since all contributions to the sum are non-negative, it also holds that the sequence of the summation does not influence the result,
%where the modified sequence must not delay summands infinitely.}

%\davidcomment{Hab es z.T. oben eingebaut, das ``delay summands infinitely'' versteh ich nicht: Ist irgendwie der Unterschied zwischen abzählbar und aufzählbar? Für die Definition braucht man m.E. nur abzählbar damit die Reihe
%(bzw. die Folge der Partialsummen) definiert ist.?} 
%
%\manfredcomment{Es ist schon ok. Mathematisch starttet man mit  Summen i = 1,2,3,...  Die Definition ist aber Summe q  ueber eine Menge, wobei die Summanden positiv und die Menge abzaehlbar ist. Also braucht man nur einmal kurz argumetieren, dass man das machen darf.   }
% Ok

\begin{example}
 The expression
 $\tletrec~x = (\lambda y.\prob{(x~id)}{\mathit{K}})~\tin~(x~id)$
 converges with expectation of 1.
Both expressions $\Omega := (\lambda x.(x~x))~(\lambda x.(x~x))$ and $\mathit{Bot} := \tletrec~x=x~\tin~x$ converge with expectation of 0 (and thus they do not converge).
The expression  $p = \tletrec~x=\prob{(\lambda y.y)}{(\prob{\Omega}{x})}~\tin~x$ has chances to converge and diverge. We obtain $\EC(p) = 1/2$.  
\end{example} 
%\davidcomment{Weitere hinzufügen und noch ausformulieren}

%\begin{remark}
We compare the expected convergence  with convergence tests for 
%We recall convergence predicates for
non-deterministic calculi 
%and analyze their relation to expected convergence
(see e.g.~\cite{schmidt-schauss-sabel-ipl:2010} for an overview\longversion{ and further pointers on may-, must- and should-convergence and the induced contextual equivalences}).
%\begin{description}
%\item[\normalfont\itshape May-Convergence.] 
An expression $s$ may-converges iff there exists a reduction sequence from $s$ to a WHNF. Thus this is exactly the same as, that there exists an evaluation. In terms of expected convergence: an expression $s$ is may-convergent iff $(1,s)\downarrow_q$ with $q > 0$.
The negation of may-convergence is \emph{must-divergence}. Thus we also know: an expression $s$ is must-divergent iff $(1,s)\downarrow_0$. Clearly, if $s$ is must-divergent, then $\EC(p,s) = 0$ for every $p\in[0,1]$.   %\davidins{}
%\item[\normalfont\itshape Must-Convergence.] 
An expression $s$ must-converges iff any reduction sequence starting from $s$ is finite and ends with a WHNF. 
Expression $s$ being must-convergent is not the same as $(1,s)\downarrow_1$:  there are expressions that converge with expectation 1, but are not must-convergent, e.g.~$\tletrec~x = (\lambda y.\prob{(x~id)}{\mathit{K}})~\tin~(x~id)$.
However, 
%must-convergence implies convergence with expectation of 1.
%or more general: 
if $s$ is must-convergent then $(p,s)\downarrow_p$ for all $p$. 
%\item[\normalfont\itshape Should-Convergence.] 

An expression $s$ is should-convergent iff for any $t$ with $s \xrightarrow{\sr,*} t$, the expression $t$ is may-convergent.
Should-convergence does not imply convergence with expectation of 1, since there are should-convergent expressions where the expected convergence is strictly smaller than 1. 
An example is the should-convergent expression
$$\begin{array}{@{}l@{~}l@{}l}
s:=&\tletrec~&cprob= \lambda i. \mathit{if}\,i=0\,\mathit{then}\,K\,\mathit{else}\,
\lambda x,y.\prob{(cprob~(i{-}1)~x~y)}{y},\\
   &      &gen = \lambda i.cprob~i~K~(gen~(i{+}1))\\
&\multicolumn{2}{l}{\tin~gen~2}
\end{array} 
$$
where numbers, and $\mathit{if}{-}\mathit{then}{-}\mathit{else}$ have to be encoded using Church numerals. 
We illustrate the executions of  $cprob~i~s_1~s_2$ and $gen~2$:
$$\begin{array}{ll}
{cprob~i~s_1~s_2:} & {gen~2:} \\
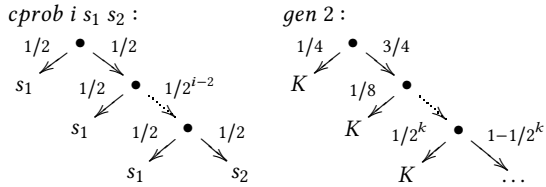
\begin{minipage}{0.45\columnwidth}
\xymatrix@C=3mm@R=2mm{
 & \bullet \ar[dl]_{1/2}  \ar[dr]^{1/2} \\
s_1 &  &  \bullet \ar[dl]_{1/2}  \ar@{.>}[dr]^{1/{2^{i-2}}} \\
 & s_1 &  &  \bullet \ar[dl]_{1/2}  \ar[dr]^{1/2} \\
 &  &s_1& & s_2
 }
 \end{minipage}
 & 
\begin{minipage}{0.45\columnwidth}
\xymatrix@C=3mm@R=2mm{
 & \bullet \ar[dl]_{1/4}  \ar[dr]^{3/4} \\
 K &  &  \bullet \ar[dl]_{1/8}  \ar@{.>}[dr]^{} \\
 & K && \bullet \ar[dl]_{1/{2^k}}  \ar@{->}[dr]^{1-1/{2^k}} \\
 & & K && \ldots
 }
\end{minipage}
\end{array}
$$
%\\\manfredcomment{Evtl eins spiegeln, damit es noch besser passt?
%}
In $cprob~i~s_1~s_2$, $s_2$ is reached with probability $1/{2^{i}}$ and $s_1$ with probability $1-1/{2^i}$:
Hence $(1,s)\downarrow_q$ with $q =  \displaystyle\sum_{i=2}^\infty\frac{1}{(1-1/{2^i})*2^i}$ which is smaller than $\displaystyle\sum_{i=2}^\infty\frac{1}{2^i} = 0.5$ and greater than 
$\displaystyle\sum_{i=2}^\infty\frac{1}{(3/4)*2^i} = 3/8 = 0,375$ 
hence $0.375 < q < 0.5$. The exact sum is easily computed using geometric sums to $q = 1/2-1/12 = 5/12$.

Since should-convergence implies may-convergence, should-convergence of $s$ implies $(1,s)\downarrow_q$ with $q > 0$.

%\end{description}
%\end{remark}

% \manfredinsert{ }
\begin{proposition}
If $(1,s) \downarrow_1$, then $s$ is should-convergent.
\end{proposition}
\begin{proof}
%Let us assume that for all expressions $e$, it holds that $e\downarrow_1$ (i.e.$\EC(1,e)=1)$. We show that this implies should-convergence. 
Suppose that for some $s$ with $\EC(s) = 1$,  $s$ is not should-convergent. Then there is some must-divergent $t$ with $s \xrightarrow{\sr,*} t$. 
But then $(1,s) \xrightarrow{\sr,*} (p,t)$ for some $p > 0$. Since $\EC(p,t)=0$ due to must-divergence of $t$, this 
% and with non-zero probability. This 
contradicts  the assumption that 
$\EC(s) = 1$.
\end{proof}

% \subsection{Contextual Equivalence}

Expression $t$ contextually approximates $s$ if whenever we replace $s$ (as a subprogram) by $t$, then the expected convergence is not decreased. If $s$ approximates $t$ and $t$ approximates $s$,
 then $s$ and $t$ are contextually equivalent:

\begin{definition}\label{def:contextual-equivalence}
\emph{Contextual approximation} $\leq_c$ on expressions of $\LNEEDPROB$ is defined as follows. 
For $s,t\in\Expr$, $s \leq_c t$ holds iff for all contexts $C$:  $C[s] \downarrow_q \implies {{C[t]\downarrow_{q'}} \wedge {q \leq q'}}$.  %%{q' \geq q}, 
\emph{Contextual equivalence} $\sim_c$ is the symmetrization of $\leq_c$, i.e. 
$s \sim_c t$ iff $s \leq_c t$ and $t \leq_c s$.
\end{definition}

The restriction to start with weight 1, is no real restriction in the definition of contextual equivalence (see also   \cref{prop:weighted-exp-formula}):
\begin{remark}
The inequation $s \leq_c t$ holds iff $\forall p\in(0,1] ,C\in\Ctxt:\EC(p,C[s]) \leq \EC(p,C[t])$.
The equation $s \sim_c t$ holds iff $\forall p\in[0,1] ,C\in\Ctxt:\EC(p,C[s]) = \EC(p,C[t])$.
\end{remark}
 
\begin{lemma}
Contextual approximation is a precongruence, and contextual equivalence is a congruence.
In addition, $s \sim_c t$ implies that
%$s\downarrow_q \iff  t\downarrow_q$.
%
% \davidins{..}
$C[s]\downarrow_q \iff C[t]\downarrow_q$ for any context $C$.
\end{lemma}
\begin{proof}
We show that $\leq_c$ is a precongruence. The other part then follows obviously. We have to show that $\leq_c$ is a preorder that is compatible with contexts. The relation $\leq_c$ is obviously reflexive ($s \leq_c s$).
For transitivity, let $r \leq_c s$ and $s \leq_c t$,
and $C$ be a context with $C[r]\downarrow_p$. From $r \leq_c s$ we have that $C[s]\downarrow_{p'}$ with $p' \geq p$. Now $s \leq_c t$ implies $C[t]\downarrow_{p''}$ with $p'' \geq p'$. Thus, in conclusion this shows $C[t]\downarrow_{p''}$ with $p'' \geq p$ and thus $r \leq_c t$.
For proving compatibility with contexts, let $s \leq_c t$ and $C$ be a context. We have to show that for every context $C'$, we have $C'[C[s]]\downarrow_p \implies C'[C[t]]\downarrow_{p'} \wedge p' \geq p$. But this follows from $s \leq_c t$ since $C'[C]$ is also a context.
\end{proof}

% \subsubsection{Examples}
% 
\begin{example}
The inequations 
\begin{inparaenum}
\item $\prob{K}{K2} \not\sim_c K$,
\item $\prob{(\prob{K}{K2})}{K2}\not\sim_c\prob{K}{K2}$, and
\item $\prob{K}{K2}\not\sim_c K2$
\end{inparaenum}
hold. The first inequation can be proved by the context $C = ([\cdot]~id~\mathit{Bot})$:
Then $C[K]\downarrow_1$, but $C[\prob{K}{K2}]\downarrow_{0.5}$ and thus $K \not\leq_c \prob{K}{K_2}$.
The second inequation can be proved by the same context: $C[\prob{(\prob{K}{K2})}{K2}]\downarrow_{0.25}$ and $0.25 < 0.5$.
Note that $\prob{(\prob{K}{K2})}{K2} \leq_c \prob{K}{K2}$ however holds.
%(where more techniques are required to prove it).
The third inequation can be proved using  the context $C=([\cdot]~\mathit{Bot}~id)$.
\end{example}
%% \davidins{..} 
While refuting contextual equivalence is possible by providing a single context as counter-example, proving contextual equivalences requires
to reason about all contexts. Hence, we develop 
techniques to enable such proofs.

% \todo[inline]{hier einfügen, bzw. die Restlichen nochmal anschauen}
% % The following equivalences hold:  %%   \manfredcomment{ loeschen:? should hold:} 
% % $\prob{s}{s} \sim_c s$, $\prob{s}{t} \sim_c \prob{t}{s}$,  $\prob{(\prob{e_1}{e_2})}{e_3} \sim_c \prob{e_1}{(\prob{e_2}{e_3})}$
% %  $\prob{K}{K} \sim_c K$ should hold.

% The expression $\tletrec~x = (\lambda y.\prob{(x~id)}{\mathit{K}})~\tin~(x~id)$ is contextually equivalent to $\mathit{K}$.

% The expressions $\Omega$ and $\bot := \tletrec~x=x~\tin~x$ are contextually equivalent, and they are smallest elements of $\leq_c$. Furthermore, all expressions with expected convergence of 0 are contextually equivalent and they are smallest elements of $\leq_c$.

% \begin{verbatim}
% 3)Für e = let cprob = \i.case i of {Zero -> \x,y.x; Succ j -> \x,y -> prob (cprob j x y) y},
%               -- Kommentar: cprob i x y liefert als x mit Ws 1/2^i und y mit Ws 1 - 1/2^i
%               gen = \i -> cprob i True (gen (Succ i))
%           in gen (Succ (Succ Zero))
% gilt: e \downarrow 0.5  denn True wird erreicht mit Erwartung \sum 1/4+1/8+1/16... = 1/2
% e sollte damit gleich sein zu prob True Omega
% (bei n.d. ohne Wahrscheinlichkeiten wäre e should-konvergent und gleich zu True)
% \end{verbatim}

%\subsubsection{Properties of the Contextual Relations}

\section{Context Lemma\label{sec:context-lemma}}
The goal of this section is to show that 
observing expected convergence in reduction contexts is sufficient to conclude contextual equivalence. Such a result is usually called a context lemma. Our formulation of the context lemma is more special, and we require some preparation to introduce it.
We first introduce
{\em multicontexts}: these are expressions
with several (or no) holes $\cdot_i$, where every hole occurs exactly once.
We write a multicontext as $C[\cdot_1,\ldots,\cdot_n]$,  and if the expressions $s_i$ for $i=1,\ldots,n$ are 
placed into the holes $\cdot_i$, then we denote the resulting expression as $C[s_1,\ldots,s_n]$. 
For a multicontext $C[\cdot_1,\ldots,\cdot_n]$,  a hole $[\cdot_i]$ is a {\em reduction hole}, iff for all expressions $s_j, j = 1,...,n$, the context $C[s_1,\ldots,s_{i-1},[\cdot]_i,s_{i+1},\ldots,s_n]$ is a reduction context.
Note that if for a multicontext $C[\cdot_1,\ldots,\cdot_n]$ and expressions $s_1,\ldots,s_n$, $C[s_1,\ldots,s_{i-1},[\cdot]_i,s_{i+1},\ldots,s_n]$  is a reduction context,
then there exists an index $j$, such that hole $[\cdot_j]$ is a reduction hole of $C[\cdot_1,\ldots,\cdot_n]$. 

\ignore{
\lucacomment{Ist der letzte Satz nicht überflüssig? Folgt das nicht direkt aus der Definition, bzw. ist genau die "Gegenrichtung" der angegebenen Definition?}
\davidcomment{Es kann sein, dass ein Kontextloch  nicht für alle Terme ein Reduktionskontext ist, aber dann gibt es immer ein anderes Loch.
Z.B. $C=\tletrec~x=[\cdot_1]~\tin~[\cdot_2]$: 
Dann ist $C[[\cdot_1],x]$ zwar ein Reduktionskontext, aber 
$C[[\cdot_1],y]$ nicht, aber zum Loch $\cdot_2$ ist immer ein Reduktionsloch (da $C[s,[\cdot_2]]$ immer ein Reduktionskontext ist).
}
Danke, jetzt verstehe ich!
}
% Thus we also say that a hole of $C$ is a reduction context. 

%For a better understanding, we assume that the holes are denoted in the left-to-right ordering.
%\davidcomment{könnte sein, dass die left-to-right-Ordnung im Beweis nicht immer eingehalten wird, wenn man reduziert, daher vielleicht eher nicht nehmen?}

\begin{definition}
%For an expression $s$ let $\EC(s)= q \iff s\downarrow_q$ be the expected convergence of $s$. %\\
The  {\em prob-length}  of an evaluation $(s \evaluation_L t)$, written $\mathit{PL}(s \evaluation_L t)$, is  the number of $\xrightarrow{\sr,\problab}$-reductions in the evaluation,
i.e.~$\mathit{PL}(s \evaluation_L t) = |L|$.
Let $\Eval(p,s,k)$ be the set of evaluations $((p,s) \evaluation_L (q,t)) \in \Eval(p,s)$   with $\mathit{PL}((p,s) \evaluation_L (q,t)) \leq k$. 
We also define $\EC(s,k)$ as the expected convergence of $s$ w.r.t. evaluations of prob-length $\le k$.
\end{definition}
It is obvious that $\displaystyle\lim_{k \to \infty} \EC(s,k) = \EC(s)$.
% DS: irgendwie klar:
%Reduction contexts and others are defined in \cref{def:contexts-A-und-S-und-R}. 
\begin{definition}
For two expressions $s,t$, we define $s \le_{\RCtxt} t$ iff for all reduction contexts $R$: 
$R[s]\downarrow_p \implies (R[t]\downarrow_{q} \wedge p \leq q)$.
%\manfredcomment{ das $p$ ist ja scharf definiert, oder?}
%\davidcomment{$\downarrow_0$ ist erlaubt, aber es macht nix, da jeder Term irgendeine expected convergence in [0,1] hat}
%and all $p \in (0,1]$:
% $R[s]\downarrow_p \implies R[t]\downarrow_{p}$.\\
 A refined definition variant  is:
 for two expressions $s,t$, and a natural number $k$, we define $s \le_{\RCtxt,k} t$ iff for all reduction contexts $R\in\RCtxt$: %%  and all $p \in (0,1]$:
 $\EC(R[s],k) \leq \EC(R[t],k)$. 
 % $\EC(R[s],k) = \EC(R[t],k)$ 
 \end{definition}

% \begin{definition}
% For two expressions $s,t$, we define $s \le_{\RCtxt} t$ iff for all reduction  contexts $R$, and all $p \in [0,1]$:
% $R[s]\downarrow_p \implies  R[t]\downarrow_{p'}$ for some $p' \geq p$.
% \end{definition} 
 
%%   \davidcomment{$p=0$ ist komisch, weil es $R[t]\Uparrow \implies R[s]\Uparrow$ bedeutet (nur als Kommentar...)}
%%  \manfredcomment{So besser?:  hast Recht, das andere war zu eng. nochmal anders. } 
%%\davidcomment{Jein, ich meinte statt $p\in [0,1]$ sollte man evtl. eher $p \in (0,1]$ fordern?}
%%\manfredcomment{Nach Diskussion: verschaerfte  Definition. }
 
%Later we will show that for certain $s,t$ with $s \to t$, we can show that $\EC(R[s],k) = \EC(R[t],k)$ for every $k$ by analysing the
%reduction sequences. 
 % \manfredcomment{ Die  Eigenschaft $\EC(R[s],k) = \EC(R[t],k)$ f   sollte bei der Überlappungsrechnungen sich ergeben!  } 
 
%  vorher: \input{neues-context-lemma}
%%  \input{neues2-context-lemma}
%%  Versuch 3, mit forall k exists d....
%\davidins{..}
The following lemma shows that preservation
of expected convergence for any bound on the prob-length implies preservation of expected convergence.
\ignore{ 
\begin{lemma}\label{lemma-limit-plus-d}
Let $s,t$ be expressions and $d \geq 0$ 
such that $\forall k \geq 0: \EC(s,k) \leq \EC(t,k+d)$.
Then $\EC(s) \leq \EC(t)$.
\end{lemma}
 %%  \ignore{% im Anhang
\begin{proof}
For real-numbers $m \in \mathbb{R}$ and $\varepsilon>0$,
we denote the \emph{$\varepsilon$-neighborhood} of $m$ 
with $N_\varepsilon(m)$ (i.e.
$N_\varepsilon(m) = \{p \in \mathbb{R}\mid |m - p|  < \varepsilon\}$).

Assume the claim of the lemma is false, i.e. let $s,t$ and $d \geq 0$ be given such that $\forall k \geq 0: \EC(s,k) \leq \EC(t,k+d)$,
but $\EC(s) > \EC(t)$.
Note that $\lim_{k\to\infty}(\EC(s,k)) = \EC(s)$
and $\lim_{k\to\infty}(\EC(t,k)) = \EC(t)$.

Let $\delta = \EC(s) - \EC(t)$ (note that $\delta > 0$)
and let  $\varepsilon = \frac{\delta}{3}$.
From $\EC(s) > \EC(t)$ we have $\EC(t)+\varepsilon < \EC(s) - \varepsilon$.
Since $\EC(s) =\lim_{k\to\infty}(\EC(s,k))$ and
$\EC(t) = \lim_{k\to\infty}(\EC(t,k))$, and the limits exist,
there exist numbers $k_0$, $k_1$ such that
\begin{align}
\forall k > k_0: \EC(s,k) \in N_\varepsilon(\EC(s))\\
\forall k > k_1: \EC(t,k) \in N_\varepsilon(\EC(t))\label{eq2}
\end{align}
\Cref{eq2}  also implies
%  Equation \ref{eq2}  also implies
$\forall k > k_1: \EC(t,k+d) \in N_\varepsilon(\EC(t))$

This shows for all $K > \max(k_0,k_1)$:
$\EC(t,K+d) < \EC(t) + \varepsilon < \EC(s) - \varepsilon < \EC(s,K)$
and thus $\EC(t,K+d) < \EC(s,K)$ which is a contradiction.
\end{proof}
 \endignore}
 % \manfredcomment{Ich sehe es, aber es ist gerad zu schwer fuer mich... Brauche etwas Zeit dafür. Soll Deine Aenderung mathematische Fehler korrigieren  oder was anderes?
 % Ich muss erst verstehen... bevor ich da etwas abaendere oder einkommentiere
%  }
% \input{lemma-3-3-strich.tex}
\begin{lemma}\label{lemma-limit-plus-d}
Let $s,t$ be expressions and 
such that $\forall k \geq 0: \exists d\geq 0: \EC(s,k) \leq \EC(t,k+d)$.
Then $\EC(s) \leq \EC(t)$.
\end{lemma}
\begin{proof}
For real-numbers $m \in \mathbb{R}$ and $\varepsilon>0$,
we denote the \emph{$\varepsilon$-neighborhood} of $m$ 
with $N_\varepsilon(m)$ (i.e.
$N_\varepsilon(m) = \{p \in \mathbb{R}\mid |m - p|  < \varepsilon\}$).

Assume the claim %of the lemma 
is false, i.e. let $s,t$  be given, and $\forall k \geq 0: \exists d\geq 0: \EC(s,k) \leq \EC(t,k+d)$,
but $\EC(s) > \EC(t)$.
Note that $\lim_{k\to\infty}(\EC(s,k)) = \EC(s)$
and $\lim_{k\to\infty}(\EC(t,k)) = \EC(t)$.
Let $\delta = \EC(s) - \EC(t)$ %(note that $\delta > 0$)
and let  $\varepsilon = \frac{\delta}{3}$.
From $\EC(s) > \EC(t)$ we have $\EC(t)+\varepsilon < \EC(s) - \varepsilon$.
Since $\EC(s) =\lim_{k\to\infty}(\EC(s,k))$ and
$\EC(t) = \lim_{k\to\infty}(\EC(t,k))$, and the limits exist,
there exist numbers $k_0$, $k_1$ such that
\begin{align}
\forall k > k_0: \EC(s,k) \in N_\varepsilon(\EC(s))\\
\forall k > k_1: \EC(t,k) \in N_\varepsilon(\EC(t))\label{33strich-eq2}
\end{align}
\Cref{33strich-eq2}  also implies for all $d \geq 0$:
%  Equation \ref{eq2}  also implies
$\forall k > k_1: \EC(t,k+d) \in N_\varepsilon(\EC(t))$
This shows for all $K > \max(k_0,k_1)$:
$\EC(t,K+d) < \EC(t) + \varepsilon < \EC(s) - \varepsilon < \EC(s,K)$
and thus $\EC(t,K+d) < \EC(s,K)$ which is a contradiction.
\end{proof}
\ignore{\davidcomment{Kein richtiger Fehler, da die Formeln vorher und nachher äquivalent sind. Ich finde nur die Formulierung schwer verständlich weil man z.B. die "`Precondition"' nicht als solche erkennt. 

Die jetztige Formulierung ist von der Form
$$\exists d,n:\forall C: Precondition \implies ExCV(C...) \leq ExCV(C...)$$
wobei $C$ aber gar nicht in der Preconditon vorkommt und das $\implies$ nicht explizit da steht (auch nicht mit if-then)

Die neue Formulierung ist:

$$\exists d,n:Precondition \implies \forall C: ExCV(C...) \leq ExCV(C...)$$

Im Beweis, wird dann aber 
$\exists D:ExCV(C...,k) \leq ExCV(C,...k+D)$ gezeigt.
Bei Formulierung I müsste man den 
Allquantor und die Precondition wiederholen
$\exists d,n:\forall C: Precondition \implies \exists D:ExCV(C...,k) \leq ExCV(C,...k+D)$ gezeigt.

Bei der anderen kann m.E. direkt anfangen mit $\exists D:ExCV(C,...k) \leq ExCV(C,...,k+D)$

============

Das rote ganz am Ende ist nur Platz-optimierung, damit die Formeln kleiner wurden.
}
}

%\manfredcomment{
%d,n:   d ist eigentlich auch forall, aber ich dachte es ist fix in der Induktion. Deswegen hatte ich es außen gelassen. Das $n$ wird aber in der Induktion benutzt und für kleinere $n$ was angenommen. 
%Deswegen sollte es mE 
%in der Induktionsbehauptung drin stehen. Das $C$ verschieben ist wohl egal.
%Sollen wir die Induktionsstellen extra markieren mit Schluesselwort "by induction"?  erstmal nur fuer uns.  Reine Vorsicht.   Die Induktionn ist in meinen Augen "knapp". 
%}

% \davidins{..}
We are now ready to prove the context lemma for $\LNEEDPROB$. 

\begin{theorem}[Context Lemma]\label{thm:context-lemma-ext}

Let $n\ge 0$, and for $1 \leq i \leq n$, let $s_i,t_i$ be expressions such that 
$\forall k \geq 0$, 
$\forall R\in\RCtxt$, 
there exists $d\geq 0:$ 
 $\EC(R[s_i],k) \leq \EC(R[t_i],k+d)$. Let $C$ be a multicontext with $n$ holes. Then  the inequation
 $\EC(C[s_1,\ldots,s_n]) \leq \EC(C[t_1,\ldots,t_n])$
holds.
\end{theorem}
%\manfredcomment{Die Induktionsbehauptung besser voll hinschreiben: Versuch:}
\begin{proof}
% We first show the following:\\
% There exists $D \geq 0$ such that
%  $\forall K\geq 0:\EC(C[s_1,\ldots,s_n],K) \leq \EC(C[t_1,\ldots,t_n],K+D)$.
%\davidcomment{Hier macht es mit der anderen Formulierung mehr Sinn, ansonsten müsste man nochmal alles wiederholen $\forall C... \implies ...$} %\\
%\manfredinsert{
We first show:
\begin{equation}
\begin{array}{c}
 \multicolumn{1}{l}{\forall K \geq 0: \exists D \geq 0 \text{ such that: }}\\
 \multicolumn{1}{r}{\qquad\EC(C[s_1,\ldots,s_n],K) \leq \EC(C[t_1,\ldots,t_n],K+D)}
\end{array} \label{clemma}
\end{equation}
  
%\davidcomment{grosse D genommen, da im Beweis auch gross und ein anderes D als das d aus der Precondition} 
% \manfredcomment{Stimmt.  Das kleine d muss im Beweis dann aber ersetzt werden durch D''. Da das kleine d keine Rolle spielt in der Induktion.}
%\davidcomment{nein, das $D''$ muss (nach aktueller Formulierung) in 2(b) doch das $d$ sein, da es aus der Precondition der Aussage kommt (mit $k = K+D'$ und $R = C[t_1,\ldots,t_{i-1},[\cdot]_i,t_{i+1},\ldots,t_n]$)}
We use induction on the lexicographically ordered triple:
\begin{enumerate}[left=5pt]
   \item[(i)] the number $K$.
    \item[(ii)]  the maximal length of the evaluations (with at most $K$
     prob-reduction steps) of $C[s_1,\ldots,s_n]$, where the measure is 0 if no such evaluation exists.\\
     (Note that the measure is well-defined, since the set of evaluations with at most $K$ prob-steps is finite.)
    \item[(iii)] the number $n$ of holes of the multicontext $C$.
\end{enumerate}

%% Ist ok so, und klarer.
%  Ist Induktion: deswegen wird entlang der Ordnung bewiesen..
%A ssume the precondition holds for 
%  $n,C,s_i,t_i,k (i=1,\ldots,n)$.
%\davidcomment{
%statt
%$C,s_i,t_i (i=1,\ldots,n)$
%eher
%$d,n,s_i,t_i (i=1,\ldots,n)$?}

As a base case, assume that $K$ is arbitrary, but $C$ has no holes. Then the claim obviously holds.

Now assume that $C$ has at least one hole.
We consider evaluations of $C[s_1,\ldots,s_n]$ and take only evaluations into account that
have $k \leq K$ prob-reductions. We distinguish two cases:

%\manfredcomment{ Die Stellen mit $\forall k$: nochmal anschauen:
%sieht aber alles ok aus.} 

\noindent\textit{Case 1:}
% \begin{enumerate}[left=0pt]
    % \item 
     The maximal length of the evaluations of
     $C[s_1,\ldots,s_n]$ that use at most $K$ prob-reductions is 0. Then either no such evaluation exists, or $C[s_1,\ldots,s_n]$ is a WHNF:
     \begin{enumerate}[labelsep=1ex,itemindent=10pt,left=0pt]
         \item   If $C[s_1,\ldots,s_n]$ is not a WHNF, then $\EC(C[s_1,\ldots,s_n],K) = 0$ and
         the claim holds.
         \item If $C[s_1,\ldots,s_n]$ is a WHNF, then  there are subcases: 
      \begin{enumerate}[itemindent=0pt,labelwidth=0pt,labelsep=1ex,left=-1pt]
          \item No hole of $C[\cdot_1,\ldots,\cdot_n]$ is a reduction hole.   %% context for $s_1,\ldots,s_n$.
          Then  $C[t_1,\ldots,t_n]$ is also a WHNF,
          $\EC(C[t_1,\ldots,t_n],K)=1$ and the claim holds.
          \item\label{cases-reduction-case-dD} For some $i$,
          hole $[\cdot_i]$ is a reduction hole %%context 
          for context $C$.
          %Then contexts $C[s_1,\ldots,s_{i-1},[\cdot]_i,s_{i+1},\ldots, s_n]$ and
          %$C[t_1,\ldots,t_{i-1},[\cdot]_i,t_{i+1},\ldots, t_n]$ are reduction contexts.
         Let $C' = C[\ldots, [s_i]_i,\ldots]$ with $n-1$ holes. By the induction assumption, we have  $D'\geq 0$ such that
          $\EC(C[s_1,\ldots,s_n],K)$ $\leq$ $\EC(C[t_1,\ldots,t_{i-1},[s_i]_i,t_{i+1},\ldots, t_n],K+D')$
          and by the assumption on $s_i,t_i$
          we have 
          $$\begin{array}{@{\qquad}l@{~}l@{}} & \EC(C[t_1,\ldots,t_{i-1},[s_i]_i,t_{i+1},\ldots, t_n],K+D') \\ \leq &
        %%  \EC(C[t_1,\ldots,t_{i-1},[t_i]_i,t_{i+1},\ldots, t_n],K+D'+d) 
           \EC(C[t_1,\ldots,t_{i-1},[t_i]_i,t_{i+1},\ldots, t_n],K+D'+D'') 
          \end{array}$$  
          since hole $[\cdot]_i$ is a reduction hole and thus
          $C[s_1,\ldots,s_{i{-}1},[\cdot]_i,s_{i+1},\ldots, s_n]$, $C[t_1,\ldots,t_{i{-}1},[\cdot]_i,t_{i+1},\ldots, t_n]$ are reduction contexts.
          Thus the claim holds for $D =  D'+D''$.
          \end{enumerate}
     \end{enumerate}
\noindent\textit{Case 2:} There is an evaluation of $C[s_1,\ldots,s_n]$ with prob-length $k \leq K$ and of length $>0$. Then 
      there are subcases:
       \begin{enumerate}[labelsep=1ex,itemindent=10pt,left=0pt]
           \item No hole of  $C$ is a reduction hole.
        Then we  consider the first standard reduction step of the expression
               $C[s_1,\ldots,s_n]$ 
               with $k \leq K$ prob-reductions. 
%                \manfredcomment{Ich denke dass ist nicht noetig:? and of 
%                of maximal length.}
               Note that this step is either unique for all evaluations of 
               $C[s_1,\ldots,s_n]$  (if it is a non-prob-reduction) or there may be two possible prob-steps.
               Hence we distinguish between non-prob and prob-steps:
             \begin{itemize}[labelsep=1ex,itemindent=0pt,left=-1pt]
                 \item  
                 If it is a non-prob-reduction, then the reduction is unique, the number $k$ remains untouched.
                 Since no hole is in  a reduction context, the reduction must be of the form
                 $C[s_1,\ldots,s_n] \xrightarrow{sr}
                 C'[s_1',\ldots,s_{m}']$
                 where there is a mapping $f$ such that $s_i' = s_{f(i)}$ and for any expressions $r_1,\ldots,r_n$ the  reduction 
                 $C[r_1,\ldots,r_n] \xrightarrow{sr}
                 C'[r_1',\ldots,r_{m}']$
                 exists where $r_i' = f(r_i)$.
                 Thus the reduction can also be done for $r_i = t_i$ for $i=1,\ldots,n$.
                 The expected convergence is not changed, i.e.
                 $\EC(C[s_1,\ldots,s_n],K)
                 =\EC(C'[s'_1,\ldots,s_{m}'],K)$
                 and for all $K'$,
                 $\EC(C[t_1,\ldots,t_n],K') =\EC(C'[t'_1,\ldots,t_{m}'],K')$
                 and we can use the induction hypothesis, since $K$ remains the same, the length
                 of a maximal evaluation using at most $K$ prob-steps is strictly smaller 
                 for $C[s'_1,\ldots,s_{m}']$,
                 and the precondition $\forall k \geq 0 \forall R \exists d$: $\EC(R[s'_i],k) \leq \EC(R[t'_i],k+d)$ still holds.
                 This shows that there exists $D'$ such that $\EC(C[s_1,\ldots,s_n],K) = \EC(C'[s_1',\ldots,s_{m}'],K)
                 \leq 
             \EC(C'[t_1',\ldots,t_{m}'],K+D') =
             \EC(C[s_t,\ldots,t_n],K+D')$.
             Hence the claim holds for $D=D'$.
             
\item If it is a prob-reduction, then there are two possibilities with probability measure $0.5$:
\begin{itemize}[left=0pt]
\item[i)]  $C[s_1,\ldots,s_n] {\xrightarrow{\!\sr,\problabr\!}} C'[s_1',\ldots,s_{m_r}']$ and    $C[t_1,\ldots,t_n] \xrightarrow{\sr,\problabr} C'[t_1',\ldots,t_{m_r}']$ and there is a mapping $f_r$ such that $s_i' = s_{f_r(i)}$, and $t_i' = t_{f_r(i)}$; for $i = 1,\ldots,m_r$, and
\item[ii)] $C[s_1,\ldots,s_n] {\xrightarrow{\sr,\problabl}} C''[s_1'',\ldots,s_{m_l}'']$ and    $C[t_1,\ldots,t_n] \xrightarrow{\sr,\problabl} C''[t_1'',\ldots,t_{m_l}'']$ and there is a mapping $f_l$ such that $s_i'' = s_{f_l(i)}$, and $t_i'' = t_{f_l(i)}$; 
                      for $i = 1,\ldots,m_r$.
\end{itemize}
%%\davidins{
Let us abbreviate the new expressions as $r'_s = C'[s_1',\ldots,s_{m_r}']$,
$r'_t = C'[t_1',\ldots,t_{m_r}']$,
$r''_s = C''[s_1'',\ldots,s_{m_l}'']$, 
$r''_t = C''[t_1'',\ldots,t_{m_l}'']$.

We apply the induction hypothesis twice: to 
$r'_s$ and $r'_t$ as well as to $r''_s$ and $r''_t$, where 
%$C'[s_1',\ldots,s_{m_r}']$
%and 
%$C'[t_1',\ldots,t_{m_r}']$ 
%as well as to
%$C''[s_1'',\ldots,s_{m_r}'']$ and $C''[t_1'',\ldots,t_{m_r}'']$, where
we use $K-1$ instead of $K$. Note that the precondition $\forall k,\forall R,\exists d:\EC(R[s'_i],k) \leq \EC(R[t'_i],k+d)$ 
($\forall k,\forall R,\exists d:\EC(R[s''_i],k) \leq \EC(R[t''_i],k+d)$, resp.)
holds for $k \geq 0$, and since $K-1 < K$ the induction measure is strictly smaller.
From the induction hypothesis we obtain that there exist $D',D''$ with
$\EC(r_s',K-1) \leq \EC(r_t',K+D'-1)$
and
$\EC(r_s'',K-1) \leq \EC(r_t'',K+D''-1)$.
\ignore{
$     \EC(C'[s_1',\ldots,s_{m_r}'],K{-}1) 
{\,\leq\,} \EC(C'[t_1',\ldots,t_{m_r}'],K{+}D'{-}1)$,

      $\EC(C''[s_1'',\ldots,s_{m_l}''],K{-}1) 
{\,\leq\,}\EC(C''[t_1'',\ldots,t_{m_l}''],K{+}D''\!{-}1).$
\endignore}
%$$
%\begin{array}{l@{~}l}
%     \EC(C'[s_1',\ldots,s_{m_r}'],K-1) 
%\leq \EC(C'[t_1',\ldots,t_{m_r}'],K+D'-1)\\ &\text{ and}\\
%     &\EC(C''[s_1'',\ldots,s_{m_l}''],K-1) \\
%\leq & \EC(C''[t_1'',\ldots,t_{m_l}''],K+D''-1)
%\end{array}
%$$ 
Clearly, the following equations hold for any $k'$:
$$\begin{array}{@{\quad}l@{}l@{}}
\EC(C[s_1,\ldots,s_n],k') &= 0.5( \EC(r'_s,k'{-}1) + \EC(r''_s,k'{-}1))
\\
\EC(C[t_1,\ldots,t_n],k') &= 0.5( \EC(r'_t,k'{-}1) + \EC(r''_t,k'{-}1))
\end{array}$$
%$$\begin{array}{l@{~}c@{~}l@{}}
%\multicolumn{3}{@{\qquad}l@{}}{
%\EC(C[s_1,\ldots,s_n],k')}\\
%\EC(C[s_1,\ldots,s_n],k')&= & 0.5 %\EC(C'[s_1',\ldots,s_{m_r}'],k'-1) \\
%&&+ 0.5 \EC(C''[s_1'',\ldots,s_{m_l}''],k'-1) %
%\\
%&&\text{ and}\\
%\multicolumn{3}{@{\qquad}l@{}}{\EC(C[t_1,\ldots,t_n],k')}
%\\
%\EC(C[t_1,\ldots,t_n],k')&= & 0.5 \EC(C'[t_1',\ldots,t_{m_r}'],k'-1) \\
%&&+ 0.5 \EC(C''[t_1'',\ldots,t_{m_l}''],k'-1)
%\end{array}
%$$
Finally, this shows
$$\begin{array}{@{\quad}l@{}}
\EC(C[s_1,\ldots,s_n],K)
=
    0.5 (\EC(r'_s,K{-}1)+\EC(r''_s,K{-}1))
    \\
\leq
    0.5 (\EC(r'_t,K+D'{-}1)+\EC(r''_t,K+D''{-}1))
    \\ 
\leq    
    0.5 (\EC(r'_t,K{+}\max(D',D''){-}1)\\
    \phantom{\leq 0.5~}+\EC(r''_t,K{+}\max(D',D''){-}1))
    \\
= \EC(C[t_1,\ldots,t_n],K+\max(D',D''))
\end{array}
$$    
\ignore{$$\begin{array}{@{}l@{~}c@{~}l@{}}
\multicolumn{3}{@{}l@{}}{\EC(C[s_1,\ldots,s_n],K)}
\\&=&
    0.5 \EC(C'[s_1',\ldots,s_{m_r}'],K{-}1) 
 + 0.5 \EC(C''[s_1'',\ldots,s_{m_l}''],K{-}1) 
\\
&\leq&  
 0.5 \EC(C'[t_1',\ldots,t_{m_r}'],K+D'-1)\\
 &&+ 0.5 \EC(C''[t_1'',\ldots,t_{m_l}''],K+D''-1)
 \\
&\leq&  
 0.5 \EC(C'[t_1',\ldots,t_{m_r}'],K+\max(D',D'')-1) \\
 &&+ 0.5 \EC(C''[t_1'',\ldots,t_{m_l}''],K+\max(D',D'')-1)
 \\
&=&
\EC(C[t_1,\ldots,t_n],K+\max(D',D''))
\end{array}$$}
Thus the claim holds for $D=\max(D',D'')$
%%% }%\enddavidins!
\end{itemize} 
               
           \item A hole of  $C$ is a reduction hole.   Then this hole is a reduction hole for $s_1,\ldots,s_n$ as well as for $t_1,\ldots,t_n$.   Now the same reasoning as in  Case 1~\cref{cases-reduction-case-dD} is valid.
       \end{enumerate}
     
% \end{enumerate}

 The final  reasoning is as follows: we apply
    \cref{lemma-limit-plus-d} to \cref{clemma}. This shows
% that for all $K$:  $\EC(C[s_1,\ldots,s_n],K) \leq \EC(C[t_1,\ldots,t_n],K)$.
%   Computing the limit for $K \to \infty$ shows 
$\EC(C[s_1,\ldots,s_n]) \leq \EC(C[t_1,\ldots,t_n])$.
 \end{proof}

\section{Program Transformations\label{sec:program-transf}}
We define the notion of correct program transformations:

\begin{definition} A program transformation $\xrightarrow{T}$ is a binary relation on expressions. It is  \emph{correct}, iff $\xrightarrow{T} ~\subseteq ~\sim_c$.
With $\xrightarrow{T,*}$ and $\xrightarrow{T,+}$ we denote the reflexive-transitive and the transitive closures of $\xrightarrow{T}$, and $\xrightarrow{T_1 \vee T_2}$ denotes the union of $\xrightarrow{T_1}$ and $\xrightarrow{T_2}$.
\end{definition}

In \cref{trans-rules} we define several program transformation. Some are generalizations of the standard reductions and can be used for partial evaluation. The rule (\xch) exchanges a variable-to-variable-binding, the rule (\ucp) means unique copying
and inlines binding if the bound variable occurs once and not below a $\lambda$-binder. The rule (\gc)  performs garbage collection, \ie~it removes (parts of) $\tletrec$-environments that are unused. 
Rules (\probid), (\probassoc) and (\probcomm) are algebraic laws (idempotence, associativity, commutativity) of the $\prob{}{}$-operator, $(\probprob)$ shifts a $\prob{}{}$-operation over another one, and
$(\probreorder)$ reorders nested $\prob{}{}$-operations.
The transformation
(\probassoc) is \emph{not correct}:
\begin{proposition}
The transformation $\probassoc$ is not correct.
\end{proposition}
\begin{proof}
%\davidins{}
Let $w$ be a WHNF.
Then $
s:=\prob{w}{(\prob{\Omega}{\Omega})}
\xrightarrow{\probassoc}
\prob{(\prob{w}{\Omega})}{\Omega}:=t$
but
$s\downarrow_{0.5}$ and
$t\downarrow_{0.25}$, and thus $s \not\sim_c t$.
%which refutes contextual equivalence.

%is incompatible with the definition of contextual 
%Consider three semantically different WHNFs $w_1,w_2,w_3$ (for instance, $w_i = \lambda x_1.\lambda x_2.\lambda x_3.x_i$).
%Then  $\prob{(\prob{w_1}{w_2})}{w_3}\downarrow_{0.5} w_1$ but 
%$\prob{w_1}{(\prob{w_2}{w_3})}\downarrow_{0.25} w_1$, which is incompatible with the definition of contextual equivalence.
\end{proof}

%% \begin{figure}[tpbh]
\begin{figure*}[tpb]
\begin{minipage}{.52\textwidth}
$\begin{array}{@{}l@{\,}l}
(\lbeta) &((\lambda x.s)~t) \to \tletr~x=t~\tin~s
\\                 
% (sr,\problabl)&R[\prob{e_1}{e_2}] \to R[e_1]
% \\                 
% (sr,\problabr)&R[\prob{e_1}{e_2}] \to R[e_2]
% \\                 
(\lapp) 
&((\tletr~\env~\tin~s)~t) \to \tletr~\env~\tin~(s~t)
\\
(\cpin)&
\tletr~x =\lambda y.s, \env~\tin~C[x]\\
&
\to  
                     \tletr~x=\lambda y.s, \env~\tin~C[\lambda y.s]
\\                 
(\cpe)&
\tletr~x = C[y], y = \lambda z.s,\env~\tin~t\\
&
\to 
\tletr~x = C[\lambda z.s], y = \lambda z.s,\env~\tin~t
 \text{ for $m {\geq} 1$}
\\
(\lletin)&
\tletr~\env_1~\tin~\tletr~\env_2~\tin~s
   \to
   \tletr~\env_1,\env_2~\tin~s
\\                 
(\llete)&
\tletr\,x = (\tletr\,\env_1\,\tin\,s),\env_2\,\tin\,t
\to
\tletr\,x = s, \env_1,  \env_2\,\tin\,t
\\
(\cpxin)& \tletr~x=y,\env~\tin~C[x] \to \tletr~x=y,\env~\tin~C[y]
\\
(\cpxe)& \tletr~x=y,z=C[x],\env~\tin~s \to \tletr~x=y,z=C[y],\env~\tin~s
\\
(\ucpone) & \tletr~x=t,\env~\tin~S[x] \to \tletr~\env~\tin~S[t]\\
&\text{if $x \not\in (FV(\env)\cup FV(S) \cup FV(t))$}
\\
(\ucptwo) & \tletr~x=t, \env,y=S[x]~\tin~s\to\tletr~\env,y=S[t]~\tin~s\\
&
\text{if $x \not\in (FV(t)\cup FV(S) \cup FV(s))$}
\\
(\ucpthree) & \tletr~x=t~\tin~S[x] \to S[t]~~\text{if $x\not\in FV(S) \cup FV(t)$}
\end{array}$
\end{minipage}~\begin{minipage}{.45\textwidth}
$\begin{array}{@{}l@{~}l}
(\xch)&\tletr\,x=y{,}y=s{,}\env\,\tin\,t \to \tletr\,x=s{,}y=x{,}\env\,\tin\,t
\\
(\gcone) &\tletr~\env_1,\env_2~\tin~s \to \tletr~\env_1~\tin~s\\
&\text{if $(LV(\env_2)\cap (FV(\env_2) \cup FV(s))) = \emptyset$}
\\
(\gctwo) &\tletr~\env~\tin~s \to s~~\text{if $LV(\env) \cap FV(s) = \emptyset$}
\\
(\probid) & \prob{s}{s} \to s 
\\
(\probcomm) & \prob{s}{t}\to \prob{t}{s}
\\
(\probassoc) & \prob{r}{(\prob{s}{t})} \to \prob{(\prob{r}{s})}{t}
\\
%{\color{blue}{Distributivgesetz?} \color{black}}   \\
%DS:ok
(\probprob) & \prob{r}{(\prob{s}{t})} \to \prob{(\prob{r}{s})}{(\prob{r}{t})}
\\
(\probreorder) & \prob{(\prob{s_1}{s_2})}{(\prob{t_1}{t_2})}
\to \prob{(\prob{s_1}{t_1})}{(\prob{s_2}{t_2})}\\[1.1ex]
\end{array}
$\\
\textbf{Unions and Variants:}\\
$\begin{array}{@{}l@{}}
\begin{array}{@{~}l@{~}ll@{~}l}
(\cpx) & (\cpxin) \cup (\cpxe) 
&
 (\llet) &  (\lletin) \cup (\llete) 
 \\
 (\lllrule) &  (\llet) \cup (\lapp) 
&
(\gc) &  (\gcone) \cup (\gctwo)
\\
(\ucp) & (\ucpone) \cup (\ucptwo) \cup (\ucpthree)
&
(\cp) &  (\cpe) \cup (\cpin) 
 \\
 \end{array}
\\
\begin{array}{@{~}l@{~}l}
(\cpd)& (\cp)  \mbox{ and  context $C$ in the rule is not an $\SCtxt$-context}
\\
(\cpS)& (\cp)  \mbox{ and  context $C$ in the rule is an $\SCtxt$-context}
\end{array}
\end{array}
$
\end{minipage}
\caption{Transformations\label{trans-rules}}
\end{figure*}

Correctness of $\xrightarrow{lbeta}$ follows with the context lemma, since applying it inside reduction context is always a $\xrightarrow{sr}$-step.

% \subsection{Correctness of Transformations (lbeta)}
\begin{proposition}\label{prop-lbeta-correct}
The transformation $\xrightarrow{lbeta}$  is correct.
\end{proposition}
% \begin{proof}
% We use the context lemma (\cref{thm:context-lemma-ext})
% for $n=1$ and $d=0$. Let $s \xrightarrow{lbeta} t$ and $R$ be a reduction context. 
% The structure of reduction contexts implies $R[s] \xrightarrow{\sr,lbeta} R[t]$.
% Thus there is
% %is an evaluation
% $(1,R[s])\evaluation_L(p,s')\in\Eval(R[s])$
% if, and only if there is 
% %an evaluation
% $(1,R[t])\evaluation_L(p,s')\in\Eval(R[t])$.
% This implies for all $k\geq0:$ $\EC(R[s],k) = \EC(R[t],k)$. 
% %The reduction rule $s \xrightarrow{\sr,lbeta} t$ also implies $\EC(R[t]) = \EC(R[s]$,
% %since this is the only standard reduction of $R[s]$.
% With the context lemma we obtain
%  $\EC(C[s]) = \EC(C[t])$ for all contexts $C$, and hence $s \sim_c t$.
% \end{proof}

Proving correctness of other transformations requires more sophisticated techniques. 
%The proof technique(s) will exploit the operational behaviour, and the context lemma for most cases to show semantical equivalence.  
In the remaining part of the section we provide two  criteria for proving correctness of transformations. For a transformation $\xrightarrow{T}$, with $\xrightarrow{\Ctxt,T}$ ($\xrightarrow{\SCtxt,T}$, resp.) we denote the closure of $T$ inside all (surface, resp.) contexts. 
% This will also be used for the closure w.r.t.~surface contexts, written $\xrightarrow{\SCtxt,T}$.

\begin{proposition}[Correctness Criterion: Same Prob-Sequences]\label{prop-surface-same-prob-seq}
Let $\xrightarrow{T}$ be a program transformation and $X \in \{\Ctxt,\SCtxt,\RCtxt\}$.
If for all $s,t$ with $s \xrightarrow{X,T} t$ the following holds:
\begin{equation}
\label{cond1} \text{for all $s \evaluation_L s' \in \Eval(s)$
 there exists $t \evaluation_L t' \in \Eval(t)$}
%\item\label{cond2} for all $t \evaluation_L t' \in \Eval(t)$ 
%there exists $s \evaluation_L s' \in Eval(s)$
\end{equation}
then ${\xrightarrow{T}} \subseteq {\leq_c}$ holds.
Correctness of $\xrightarrow{T}$ can be shown by applying the criterion for $\xrightarrow{T}$ and its inverse ${\xleftarrow{T}} := \{(t,s) \mid s \xrightarrow{T} t\}$.
%\davidcomment{Aussage und Beweis angepasst auf $\leq_c$ statt $\sim_c.$}
\end{proposition}
\begin{proof}
We use expressions without weights, but keep track of the prob-sequences: for $s\in\Expr$ and an evaluation $s \evaluation_L s'$,  the resulting probabilty of $(1,s)$ is $\frac{1}{2^{|L|}}$, where $|L|$ is the length of $L$.
%(i.e. the prob-length of the evaluation).

%We show $\xrightarrow{T}~{\subseteq}~{\leq_c}$.

We show  $\forall R \in \RCtxt,\forall k \geq 0: \EC(R[s],k) \leq \EC(R[t],k)$.
The context lemma (\cref{thm:context-lemma-ext}) with
$n=1$ (and $d=0$ for all reduction contexts) then shows $\forall C\in \Ctxt:\EC(C[s]) \leq \EC(C[t])$ and hence $s \leq_c t$.

 Let $R,k$ be arbitrary but fixed and $\EC(R[s],k) = q$, and let $ev$ be the set of evaluations of $R[s]$ of prob-length $\leq k$. This set is finite and thus $\EC(R[s],k) = \sum_{(s \evaluation_L s' \in ev)} \frac{1}{2^{|L|}} = q$. Now apply Condition~\eqref{cond1} of the claim to every 
  $s \evaluation_L s' \in ev$ (this is possible for $X \in \{\RCtxt,\SCtxt,\Ctxt\}$ (surface) contexts include all reduction contexts). This results in a finite set of evaluations $ev' = \{t \evaluation_L t' \mid t \evaluation_L t' \in \Eval(R[t],k)\}$.
  Since the prob-labels are kept (from $L$ to $L$), $|ev| = |ev'|$
  and since also the prob-lengths are same, we have $\sum_{
  t \evaluation_L T' \in ev'} \frac{1}{2^{|L|}} = q$. Since $ev' \subseteq \EC(R[t],k)$, the claim is shown.
 %
%The other part $\xrightarrow{T}~\subseteq~{\geq_c}$ follows in the same way, by proving $\xleftarrow{T}~\subseteq~{\leq_c}$ and using \cref{cond2} instead of \cref{cond1}. 
\end{proof}

%\manfredcomment{Anfang der Einfügung zu den prob-Axiomen} 
%\manfredcomment{Alle alten Kommentare in einem Extra-File}
  
 \begin{definition}A {\em frontier} is a set of words over \{\problabl,\problabr\} generated by
 starting with $\{\epsilon\}$ ($\epsilon$ denotes the empty string) and then applying the following operation multiple times to
 the set: take a string $r$ and replace it by two words $r;\problabl$
 and $r;\problabr$. 
 \end{definition}
 For example $\{\epsilon\}$, $\{\problabl,\problabr\}$
  and $\{\problabl;\problabl$, $\problabl;\problabr$, $\problabr\}$ are frontiers.
 \begin{definition}
Let $s\in\Expr$, $d \geq 1$, and $F = \{L_1,\ldots,L_n\}$ be a frontier with $|L_i| \leq d$ for all $i$.  Then 
a {\em frontier-evaluation}  of $s$ w.r.t. $(F,d)$ is a multiset of the form $\{(L_1,q_1,s_1),\ldots,(L_n,q_n,s_n)\}$ such that for all $i = 1,\ldots,n$: 
$(1,s) \xrightarrow{wsr,\problabl \vee \problabr,*} (q_i,s_i)$ 
with $L_i = \PS((1,s) \xrightarrow{wsr,\problabl \vee \problabr,*} (q_i,s_i))$
(i.e. there is a reduction sequence from  $s$ to $s_i$ resulting in weight $q_i$, that uses only prob-reductions  
and the prob-sequence is $L_i$).
Let the multiset  $\{(q_1,s_1),\ldots,(q_n,s_n)\}$  be the {\em frontier evaluation result}.
In a frontier evaluation result, the sum of all probabilities is always $1$. 
%
%\manfredcomment{bot weglassen ist schlecht, da es einen Unterschied macht wenn man die nur teilweise weglässt.rechts bzw links. }   
%%Entries $s_i$ with $\EC(s_i) = 0$ can be removed, since they do not play a role in the notions and criteria below. }
%
\end{definition}

We compare two multisets of frontier evaluation results   $A$  and $B$ in order to reconstruct the contextual equivalence or preorder resp.

\begin{definition}\label{def:leq-criteria-on-sets}
Let $A,B$ be two  frontier evaluation results. We define the following criteria:
% will later be used to show contextual equivalence and preorder:
\begin{description}
    
    \item[EqCr1] For every $(q,s) \in A$ there is some $(q',s) \in B$  with $q \leq q'$.
    \item[EqCr2] For every $(q,s) \in A$, let $q_{s,A}$ be the sum of all $q'$ such that $(q',s)$ is an entry in $A$, and let $q_{s,B}$ be the sum of all $q'$ such that $(q',s)$ is an entry in $B$,
    then the inequation $q_{s,A} \leq q_{s,B}$ must hold.
     \item[EqCr3] This is applicable in case that $\Omega$ may appear as expression $s$. For every $(q,s) \in A$, with $s \not= \Omega$,
     % \davidcomment{$\not=\bot$ ist undefiniert? $s \not\sim_c \bot$?}
     let $q_{s,A}$ be the sum of all $q'$ such that $(q',s)$ is an entry in $A$, and let $q_{s,B}$ be the sum of all $q'$ such that $(q',s)$ is an entry in $B$.
    Then the inequation $q_{s,A} \leq q_{s,B}$ must hold.
   % And: Let $q_{A}$ be the sum of all $q'$ such that $(q',s)$ is an entry % in $A$, and let $q_{s}$ be the sum of all $q'$ such that $(q',s)$ is an % entry in $B$, then the inequation $q_{A} \leq q_{B}$ must hold.
  %  \davidcomment{Ich verstehe den Unterschied nicht zwischen dem ersten Teil und dem Teil nach And: ...}
\end{description}
 \end{definition}

 % DS: den hier gestrichen, da er dem Leser nicht viel hilft,
 % er fragt sich eher, warum wir das nicht beweisen koennen
\longversion{\begin{remark}
An extended criterion  for equivalence that we conjecture to hold, but could not  be proved with the current techniques is: 
%  $\bullet$: 
 For every $(q,s) \in A$, let $q_{s,A}$ be the sum of all $q'$ such that $(q',s')$ with $s \leq_c s'$ is an entry in $A$, and let $q_{s,B}$ be the sum of all $q'$ such that $(q',s')$ with $s \leq_c s'$ is an entry in $B$,
    then the inequation $q_{s,A} \leq q_{s,B}$ must hold.
\end{remark}
 }

\begin{example}
An example is the evaluation of $\prob{(\prob{s_1}{s_2})}{(\prob{s_1}{s_3})}$
w.r.t. $\{\problabl;\problabl$, $\problabl;\problabr$; $\problabr;\problabl$, $\problabr;\problabr\}$ which results in 
$\{(\problabl;\problabl,0.25,s_1)$, $(\problabl;\problabr,0.25,s_2)$,  $(\problabr;\problabl,0.25,s_1)$, $(\problabr;\problabr,0.25,s_3)\}$. 
 The  frontier-evaluation result is 
$\{(0.25,s_1), (0.25,s_1), (0.25,s_2),(0.25,s_3)\}$.   
% The compacted  frontier-evaluation result is 
% $\{(0.5,s_1), (0.25,s_2),(0.25,s_3)\}$.   
The other side of the $\probprob$-rule is $\prob{s_1}{(\prob{s_2}{s_3})}$, which has a frontier evaluation result $\{(0.5,s_1), (0.25,s_2),(0.25,s_3)\}$, and satisfies criterion (EqCr2) of \cref{def:leq-criteria-on-sets} in both directions.
\end{example}

\begin{example}
Further examples for the criteria in \cref{def:leq-criteria-on-sets}.
%% $\leq_c$ and $\not\leq_c$  on sets. Where we assume that the sum of the probabilities is $1$.  
% \manfredcomment{Mehr Beispiele zum Verstehen, kann man ja wieder auskommentieren, wenn es zuviel ist.}
\begin{enumerate}
\item
 Let $M_1 := \{(0.5,a), (0.5,\Omega)\}$ and $M_2 := \{(0.7,a), (0.3,\Omega)\}$.
Then  for $(0.5,a)$, the element $(0.7,a)$ is sufficient to detect $\leq_c$, for  $(0.5,\Omega)$, we obtain  $1 \leq 1$. This satisfies %criterion 
(EqCr3).  %%%%% Thus  $M_1 \leq_c M_2$.  
\item For $M_1 := \{((0.1,\Omega),(0.6,a), (0.3,b),\}$ and $M_2 := \{(0.2,\Omega),(0.5,a), (0.3,b) \}$,
no criterion is satisfied. 
\end{enumerate}
\end{example}

The idea of the 
%contextual equality (or preorder) criterion
criteria is to evaluate the two expressions in several ways and to get intermediate resulting sets that can be compared with the criteria.
For example, consider 
$\probreorder$: 
Evaluating the left-hand side to a  frontier-evaluation result:   $\prob{(\prob{s_1}{s_2})}{(\prob{t_1}{t_2})}$ results in 
$\{(0.25,s_1),(0.25,s_2),(0.25,t_1),(0.25,t_2)\}$, and 
$\prob{(\prob{s_1}{t_1})}{(\prob{s_2}{t_2})}$ evaluates also to the same   frontier-evaluation result. For $\probprob$ the left-hand side
 $\prob{r}{(\prob{s}{t})}$  has a frontier-evaluation result
 $\{(0.5,r),(0.25,s),(0.25,t)\}$, and  from
 $\prob{(\prob{r}{s})}{(\prob{r}{t})}$ we obtain $\{(0.25,r),(0.25,r),(0.25,s),(0.25,t)\}$. The criterion (EqCr2) is satisfied in all cases.

 \begin{proposition}[Correctness Criterion: Same
 distribution after prob-reduction]\label{prop-surface-almost-same-prob-seq}
Let $\xrightarrow{T}$ be a program transformation. The following claim holds for $X \in \{\Ctxt,\SCtxt,\RCtxt\}$.
Transformation $\xrightarrow{T}$ is included in $\leq_c$, i.e. $\xrightarrow{T}~ \subseteq ~\leq_c$,  if for all $s,t$ with $s \xrightarrow{X,T} t$ the following holds:
There is a frontier-evaluation result $R_s$ of $s$ and a 
   frontier-evaluation result $R_t$ of $t$ such that criteria (EqCr1), (EqCr2), or (EqCr3) holds.
 If $R_s \sim_c R_t$, then correctness of $\xrightarrow{T}$ follows by symmetry.
  \end{proposition}
 \begin{proof}
 % \davidins{We only show $\xrightarrow{T} \subseteq \leq_c$, since the other direction follows by symmetry.}
 We verify the preconditions of  context lemma (\cref{thm:context-lemma-ext}) for $n=1$.
%  Therefore the preconditions have to  be verified. 
 %\davidins{} 
 Since $\xrightarrow{X,T}$ already covers $\xrightarrow{T}$-steps in all reduction contexts, it suffices to show that there exists $d \geq 0$ such that $\forall k \geq 0$ we have 
 $\EC(s,k) \leq \EC(t,k+d)$.
This holds, since the frontier-evaluations results satisfy criteria (EqCr1), (EqCr2), or (EqCr3) hold, where
% are the same\davidcomment{are $\leq_c$ on sets?}, and the % if evaluated in reduction contexts. %
the difference $d$ can be chosen as the maximum of 
 the prob-reduction depth of the frontier-evaluations.
 \end{proof}
 
%  For using the criteria (EqCr4) more effort is required.
%% \davidcomment{Hm, im Grunde EqCr4 streichen oder in einen Remark verschieben?}  S.O.
 %

 \begin{corollary}\label{cor:problaws}
 The program transformations $\xrightarrow{\probid}$, 
 $\xrightarrow{\probcomm}$, $\xrightarrow{\probprob}$, and $\xrightarrow{\probreorder}$
 are correct. 
%  Also generalizations of these program transformations like an arbitrary reordering where the expressions remain on the same level within a $\problab$-expression is correct.
% DS; das hab ich gekürzt, weil...
 \end{corollary}
 \ignore{
 Ist nicht noetig, da dir Kriterien syntaktsch sind...
 \begin{proof}
 For proving the correctness using \cref{prop-surface-almost-same-prob-seq},  we have to consider all possibilities of equality/dis-equality of the parameter expressions. 
 For example consider the parameters
 $r,s,t$ in the expressions
 $\prob{r}{(\prob{s}{t})}$  and  $\prob{(\prob{r}{s})}{(\prob{r}{t})}$. 
 We have to consider all the 5 equivalence classes possibilities separately, which makes 5 
  applications of \cref{prop-surface-almost-same-prob-seq}.
  Clearly for all these cases, the correctness conclusions hold.
 \end{proof}
 }
 
 We also obtain inequations like $\prob{s}{\bot} \leq_c s$, since this corresponds to the resulting sets $\{(0.5,s), (0.5,\bot)\}$ and  $\{(1,s)\}$.
 % \manfredcomment{Ende der Einfügung zu den prob-Axiomen} 
 % XXXXManfred\\
 
\section{Correctness by Diagrams
\label{sec:diagram}}
In this section we want to show that the remaining transformations in \cref{trans-rules} are correct.
%a large set of program transformations of $\LNEEDPROB$ is correct w.r.t. the defined semantics, i.e. is invariant w.r.t. $\sim_c$ (see \cref{def:contextual-equivalence}). This large set of transformations can then be used for (correct) program manipulations and optimizations.
%
\ignore{  Ist jetzt weiter oben...
To be more precise, we define:
\begin{definition} A transformation $\xrightarrow{T}$ is a binary relation on expressions. It is  \emph{correct}, iff $\xrightarrow{T} ~\subseteq ~\sim_c$.
\end{definition}
We use $\xrightarrow{T,*}$ to notate the reflexive-transitive closure of $\xrightarrow{T}$, and $\xrightarrow{T_1 \vee T_2}$ denotes the union of $\xrightarrow{T_1}$ and $\xrightarrow{T_2}$.

The proof technique will exploit the operational behaviour, and the context lemma for most cases to show semantical equivalence.  
}
%
%
%
%\subsection{Explanations of the Diagram-based Proof-Method} 
%
% \manfredcomment{Unten das ist ne sehr alte Erinnerung, bin nicht mehr sicher was da genau schiefgeht:}
% The diagrams are computed for $\SCtxt$-contexts, since the computation
% for $\RCtxt$-contexts leads to more complex diagrams and sometimes is not successful, i.e. needs the surface-contexts, anyway.
%   \manfredcomment{Ich hab jetzt einige DInge geändert ohne Spuren zu  hinterlassen, da alles noch etwas vorläufig ist. Später oder wenn Dinge fix sind, werde ich markieren was ich geändert hab}
%
%
%\davidins{ }  
We use the so-called \emph{diagram method} \cite{schmidt-schauss-schuetz-sabel:08,sabel-schmidt-schauss-MSCS:08,rau-sabel-schmidtschauss:12,sabel:wpte:2019} to prove their correctness. 
% of program transformations.
%The method for showing contextual equivalence of the above-mentioned transformation rules (with the exception of \problabl and \problabr).
% The method can be summarized as follows:
% first compute complete sets of forking diagrams and commuting commuting diagrams and, as a base case, inspect  the case where a transformation step is applied or leads to a WHNF. 
% 
Given a transformation  $\xrightarrow{T}$, 
the method uses \cref{prop-surface-same-prob-seq} 
to show correctness:
for a step $s \xrightarrow{\SCtxt,T} t$ (note that we work with the closure of $\xrightarrow{T}$ w.r.t.~surface contexts) 
it shows that for every evaluation of $s$ there is an
evaluation of $t$ and vice versa, where the prob-sequences are the same.
% 
% Since the structure of reduction contexts is quite complex and sometimes the property ``a transformation is applied inside a reduction context'' gets lost by applying a standard reduction, and since the restriction to reduction contexts may lead to deviations, we use the simplification to consider all surface contexts instead of all reduction contexts. Thus our goal is to prove $\EC(S[s],k) = \EC(S[t],k)$ for all $S\in\SCtxt$ and all $k$.
% This mainly means, that we have to show that for every evaluation of $S[s]$ there is an evaluation of $S[t]$ (of the same prob-length) and the reverse: that for every evaluation of $S[t]$ there is an evaluation of $S[s]$ (of the same prob-length). 

% To ease notation, we close the transformation under consideration by surface contexts.
%and thus write $s \xrightarrow{\SCtxt,T} t$ meaning that transformation $T$ is applied inside a surface context of $s$, i.e. $s=S[s'], t=S[t']$ and $s' \xrightarrow{T} t'$. 

% Thus it suffices to show the following for $s \xrightarrow{\SCtxt,T} t$: For every evaluation of $s$, there is an evaluation of $t$ (of the same prob-length), and  for every evaluation of $t$, there is an evaluation of $s$ (of the same prob-length). 

Base cases cover the cases that $s$ ( or $t$, resp.) already is a WHNF, and show that $t$ ( or $s$, resp.) can then be evaluated to a WHNF. In the general case, so-called forking or commuting diagrams are used. 
The forking diagrams are for constructing
an evaluation for $t$, from a given evaluation for $s$. The commuting diagrams serve the same purpose for the other direction. 
A single forking diagram describes a \emph{fork overlap} $s' \xleftarrow{\sr} s \xrightarrow{\SCtxt,T} t$  (using solid arrows) and how the pair $s'$ and $t$ can be joined using transformations and reductions (dashed arrows).
A set of forking diagrams is complete if each fork overlap that occurs is covered by at least one of the diagrams.
The base case and the forking diagrams are \emph{inductively} applied to construct an evaluation for $t$, starting from a (given) evaluation for $s$.

The commuting diagrams cover \emph{commuting overlaps} $s \xrightarrow{\SCtxt,T} t \xrightarrow{\sr} t'$
(solid arrows) and show how the pair $t$ and $t'$ can be joined by transformations and standard reductions (written as dashed arrows).
A set of commuting diagrams is complete if it covers all commuting overlaps. Such a set is used to (inductively) show that given an evaluation for $t$, there is an evaluation for $s$.

In both directions, the preservation of the prob-sequences has to be proved by inspecting the diagrams.

The diagrams abstract from the concrete expressions (they do not occur in the diagrams), since a single diagram usually represents infinitely many concrete overlaps of concrete expressions (and thus the complete set of diagrams is a finite representation of all concrete overlaps). Information that is kept is the labels of the transformations and reductions.
Thus the (forking and commuting) diagrams are interpreted and used as (non-deterministic) rewrite rules on reduction sequences. The non-determinism reflects the missing information on the concrete expressions that are manipulated.

In earlier work (e.g.~\cite{schmidt-schauss-schuetz-sabel:08,sabel-schmidt-schauss-MSCS:08}), diagrams were computed manually, but nowadays they are computed automatically by a tool, that unifies the left-hand sides of the standard reductions with left- (forking) and also right-hand sides (commuting) of the transformation, and then searches to show joinability (the unification algorithm is described in \cite{schmidtschauss-sabel-PPDP:2016}, the tool is described in \cite{sabel:wpte:2019}).
If not stated otherwise, the diagrams in this paper were computed by this tool.
% \end davidins{}

%First we explain the use of forking diagrams. For $s \xrightarrow{T} t$ for some of the rules $r$, we show the precondition of the  context lemma (\cref{thm:context-lemma}). This is done by showing that for every transformation step $S[s] \xrightarrow{T} S[t]$, and every  evaluation $S[s] \xrightarrow{\sr,*} w_s$, (remember that $w_s$ is a WHNF),  there is also an evaluation $S[t] \xrightarrow{\sr,*} w_t$, (where $w_t$ again must be a WHNF) with 
% DS: unklar: $\xrightarrow{\SCtxt,U,*} w_t$ and $U$ is a  set of transformation rules not containing \problabr,\problabl. Thus every evaluation of $S[s]$ can be translated into one of $S[t]$ with the same final probability.

%The so-called commuting diagrams are used for arguing for the converse:
%that every evaluation of $S[t]$ can be translated into one of $S[s]$ with the symmetric properties as above and the same final probability.
\longversion{
%\manfredinsert{..}
Technically, two cases have to be considered:
\begin{description}
\item[Forking] From  $s \xleftarrow{\SCtxt,T} t \xrightarrow{\sr,*} t'$ where the latter s an evaluation, construct a (related) evaluation of $s$. Here the forking diagrams are used.
\item[Commuting] From  $s \xrightarrow{\SCtxt,T} t \xrightarrow{\sr,*} t'$  where the latter s an evaluation, construct a (related) evaluation of $s$. Here the commuting diagrams are used.
\end{description} 

In several cases we will view the reduction sequences as objects like a list, interpret the forking and commuting diagrams as operations on the full reduction sequences. In particular 
 various measures are employed to compare reduction sequences and to show termination of the manipulations of these sequences. 
% \endmanfredins}
}

% \subsection{Correctness of let-Shiftings by (\lllrule)}\label{subsec:let-correct}
%%\davidins{
% For a transformation $\xrightarrow{T}$, we denote by $\xrightarrow{\Ctxt,T}$ % the closure of $T$ inside alle contexts. This will also be use for the % closure w.r.t.~surface contexts (written $\xrightarrow{\SCtxt,T}$).
% }
%%  \davidcomment{evtl. irgendwo anders platzieren.}
%%    oben in Explanationa dn Motivation    
\ignore{
We show that applying $(\lllrule)$-steps in arbitrary contexts terminates:
\begin{lemma}\label{lemma:lll-terminates}
Every sequence of $\xrightarrow{(\Ctxt,\lllrule)}$-steps is finite.
\end{lemma}
\begin{proof}
Let the measure $\LMP(s)$ be defined as the pair $(\#(\tletr,s), \LM(s))$, ordered lexicographically, where $\#(\tletr,s)$ is the number of occurrences of {\tletr} in $s$, and $\LM$ is a polynomial measure as
$\LM(x) %= \LM(c) 
= 1$;
$\LM(\lambda x.s) = 1+\LM(s)$; $\LM(s~t) =  2\LM(s) + \LM(t)$;
$\LM(\tletr~\env~\tin ~s) =  2\LM(\env) + \LM(s)$; $\LM(x_1 = s_1;\ldots; x_n = s_n)$ $=$ $\LM(s_1) + \ldots + \LM(s_n)$. 

The definitions show that $\xrightarrow{\Ctxt,\llete}$- and $\xrightarrow{\Ctxt,\lletin}$-steps strictly reduce the number of let-expressions and thus, the first component of the measure is strictly decreased. For the $\xrightarrow{\Ctxt,\lapp}$-steps,
the first component is not changed, but the measure $\LM(\cdot)$ is strictly decreased.
%let $C[(\tletrec~\env~\tin~s)~t)] \to C[(\tletrec~\env~\tin~(s~t)]$. Then 
%$\LM(C[(\tletrec~\env~\tin~s)~t)]) = 
%x + 2^k\cdot \LM((\tletrec~\env~\tin~s)~t)
%=
%x + 2^{k+2}(\LM(\env)) + 2^{k+1}\LM(s)+2^k\LM(t)$
%for some $k \geq 0, x \geq 0$.
%Then  $\LM(C[(\tletrec~\env~\tin~(s~t))]
%= x+2^k\LM(\tletrec~\env~\tin~(s~t) = x+2^{k+1}\LM(\env)+2^{k+1}\LM(s) + 2^k(\LM(t))$.
%Since $\LM(t) > 0 $ in any case, this shows that the measure is strictly decreased.
%
%The rule $\lletin$ is defined as 
%$\tletr~\env_1~\tin~\tletr~\env_2~\tin~e$  $\to$
%   $\tletr~\env_1,\env_2~\tin~e$ strictly decreases the second part of the measure: $\LM(e)$. 
Hence there are no infinite reduction sequences consisting only of $\xrightarrow{\Ctxt,\lllrule}$-steps.
\end{proof}
}

% \davidcomment{(bei llet-in wirds nicht kleiner oder:

%%  $\LM(\tletr~\env_1~\tin~\tletr~\env_2~\tin~e)
%% = 2\LM(\env_1)+\LM(\tletr~\env_2~\tin~e)
%% = 2\LM(\env_1)+2\LM(\env_2) + \LM(e)
%% $

%% $\LM(\tletr~\env_1,\env_2~\tin~e) = 2\LM(\env_1,\env_2)+\LM(e) =   2\LM(\env_1) + 2\LM(\env_2)+\LM(e)$
% }
% \manfredcomment{ist korrigiert  mit lexicographic measure}
\begin{figure*}[t]
\begin{minipage}{.51\textwidth}
\textbf{Forking diagrams:}\\
$
\begin{array}[t]{@{}c@{}}
\\[-3ex]
\xymatrix@R=5mm@C=5mm{
 \cdot \ar@/^-.5pc/[r]_{\sr,a}\ar@/^.5pc/[r]^{\SCtxt,\lllrule}  & \cdot
}
\\
a \in \{\lapp,\llet\}
\end{array}\!\!\!
\begin{array}[t]{@{}c@{}}
\xymatrix@R=5mm@C=5mm{
 \cdot \ar[d]_{\sr,a}\ar[r]^{\SCtxt,lll}  & \cdot \ar@{-->}[d]^{\sr,a}\\
 \cdot \ar@{-->}[r]_{\SCtxt,lll}         & \cdot \\
}
\end{array}\!\!\!\!\!\!\!
%%%%%%%%%%%%%%%%%%%%%%%%%%%%%%%%%%%
\begin{array}[t]{@{}c@{}}
\xymatrix@R=5mm@C=5mm{
 \cdot \ar[d]_{\sr,a}\ar[r]^{\SCtxt,lll}  & \cdot \ar@{-->}[dl]^{\sr,a}\\
 \cdot \\
}
\\
a \in \{\problabl,\problabr\}
\end{array}\!\!\!\!\!
%%%%%%%%%%%%%%%%%%%%%%%%%%%%%%%%%%%
\begin{array}[t]{@{}c@{}}
\xymatrix@R=5mm@C=5mm{
 \cdot \ar[d]_{\sr,\lapp}\ar[rr]^{\SCtxt,\llet}  && \cdot \ar@{-->}[d]^{\sr,\lapp}\\
 \cdot \ar@{-->}[r]_{\SCtxt,\lapp}&\cdot\ar@{-->}[r]_{\SCtxt,\llet}     & \cdot \\
}
\end{array}
$
\textbf{Commuting diagrams:}\\
$
\begin{array}{@{}c@{}}
\begin{array}[t]{@{\!\!}c@{}}
\xymatrix@R=5mm@C=5mm{
 \cdot \ar@{-->}[d]_{\sr,a}\ar[r]^{\SCtxt,b}  & \cdot \ar[d]_{\sr,a}\\
 \cdot \ar@{-->}[r]_{\SCtxt,b}         & \cdot \\
}\\
b \in \{\lapp,\\\quad\llet\}
\end{array}\!\!\!
%\ignore{
%Das war doppelt und hatte nicht-standard Form
\begin{array}[t]{@{}c@{}}
\xymatrix@R=4mm@C=0mm{
 \cdot \ar@{-->}[dr]_{\sr,a}\ar[rr]^{\SCtxt,lll}  && \cdot \ar[dd]^{\sr,a}\\
 &\cdot \ar@{-->}[dr]_{\sr,lll}         \\
 &&\cdot \\
}
\\
\end{array}
%\endignore}
%
%%%%%%%%%%%%%%%%%%%%%%%%%%%%%%%%%%%
\begin{array}[t]{@{}c@{}}
\xymatrix@R=5mm@C=5mm{
 \cdot \ar@{-->}[dr]_{\sr,a}\ar[r]^{\SCtxt,lll}  & \cdot \ar[d]^{\sr,a}\\
 &\cdot \\
}
\\
a \in \{\problabl,\\\quad\problabr\}
\end{array}\!\!\!
%%%%%%%%%%%%%%%%%%%%%%%%%%%%%%%%%%%
\begin{array}[t]{@{}c@{}}
\xymatrix@R=5mm@C=5mm{
 \cdot \ar@{-->}[d]_{\sr,\lapp}\ar[rr]^{\SCtxt,\llet}  && \cdot \ar[d]_{\sr,\lapp}\\
 \cdot \ar@{-->}[r]_{\SCtxt,\lapp}&\cdot\ar@{-->}[r]_{\SCtxt,\llet}     & \cdot \\
}
\end{array}
\begin{array}[t]{@{}c@{}}
\xymatrix@R=4mm@C=6mm{
 \cdot \ar@{-->}[d]_{\sr,\lapp}\ar[r]^{\SCtxt,\llet}  & \cdot \ar[dd]_{\sr,\lapp}\\
 \cdot \ar@{-->}[d]_{\sr,\lapp}\\
 \cdot\ar@{-->}[r]_{\SCtxt,\llet}     & \cdot \\
}
\end{array}
\end{array}
%%%%%%%%%%%%%%%%%%%%%%%%%%%%%%%%%%%
$

\end{minipage}~~~\begin{minipage}{.5\textwidth}
\textbf{Base cases:}
Let $s \xrightarrow{\SCtxt,lll} s'$. Then 
$s$ is a WHNF, then $s'$ is a WHNF, and if $s'$ is a WHNF, 
or $s \xrightarrow{\sr,llet,0\vee 1} s''$ where $s''$ is a WHNF.

\textbf{TRS $R_1$ for forking diagrams:}
\small$$\begin{array}{r@{}c@{}lr@{}c@{}lr@{}c@{}l}
\Slll(\SRlll(x)) &\to& x
&\Slll(\SR(x)) &\to& \SR(x)
\\
\Slll(\SR(x)) &\to& \SR(\Slll(x))
&\Slll(\SRlll(x)) &\to& \SRlll(\Slll(x))
\\
\Slll(\SRlll(x)) &\to& \SRlll(\Slll(\Slll(x)))
\end{array}
$$\normalsize

\textbf{TRS $R_2$ for commuting diagrams:}
\small
$$\begin{array}{r@{}c@{}l@{}r@{}c@{}l}
\Sllet(\SR(x)) &\to& \SR(\Sllet(x))
&\Sllet(\SR(x)) &\to& \SRlll(\SR(x))
\\
\Sllet(\SR(x)) &\to& \SR(x)
&\Sllet(\SRlll(x)) &\to& \SRlll(\Sllet(x))
\\
\Sllet(\SRlll(x)) &\to& \SRlll(\Slapp(\Sllet(x)))
&\Sllet(\SRlll(x)) &\to& \SRlll(\SRlll(x))
\\
\Slapp(\SR(x)) &\to& \SR(\Slapp(x))
&\Slapp(\SR(x)) &\to& \SRlll(\SR(x))
\\
\Sllet(\SRlll(x)) &\to& \SRlll(\SRlll(\SRlll(x)))
&\Slapp(\SR(x)) &\to& \SR(x)
\\
\Slapp(\SRlll(x)) &\to& \SRlll(\Slapp(x))
&\Slapp(\SRlll(x)) &\to& \SRlll(\SRlll(x))
\end{array}$$
\end{minipage}
\normalsize
\caption{Diagrams, Base Cases and TRSs for $(\lllrule)$\label{fig:lll-diagrams}}
\end{figure*}

We prove the correctness of transformation $\xrightarrow{\lllrule}$ by the diagram method.
% using diagrams as described before.
Forking and commuting diagrams for $\xrightarrow{\SCtxt,lll}$ were computed by the automated tool. Additionally, unifying WHNFs with left or right-hand sides of transformation $\xrightarrow{\SCtxt,lll}$, then applying the unifier to the other side of the rule,
and then applying standard reductions to 
% applied
%(the right or the left-hand side).
% Applying standard reduction to 
% to
the obtained expressions allows to 
 compute the base case automatically. We show the diagram and the base case in \cref{fig:lll-diagrams}. 
%  and use them in the following proofs.

\begin{proposition}
The transformation $\xrightarrow{\lllrule}$ is correct.
\end{proposition}
\begin{proof}
We first show $\xrightarrow{\SCtxt,\lllrule} \subseteq \leq_c$.
Let $s,t \in \Expr$ with $s \xrightarrow{\SCtxt,\lllrule} t$. We show that  for every evaluation $s \evaluation_L s'\in\Eval(s)$, there exists  $t \evaluation_L t' \in \Eval(t)$.
The technique to show that any reduction sequence $s'\xleftarrow{\sr,*} \cdot \xleftarrow{\sr,a} s \xrightarrow{\SCtxt,\lllrule} t$ where $s'$ is a WHNF can be 
transformed in a finite number of steps into  $s'\xrightarrow{\SCtxt,\lllrule,*}t'\xleftarrow{\sr,*} t$
by applying the forking diagrams.
As a base case, we have that $t'$ is a WHNF, or can be reduced to a WHNF not changing the prob-sequence.
Termination of transforming the reduction sequence can be shown by 
proving termination of the rewrite system induced by the diagrams: they are encoded as the term rewrite system (TRS) $R_1$ shown in \cref{fig:lll-diagrams} and a termination prover is applied
to show innermost termination (in \cite{rau-sabel-schmidtschauss:12} this technique is explained in detail)\footnote{An alternative manual proof is given in \ref{appendix-lll}, \cref{lemma-lll-fork}.}.
In $R_1$ the function symbol $\Slll$ represents any $\xrightarrow{\SCtxt,\lllrule}$-step, $\SR$ represents any $\xleftarrow{\sr,a}$-step with $a \not\in \{\lapp,\llet\}$, and $\SRlll$ represents a step $\xleftarrow{\sr,llet}$ or $\xleftarrow{\sr,lapp}$,
and $x$ is a variable. The termination prover TTT \cite{ttt,ttt-05,ttt2,ttt-09} delivers the Knuth-Bendix-order (KBO) with weight function $w(\Slll) = 0$, $w(\SRlll) = w(\SR) = 1$ and precedence $\Slll > \SR, \Slll > \SRlll$ and $\Slll > c$ (for any constant $c$) proving innermost-termination. 
The diagrams show that prob-sequences are kept by the construction.

For $\xrightarrow{\lllrule}\subseteq\geq_c$, we show that if $s \xrightarrow{\SCtxt,\lllrule} t$,
then for every evaluation $t \evaluation_L t'\in\Eval(t)$,
there exists $s \evaluation_L s' \in \Eval(s)$.
We apply the commuting diagrams starting with an evaluation for $t$ to derive an evaluation for $s$. Termination of the rewriting on the sequences can be shown by proving innermost termination of the TRS
$R_2$ shown in \cref{fig:lll-diagrams}\footnote{An alternative manual proof is given in \ref{appendix-lll}, \cref{lemma:lll-commuting}}.
Here $\SRlll$ represents a $\xrightarrow{\sr,a}$-step with $a\in\{\llet,\lapp\}$, $\SR$ represents a $\xrightarrow{\sr,a}$-step
with $a\not\in\{\llet,\lapp\}$, $\Slapp$ represents $\xrightarrow{\SCtxt,\lapp}$, and $\Sllet$ represents a $\xrightarrow{\SCtxt,\llet}$-step (and $x$ is a variable).
Applying TTT to $R$ shows (innermost) termination using the lexicographic path order (LPO): $\Sllet > \Slapp > \SR$ and $\Slapp > \SRlll$. 
The diagrams also show that prob-sequences are kept by the construction.

Finally, \cref{prop-surface-same-prob-seq} shows that $\xrightarrow{\lllrule}$ is correct.
\end{proof}

% Applying 
% \cref{lemma:lll-forking,lemma:lll-commuting} to \cref{prop-surface-same-prob-seq}  shows:
% \begin{proposition}
% The transformation $\xrightarrow{\lllrule}$ is a correct program transformation.
% \end{proposition}

% \subsection{Correctness of Copying Abstractions by (\cp)}
For transformation $\xrightarrow{\cp}$, we computed the diagrams and the base case automatically, 
the results are shown in \cref{fig:cp-diagrams}.
where $(\cp)$ is split into  $(\cpS)$ and $(\cpd)$,
distinguishing whether the target of the copy-operation is inside surface context, or it is inside an abstraction. 
%This case-distinction later helps to prove termination of the diagram application.
\begin{figure*}[t]
\begin{minipage}[t]{.37\textwidth}
\textbf{Forking diagrams:}\\
$
\begin{array}[t]{@{}c@{}}
\begin{array}[t]{@{}c@{}}
\\[-3ex]
\xymatrix@R=5mm@C=5mm{
 \cdot \ar@/^-.5pc/[r]_{\sr,cp}\ar@/^.5pc/[r]^{\SCtxt,\cps}  & \cdot
}
\end{array}
\begin{array}[t]{@{}c@{}}
\xymatrix@R=5mm@C=5mm{
 \cdot \ar[d]_{\sr,a}\ar[r]^{\SCtxt,\cps}  & \cdot \ar@{-->}[d]_{\sr,a}\\
 \cdot \ar@{-->}[r]_{\SCtxt,\cps}         & \cdot \\
}
\end{array}
%%%%%%%%%%%%%%%%%%%%%%%%%%%%%%%%%%%
\begin{array}[t]{@{}c@{}}
\xymatrix@R=5mm@C=5mm{
 \cdot \ar[d]_{\sr,a}\ar[r]^{\SCtxt,\cps}  & \cdot \ar@{-->}[dl]^{\sr,a}\\
 \cdot \\
}
\\
a \in \{\problabl,\problabr\}
\end{array}
%%%%%%%%%%%%%%%%%%%%%%%%%%%%%%%%%%%
\begin{array}[t]{@{}c@{}}
\xymatrix@R=5mm@C=5mm{
 \cdot \ar[d]_{\sr,a}\ar[r]^{\SCtxt,\cpd}  & \cdot \ar@{-->}[d]_{\sr,a}\\
 \cdot \ar@{-->}[r]_{\SCtxt,\cpd}         & \cdot \\
}
\end{array}
\\
%%%%%%%%%%%%%%%%%%%%%%%%%%%%%%%%%%%
\begin{array}[t]{@{}c@{}}
\xymatrix@R=5mm@C=7mm{
 \cdot \ar[d]_{\sr,\lbeta}\ar[r]^{\SCtxt,\cpd}  & \cdot \ar@{-->}[d]_{\sr,\lbeta}\\
 \cdot \ar@{-->}[r]_{\SCtxt,\cps}         & \cdot \\
}
\end{array}
%%%%%%%%%%%%%%%%%%%%%%%%%%%%%%%%%%%
\begin{array}[t]{@{}c@{}}
\xymatrix@R=5mm@C=5mm{
 \cdot \ar[d]_{\sr,a}\ar[r]^{\SCtxt,\cpd}  & \cdot \ar@{-->}[dl]^{\sr,a}\\
 \cdot \\
}
\\
a \in \{\problabl,\problabr\}
\end{array}
%%%%%%%%%%%%%%%%%%%%%%%%%%%%%%%%%%%
\begin{array}[t]{@{}c@{}}
\xymatrix@R=5mm@C=5mm{
 \cdot \ar[d]_{\sr,\cp}\ar[rr]^{\SCtxt,\cpd}&                            & \cdot \ar@{-->}[d]_{\sr,\cp}\\
 \cdot \ar@{-->}[r]_{\SCtxt,\cpd}         & \cdot \ar@{-->}[r]_{\SCtxt,\cpd} & \cdot\\
}
\end{array}
\end{array}
$

%\begin{verbatim}
% <-sr,a- . -S,\cps-> ~~> -S,\cps-> . <-sr,a-
% <-sr,probl- . -S,\cps-> ~~> <-sr,probl-
% <-sr,probr- . -S,\cps-> ~~> <-sr,probr-
%\end{verbatim}

% Answer diagram.
%\begin{verbatim}
% ANSWER . -S,\cps-> ~~> ANSWER
%\end{verbatim}

%\subsubsection{Set of forking diagrams for $\xrightarrow{\SCtxt,\cpd}$}
%\begin{verbatim}
% <-sr,a- . -S,\cpd-> ~~> -S,\cpd-> . <-sr,a-
% <-sr,lbeta- . -S,\cpd-> ~~> -S,\cps-> . <-sr,lbeta-
% <-sr,probl- . -S,\cpd-> ~~> <-sr,probl-
% <-sr,probr- . -S,\cpd-> ~~> <-sr,probr-
% <-sr,cp- . -S,\cpd-> ~~> -S,\cpd-> . -S,\cpd-> . <-sr,cp-
%\end{verbatim}

%Answer diagram: 
%\begin{verbatim}
% ANSWER . -S,\cpd-> ~~> ANSWER
%\end{verbatim}
\end{minipage}\begin{minipage}[t]{.4\textwidth}
\textbf{Commuting diagrams:}
\[
\begin{array}[t]{@{}l@{}}
\begin{array}[t]{@{}c@{}}
\xymatrix@R=5mm@C=5mm{
 \cdot \ar@{-->}[d]_{\sr,a}\ar[r]^{\SCtxt,\cps}  & \cdot \ar[d]_{\sr,a}\\
 \cdot \ar@{-->}[r]_{\SCtxt,\cps}         & \cdot \\
}
\end{array}
%%%%%%%%%%%%%%%%%%%%%%%%%%%%%%%%%%%
\begin{array}[t]{@{\!\!\!}c@{}}
\xymatrix@R=5mm@C=5mm{
 \cdot \ar@{-->}[dr]_{\sr,a}\ar[r]^{\SCtxt,\cps}  & \cdot \ar[d]^{\sr,a}\\
 &\cdot \\
}
\\
a \in \{\problabl,\problabr\}
\end{array}
%%%%%%%%%%%%%%%%%%%%%%%%%%%%%%%%%%%
\begin{array}[t]{@{\!\!\!\!\!\!\!\!}c@{}}
\xymatrix@R=5mm@C=5mm{
 \cdot \ar@{-->}[d]_{\sr,a}\ar[r]^{\SCtxt,\cpd}  & \cdot \ar[d]_{\sr,a}\\
 \cdot \ar@{-->}[r]_{\SCtxt,\cpd}         & \cdot \\
}
\end{array}
\\
%%%%%%%%%%%%%%%%%%%%%%%%%%%%%%%%%%%
\begin{array}[t]{@{}c@{}}
\xymatrix@R=5mm@C=7mm{
 \cdot \ar@{-->}[d]_{\sr,\lbeta}\ar[r]^{\SCtxt,\cpd}  & \cdot \ar[d]_{\sr,\lbeta}\\
 \cdot \ar@{-->}[r]_{\SCtxt,\cps}         & \cdot \\
}
\end{array}
%%%%%%%%%%%%%%%%%%%%%%%%%%%%%%%%%%%
\begin{array}[t]{@{\!\!\!\!\!}c@{}}
\xymatrix@R=5mm@C=5mm{
 \cdot \ar@{-->}[dr]_{\sr,a}\ar[r]^{\SCtxt,\cpd}  & \cdot \ar[d]^{\sr,a}\\
                                     & \cdot \\
}
\\
a \in \{\problabl,\problabr\}
\end{array}
%%%%%%%%%%%%%%%%%%%%%%%%%%%%%%%%%%%
\begin{array}[t]{@{\!\!\!\!\!\!\!\!\!}c@{}}
\xymatrix@R=5mm@C=5mm{
 \cdot \ar@{-->}[d]_{\sr,\cp}\ar[rr]^{\SCtxt,\cpd}&                            & \cdot \ar[d]_{\sr,\cp}\\
 \cdot \ar@{-->}[r]_{\SCtxt,\cpd}         & \cdot \ar@{-->}[r]_{\SCtxt,\cpd} & \cdot\\
}
\end{array}
\end{array}\!\!\!\!\!\!\!\!\!\!\!\!\!\!\!\!\!\!\!\!
\begin{array}[t]{@{}c@{}}
%%%%%%%%%%%%%%%%%%%%%%%%%%%%%%%%%%%
\begin{array}[t]{@{\!\!\!\!\!}c@{}}
\xymatrix@R=5mm@C=2mm{
 \cdot \ar@{-->}[dr]_{\sr,\lbeta}\ar[rr]^{\SCtxt,\cps}  && \cdot \ar[dd]^{\sr,\lbeta}\\
 &\cdot\ar@{-->}[dr]_{\sr,\cp} \\
 &&\cdot
}
\end{array}
\end{array}\]

%\begin{verbatim}
% <-sr,a- . <-S,\cps- ~~> <-S,\cps- . <-sr,a-
% <-sr,lbeta- . <-S,\cps- ~~> <-sr,lbeta- . <-sr,\cpd-
% <-sr,probl- . <-S,\cps- ~~> <-sr,probl-
% <-sr,probr- . <-S,\cps- ~~> <-sr,probr-
%\end{verbatim}

% Answer diagram: 
%\begin{verbatim}
% ANSWER . <-S,\cps- ~~> ANSWER
% ANSWER . <-S,\cps- ~~> ANSWER . <-sr,cp-
%\end{verbatim}
\end{minipage}\begin{minipage}[t]{.22\textwidth}
\textbf{TRS $R_1$ for forking diagrams:}\\
{\small$\begin{array}{@{\quad}r@{~}c@{}lr@{}c@{}lr@{}c@{}l}
\Scp(\SR(x)) &\to&  \SR(\Scp(x))
\\
\Scp(\SR(x)) &\to& \SR(\Scp(x))
\\
\Scp(\SR(x)) 	 &\to& \SR(x)
\\
\Scp(\SR(x)) 	 &\to& x
\\
\Scp(\SR(x)) &\to& \SR(\Scp(x))
\\
\Scp(\SR(x)) 	 &\to& \SR(x)
\\
\Scp(\SR(x)) 	 &\to& \SR(\Scp(\Scp(x)))
\end{array}
$}
\end{minipage}
\\
\begin{minipage}[t]{.24\textwidth}
\textbf{Base cases:}
Let $s \xrightarrow{\SCtxt,\cps\vee\cpd} t$
If $s$ is a WHNF, then $t$ is a WHNF.
If $t$ is a WHNF then $s$ is a WHNF or
$s \xrightarrow{\sr,cp} s'$ and $s'$ is a WHNF.
\end{minipage}~\hfill\begin{minipage}[t]{.74\textwidth}
\textbf{TRS $R_2$ for commuting diagrams:}\\
{\centering\small
$\begin{array}{@{}r@{}c@{}lr@{}c@{}lr@{}c@{}l@{}}
\Scpd(\SR(x)) &\to& \SR(x) 
&\Scpd(\SRlbeta(x)) &\to& \SRlbeta(\Scpd(x)) 
&\Scpd(\SRlbeta(x)) &\to& \SRlbeta(\ScpS(x)) 
\\
\Scpd(\SR(x)) &\to& \SR(\Scpd(x))
&\Scpd(\SR(x)) &\to& \SR(\Scpd(\Scpd(x)))
&\ScpS(\SRlbeta(x)) &\to& \SRlbeta(\ScpS(x))
\\
\ScpS(\SR(x)) &\to& \SR(x) 
&\ScpS(\SRlbeta(x)) &\to& \SR(\SRlbeta(x))
&\ScpS(\SR(x)) &\to& \SR(\ScpS(x))
\end{array}$}
\end{minipage}
\caption{Diagrams and Base Cases for $(cp)$\label{fig:cp-diagrams}}
\end{figure*}

If we do not distinguish between $(\cpS)$ and $(\cpd)$, then termination of the diagrams cannot be proved.
% \davidcomment{todo beispiel einfügen}

\ignore{

 We use a measure $\#(\iRed, (M_1,\ldots,M_n))$ that counts the number of reductions of $\iRed$ that are in $M_i$ in the standard reduction $\iRed$.
 A lexicographic comparison  may be used for tuples. 
We will be more explicit in the  respective proofs. Note that these measures are well-founded. i.e, there are no infinite properly decreasing sequences.

\begin{lemma}\label{lemma:diagram-forking-cpS}[cpS forking]
  Let $s$ be an expression, $S$ be a surface context, and $S[s] \xrightarrow{\SCtxt,\cpS} S[t]$ be a transformation. Then an evaluation of $S[s]$ with probability  $q$ can be transformed into a standard reduction of $s'$ having the same probability  $q$, the same prob-length, and the length of the given sr-reduction is increased at most by 1.
\end{lemma}
\begin{proof}
Let $s$ be an expression and $S[s]$ be the start expression, where $S$ is surface context and $s \xrightarrow{\cpS} t$ is the transformation. This implies  $S[s] \xrightarrow{\SCtxt,\cpS} S[t]$. 
If $S[s]$ is a WHNF, then also $S[t]$ is a WHNF. 
In the other case, let $S[s] \xrightarrow{\sr,+} s_0$ be an evaluation.
We apply the forking diagrams for (cpS) in \cref{subsec:diagrams-cp}, where we in addition have to consider the possibility that $\xrightarrow{\SCtxt,\cpS}$ is already an sr-reduction. 
\begin{enumerate}
    \item If $S[s] \xrightarrow{\cpS} S[t]$ is an sr-reduction, then the transformation is finished, since the given standard reduction of $S[s]$   is the result. Moreover, the length of the resulting sequence is the length of the given sr-sequence minus 1. 
    \item The first or second diagram is applicable, or the modification of $\cpS$ into sr. In all the three cases, we can use induction on the length of the evaluation.
\end{enumerate}
Since all \prob{}-reductions are transformed unchanged, the probability and the prob-length is not changed. 
 \end{proof}
 \manfredcomment{Hier ist der Fall dass cpS zu sr wird, beruecksichtigt.}

\begin{lemma}\label{lemma:diagram-commuting-cpS}[cpS commuting]
Let $s$ be an expression, $S$ be a surface context, and   $S[s] \xrightarrow{\SCtxt,\cpS} S[t]$ be a transformation. Then an evaluation of $S[t]$ with probability  $q$ can be transformed into a standard reduction  of  the same probability  $q$, the same prob-length, and  having at most the same length. {\davidins{\bfseries the length is increased by at most 1?}}
\end{lemma}
\begin{proof}
Let $s$ be an expression and $S[s]$ be the start expression, where $S$ is surface context and $s \xrightarrow{\cpS} t$ be the transformation. This implies  $S[s] \xrightarrow{\SCtxt,\cpS} S[t]$. 
If $S[t]$ is a WHNF, then also $S[s]$ is already a WHNF, or the reduction step is already a standard reduction step. In both cases the claim of this lemma holds.\\
In the other case, let $S[t] \xrightarrow{\sr,+} t_0$ be an evaluation.
We apply the commuting diagrams  for (cpS) in \cref{subsec:diagrams-cp}
where we in addition have to consider the case that $\cpS$ is turned into an sr-reduction. 
This is now an easy induction on the length of the evaluation $S[t] \xrightarrow{\sr,+} t_0$, which is a standard reduction.
\end{proof}

% \davidcomment{Das erste Diagramme kann auch $\xrightarrow{\sr,a} \xrightarrow{\sr,cp}$ erzeugen, wenn die existentielle $\xrightarrow{\SCtxt,cpS}$ eine Standardreduktion ist???}
% \manfredcomment{Hatte ich auf dem Schirm, aber nochmal checken.
% OK, hat doch gefehkt in der Indutkion, ist jetzt berücksichtigt} 
\davidcomment{M.E. ist die Aussage des Lemmas damit aber schon verletzt:
$S[s] \xrightarrow{\SCtxt,cpS} S[t] \xrightarrow{\sr} t'$ und $t'$ is WHNF. Wenn danach, das Diagramm angewendet wird: $S[s] \xrightarrow{\sr,2} s'$ mit $s'$ WHNF. Dann ist die Länge der $sr$-Reduktion um 1 groesser?}

\manfredcomment{OK, ich checke bald}
%\manfredcomment{Der Fall war einigermassen einfach...  Jetzt cpd} 

\manfredcomment{Unten die Fälle jetzt nochal durchgegangen} 
\begin{lemma}[cpd forking]
  Let $s,t$ be  expressions, $S$ be a surface context,   $s \xrightarrow{\cpd} s'$, resp.  $S[s] \xrightarrow{\SCtxt,\cpd} S[t]$ be a transformation. Then an evaluation of $S[s]$ with probability  $q$ can be transformed into a standard reduction of $S[t]$, which has  length $\leq$ of the given one, having the same probability and the same prob-length.
\end{lemma}
\begin{proof}
Let $s,t$ be  expressions and $S[s]$ be the start expression, where $S$ is a surface context and $s \xrightarrow{\cpd} t$, resp.  $S[s] \xrightarrow{\SCtxt,\cpd} S[t]$  be the transformation.  
If $S[s]$ is a WHNF, then also $S[s]$ is a WHNF. 
In the general case, let $S[s] \xrightarrow{\sr,+} s_0$ be an evaluation.
We apply the set of 4 forking diagrams for (cpd) in \cref{subsec:diagrams-cp}. The main induction measure is the length of the evaluation. The claim is that an evaluation of $S[t]$ can be constructed of length at most the length of $S[s]$, the same prob-length and the same probability.\\ 
We consider the four forking diagrams for (cpd): 
 \begin{enumerate}
\item This is standard.
\item The second diagrams: this follows from Lemma \ref{lemma:diagram-forking-cpS}.
\item This is standard.
\item In the fourth diagram we get a sr-reduction in the first (cpd) step of  strictly smaller length. Then we use this sr-reduction for the second  (cpd)-step, and obtain the  desired sr-reduction of strictly smaller length.
\end{enumerate}
In all cases the probability is retained and also the prob-length.
\end{proof}

\begin{lemma} [cpd commuting]
  Let $s$ be an expression, $S$ be a surface context, and   $s \xrightarrow{\cpd} t$, resp.   $S[s] \xrightarrow{\SCtxt,\cpd} S[t]$ be a transformation. Then an evaluation of $S[t]$ can be transformed into a standard reduction of $S[s]$ having  the same probability and the same prob-length. 
\end{lemma}
\begin{proof}
The method is similar as in the proof w.r.t. (\cpS). We will view the (mixed) reduction sequence as a data structure consisting of $sr$-reductions and $(\SCtxt,\cpd)$- and $(\SCtxt,cpS)$-reductions. where we distinguish the sr-reductions as \lbeta,\llet,\cp, \prob{}-reductions.
For the proof a measure $\mu$ on mixed reduction sequences of is defined as follows: $(\mu_1,\mu_2)$, where $\mu_1$ is the number of $(sr,\lbeta)$-reduction steps; and $\mu_2$
is the following sum: every $\cpd$-reduction step before the first $(sr,\lbeta)$ contributes $3^m$, where $m$ is the number of reductions in the set $\{(sr,\cp), (sr,\lllrule), \prob{}, \cpS\}$ between the $\cpd$-occurrence and the next $\lbeta$-occurrence. 

We scan the 4 commuting diagrams for {\cpd} in \cref{lemma:diagram-commuting-cpS}.
\begin{enumerate}
    \item The first $\cpd$-diagram shifts a $\cpd$ to the right, which   strictly reduces $\mu_2$.
    \item  The second $\cpd$-diagram shifts a $\cpd$ behind an $\lbeta$, 
    which strictly reduces $\mu_2$.
    \item The third diagram obviously strictly reduces $\mu$
    \item The fourth diagram: The $\cpd$ before the application contributes $3^m$, and the two cpd-reductions after the application
    contributes $3^{m-1}+3^{m-1}+3$,  and hence is strictly smaller than $3^m$. This strictly decreases $\mu_2$.
\end{enumerate}
Hence the translation terminates and produces the desired sr-reduction sequence: it leaves  the probability and the prob-length invariant.
\end{proof}
}

% \davidcomment{Warum cpS und cpd getrennt, man sollte die zusammen beweisen können?
\begin{proposition}
%[Correctness of (cp)]
\label{prop:correctness-cp}
The transformation $(cp)$ is correct.
\end{proposition}
\begin{proof}
 We use \cref{prop-surface-same-prob-seq} and  show
for all  $s,t$ with $s \xrightarrow{\SCtxt,\cp} t$: 
\begin{enumerate}
 \item\label{fork-cp} For every  $s \evaluation_L s'\in\Eval(s)$,
there exists $t \evaluation_L t' \in \Eval(t)$.

 \item\label{commute-cp} For every  $t \evaluation_L t'\in\Eval(t)$,
there exists  $s \evaluation_L s' \in \Eval(s)$.
\end{enumerate}
For proving \cref{fork-cp}, first assume that a sequence $s'\xleftarrow{\sr,*} \cdot \xleftarrow{\sr,a} s \xrightarrow{\SCtxt,\cp} t$ is given, where $s'$ is a WHNF.
This is transformed into a sequence 
$s'\xrightarrow{\SCtxt,cp,*}t'\xleftarrow{\sr,*} t$ by applying the forking diagrams. Termination of this rewriting on the sequences can be shown automatically by proving innermost termination of the TRS $R_1$ shown in \cref{fig:cp-diagrams}.
Here $\SR$ represents an $\xleftarrow{\sr}$-step, $\Scp$ represents an $\xrightarrow{\SCtxt,cpS}$- or $\xrightarrow{\SCtxt,cpd}$-step.
Termination prover TTT shows innermost termination of $R_1$ by using  the KBO with weight function $w(\Scp) = 0$ and $w(\SR) = 1$ and precedence $\Scp > \SR$ and $\Scp > c$. 
After obtaining $s'\xrightarrow{\SCtxt,cp,*}t'\xleftarrow{\sr,*} t$,
the base case for $\xrightarrow{\SCtxt,cp}$ is applied which shows that $t'$ must be a WHNF.
Finally, we observe that the diagrams
do not change the  prob-sequences and/or -length.

For proving \cref{commute-cp}, first assume that a sequence 
$s \xrightarrow{\SCtxt,\cp} t \xrightarrow{\sr,a}   \cdot \xrightarrow{\sr,*}t'$
is given, where $t'$ is a WHNF.
This is transformed into a sequence 
$s \xrightarrow{\sr,*} s' \xrightarrow{\SCtxt,\cp,*} t'$ by applying the commuting diagrams. Termination of  rewriting the sequences is shown automatically by proving innermost termination of the TRS
$R_2$ shown in \cref{fig:cp-diagrams}.
 Here $\SR$ represents any $\xrightarrow{\sr,a}$-reduction with $a\not=\lbeta$, $\SRlbeta$ represents a $\xrightarrow{\sr,lbeta}$-step, and $\Scpd$ and $\ScpS$ represent $\xrightarrow{\SCtxt,cpd}$ and $\xrightarrow{\SCtxt,cpS}$ steps.
TTT or AProVE \cite{aprove,FBEFFOPSKSST:14} are able to show innermost termination of $R_2$ (verifier CeTA \cite{ceta,Thiemann-Sternagel:09} can certify these proofs).
After obtaining $s \xrightarrow{\sr,*} s' \xrightarrow{\SCtxt,\cp,*} t'$ 
the base case for $\xrightarrow{\SCtxt,cp}$ shows that $s'$ can be standard reduced to a WHNF.
The diagrams and base cases show that the prob-sequences  are not changed.
\end{proof}

\ignore{ 
\begin{remark}
Termination of rewriting the reduction sequences for the forking case can be shown automatically by proving innermost termination of the TRS
$R= \{\Scp(\SR(x)) \to  \SR(\Scp(x)),
 \Scp(\SR(x)) 	 \to \SR(\Scp(x)),
 \Scp(\SR(x)) 	 \to \SR(\Scp(x)),
 \Scp(\SR(x)) 	 \to \SR(x),
 \Scp(\SR(x)) 	 \to \SR(x),
 \Scp(\SR(x)) 	 \to \SR(\Scp(\Scp(x)))
 \Scp(\SR(x)) 	 \to x\}$.
 Here $\SR$ represents an $\xleftarrow{\sr}$-step, $\Scp$ represents an $\xrightarrow{\SCtxt,cpS}$- or $\xrightarrow{\SCtxt,cpd}$-step.
 TTT shows innermost termination of $R$ by using  the Knuth-Bendix order with weight function $w(\Scp) = 0$ and $w(\SR) = 1$ and precedence $\Scp > \SR$ and $\Scp > [c]$. 
\end{remark}

\begin{remark}
(Innermost) termination of rewriting the reduction sequence for the commuting case can be shown automatically by TTT or AProVE
applied to the TRS
$R=\{
\Scpd(\SRlbeta(x)) \to \SRlbeta(\Scpd(x))$, 
$\Scpd(\SRlbeta(x)) \to \SRlbeta(\ScpS(x))$, 
$\Scpd(\SR(x)) \to \SR(x)$, 
$\Scpd(\SR(x)) \to \SR(\Scpd(x))$, 
$\Scpd(\SR(x)) \to \SR(\Scpd(\Scpd(x)))$, 
$\ScpS(\SRlbeta(x)) \to \SRlbeta(\ScpS(x))$, 
$\ScpS(\SRlbeta(x)) \to \SR(\SRlbeta(x))$, 
$\ScpS(\SR(x)) \to \SR(x)$, 
$\ScpS(\SR(x)) \to \SR(\ScpS(x))\}
\}$. Here $\SR$ represents any $\xrightarrow{\sr,a}$-reduction with $a\not=\lbeta$, $\SRlbeta$ represents a $\xrightarrow{\sr,lbeta}$-step, and $\Scpd$ and $\ScpS$ represent $\xrightarrow{\SCtxt,cpd}$ and $\xrightarrow{\SCtxt,cpS}$ steps.
\end{remark}

 \manfredcomment{ich finde Dein Argument oben überzeugend, und würde keinen eigenen extra Beweis fü cp machen. bzw meine Versuche zu cp eliminieren. 
 Die Remarks zu Lemmas machen und noch extra  Propositions für forking und commuting machen, um die Terminierung auszunutzen. dh die Normalformen aufzeigen und fertig.
 
  Du Kannst meine cp-Versuche oben wegwerfen. 
}
 \davidcomment{ok, mach ich bald, auch die restlichen Transformationen...}
}
% \begin{lemma} 
%
 % The transformation $\cpd$ transforms a standard reduction $\iRed$ into
 %  a standard reduction $\iRed'$ with $\mu(\iRed') \leq \mu(\iRed)$,   where $\mu(r) =  \#(r,(\{\lbeta,\prob\}, \{llet,\cp\})$.
% \end{lemma}

% \subsection{Correctness of (\cpx) and (\xch)}
The correctness proofs for $\xrightarrow{\cpx}$ and $\xrightarrow{\xch}$ are straight-forward and we omit them 
(they are given in the appendix).
\begin{proposition}\label{prop-cpx-xch-correct}
The transformations $\xrightarrow{\cpx}$ and $\xrightarrow{\xch}$
are correct.
\end{proposition}
The diagrams and base cases for (gc) and (ucp) are in \cref{fig:ucp-diagrams}.
\begin{figure*}[t]
\begin{minipage}[t]{.27\textwidth}
\textbf{Forking diagrams:}
\[
\begin{array}{@{}c@{}}
\begin{array}[t]{@{\!\!\!}c@{}}
\xymatrix@R=4mm@C=4mm{
 \cdot \ar[d]_{\sr,a}\ar[r]^{\SCtxt,b}  & \cdot \ar@{-->}[d]^{\sr,a}\\
 \cdot \ar@{-->}[r]_{\SCtxt,b}         & \cdot \\
}
\\
b{\in}\{\gc,\\\qquad\ucp\}
\end{array}
%%%%%%%%%%%%%%%%%%%%%%%%%%%%%%%%%%%
\begin{array}[t]{@{\!\!\!\!\!}c@{}}
\xymatrix@R=4mm@C=4mm{
 \cdot \ar[d]_{\sr,a}\ar[r]^{\SCtxt,\gc}  & \cdot \ar@{-->}[dl]^{\sr,a}\\
 \cdot \\
}
\\
a {\in} \{\llet,\lapp,\\
~~\problabl,\problabr\}
\end{array}
% \begin{array}[t]{@{\!\!}c@{}}
% \xymatrix@R=4mm@C=4mm{
%  \cdot \ar[d]_{\sr,a}\ar[r]^{\SCtxt,\ucp}  & \cdot \ar@{-->}[d]^{\sr,a}\\
%  \cdot \ar@{-->}[r]_{\SCtxt,\ucp}         & \cdot \\
% }
% \end{array}
%%%%%%%%%%%%%%%%%%%%%%%%%%%%%%%%%%%
\begin{array}[t]{@{\!\!\!}c@{}}
\xymatrix@R=4mm@C=4mm{
 \cdot \ar[d]_{\sr,a}\ar[r]^{\SCtxt,\ucp}  & \cdot \ar@{-->}[dl]^{\sr,a}\\
 \cdot \\
}
\\
a {\in} \{\problabl,\\\quad\problabr\}
\end{array}
\\
%%%%%%%%%%%%%%%%%%%%%%%%%%%%%%%%%%%
\begin{array}[t]{@{\!\!\!\!}c@{}}
\xymatrix@R=4mm@C=4mm{
 \cdot \ar[d]_{\sr,a}\ar[r]^{\SCtxt,\ucp}  & \cdot \ar@{-->}[d]^{\sr,a}\\
 \cdot \ar@{-->}[r]_{\SCtxt,\gc}         & \cdot \\
}
\\
a {\in} \{\problabl,\\\quad\problabr,\cp\}
\end{array}
%%%%%%%%%%%%%%%%%%%%%%%%%%%%%%%%%%%
\begin{array}[t]{@{\!\!\!}c@{}}
\xymatrix@R=4mm@C=4mm{
 \cdot \ar[d]_{\sr,\cp}\ar[r]^{\SCtxt,\ucp}  & \cdot \\
 \cdot \ar@{-->}[ru]_{\SCtxt,\gc}         \\
}
\end{array}
%%%%%%%%%%%%%%%%%%%%%%%%%%%%%%%%%%%
\begin{array}[t]{@{\!\!\!\!\!}c@{}}
\xymatrix@R=4mm@C=7mm{
 \cdot \ar[d]_{\sr,\llet}\ar[r]^{\SCtxt,\ucp}  & \cdot \ar@{-->}[d]_{\sr,\lllrule,+}\\
 \cdot \ar@{-->}[r]_{\SCtxt,\ucp}         & \cdot \\
}
\end{array}
\end{array}
\]
\end{minipage}%%%%%%%%%%%%%%%%%%
\begin{minipage}[t]{.73\textwidth}
\textbf{Commuting diagrams:}
\[
\begin{array}[t]{@{}c@{}}
\begin{array}[t]{@{}c@{}}
\begin{array}[t]{@{}c@{}}
\xymatrix@R=4mm@C=4mm{
 \cdot \ar@{-->}[d]_{\sr,a}\ar[r]^{\SCtxt,b}  & \cdot \ar[d]^{\sr,a}\\
 \cdot \ar@{-->}[r]_{\SCtxt,b}         & \cdot \\
}
\\
b{\in}\{\gc,\\\qquad\ucp\}
\end{array}
\\
\begin{array}[t]{@{}c@{}}
\xymatrix@R=4mm@C=4mm{
 \cdot \ar@{-->}[d]_{\sr,\lllrule,+}\ar[r]^{\SCtxt,\ucp} & \cdot \ar[ddd]^{\sr,\lbeta}\\
 \cdot\ar@{-->}[d]_{\sr,\cp}\\
 \cdot\ar@{-->}[d]_{\sr,\lbeta}\\
 \cdot\ar@{-->}[r]_{\SCtxt,\gc}                   &\cdot 
}
\end{array}
\end{array}
%%%%%%%%%%%%%%%%%%%%%%%%%%%%%%%%%%%
\begin{array}[t]{@{\!\!\!\!\!\!\!\!}c@{}}
\begin{array}[t]{@{}c@{}}
\xymatrix@R=4mm@C=4mm{
 \cdot \ar@{-->}[dr]_{\sr,a}\ar[r]^{\SCtxt,\gc\vee\ucp}  & \cdot \ar[d]^{\sr,a}\\
 &\cdot \\
}
\\
a {\in} \{\problabl,\\\quad\problabr\}
\end{array}
\\
\begin{array}[t]{@{}c@{}}
\xymatrix@R=4mm@C=4mm{
 \cdot \ar@{-->}[d]_{\sr,\lllrule,+}\ar[r]^{\SCtxt,\ucp} & \cdot \ar[dd]^{\sr,a}\\
 \cdot\ar@{-->}[d]_{\sr,a}\\
 \cdot\ar@{-->}[r]_{\SCtxt,\ucp \vee \gc}                   &\cdot 
}
\end{array}
\end{array}
\begin{array}[t]{@{\!\!\!\!\!\!\!\!}c@{}}
\begin{array}[t]{@{}c@{}}
\xymatrix@R=4mm@C=4mm{
 \cdot \ar@{-->}[d]_{\sr,a}\ar[r]^{\SCtxt,\gc}& \cdot \ar[dd]^{\sr,a}\\
 \cdot\ar@{-->}[d]_{\sr,\lllrule,+}
 \\
 \cdot \ar@{-->}[r]_{\SCtxt,\gc} & \cdot\\
}
\\
\begin{array}[t]{@{}c@{}}
\xymatrix@R=4mm@C=4mm{
 \cdot \ar@{-->}[d]_{\sr,\cp}\ar[r]^{\SCtxt,\ucp} & \cdot \ar[dd]^{\sr,\lbeta}\\
 \cdot\ar@{-->}[d]_{\sr,\lbeta}\\
 \cdot\ar@{-->}[r]_{\SCtxt,\gc}                   &\cdot 
}
\end{array}
\end{array}
\end{array}
%%%%%%%%%%%%%%%%%%%%%%%%%%%%%%%%%%%
% \begin{array}[t]{@{}c@{}}
% \xymatrix@R=4mm@C=4mm{
%  \cdot \ar@{-->}[d]_{\sr,a}\ar[r]^{\SCtxt,\ucp}  & \cdot \ar[d]^{\sr,a}\\
%  \cdot \ar@{-->}[r]_{\SCtxt,\ucp}         & \cdot \\
% }
% \end{array}
%%%%%%%%%%%%%%%%%%%%%%%%%%%%%%%%%%%
% \begin{array}[t]{@{}c@{}}
% \xymatrix@R=4mm@C=4mm{
%  \cdot \ar@{-->}[dr]_{\sr,a}\ar[r]^{\SCtxt,\ucp}  & \cdot \ar[d]^{\sr,a}\\
%  &\cdot \\
% }
% \\
% a \in \{\problabl,\\\problabr\}
% \end{array}
% \!\!\!
%%%%%%%%%%%%%%%%%%%%%%%%%%%%%%%%%%%
\begin{array}[t]{@{\!\!\!\!}c@{}}
\begin{array}[t]{@{}c@{}}
\xymatrix@R=4mm@C=4mm{
 \cdot \ar@{-->}[d]_{\sr,a}\ar[r]^{\SCtxt,\ucp}  & \cdot \ar[d]^{\sr,a}\\
 \cdot \ar@{-->}[r]_{\SCtxt,\gc}         & \cdot \\
}
\\
a{\in}\{\problabl,\\\quad\problabr,\cp\}
\\
\begin{array}[t]{@{}c@{}}
\xymatrix@R=4mm@C=4mm{
 \cdot \ar@{-->}[d]_{\sr,\lllrule}\ar[r]^{\SCtxt,\ucp} & \cdot \ar[d]^{\sr,\lllrule,+}\\
 \cdot\ar@{-->}[r]_{\SCtxt,\ucp}                   &\cdot 
}
\end{array}
\end{array}
\end{array}
\begin{array}[t]{@{}c@{}}
\begin{array}[t]{@{}c@{}}
\xymatrix@R=4mm@C=4mm{
 \cdot \ar@{-->}[d]_{\sr,lbeta}\ar[r]^{\SCtxt,\ucp} & \cdot \ar[dd]^{\sr,\lbeta}\\
 \cdot\ar@{-->}[d]_{\sr,\llet}\\
 \cdot\ar@{-->}[r]_{\SCtxt,\ucp}                   &\cdot 
}
\end{array}
\\
\begin{array}[t]{@{}c@{}}
\xymatrix@R=4mm@C=4mm{
 \cdot \ar@{-->}[d]_{\sr,\lllrule}\ar[r]^{\SCtxt,\ucp} & \cdot \ar[dd]^{\sr,\lllrule,+}\\
 \cdot\ar@{-->}[d]_{\sr,\lllrule,+}\\
 \cdot\ar@{-->}[r]_{\SCtxt,\ucp}                   &\cdot 
}
\end{array}
\end{array}
\begin{array}[t]{@{\!\!\!\!\!}c@{}}
\xymatrix@R=4mm@C=4mm{
 \cdot \ar@{-->}[d]_{\sr,\lllrule}\ar[r]^{\SCtxt,\gc}& \cdot \ar[ddd]^{\sr,a}\\
 \cdot\ar@{-->}[d]_{\sr,a}
 \\
 \cdot\ar@{-->}[d]_{\sr,\lllrule}
 \\
 \cdot \ar@{-->}[r]_{\SCtxt,\gc} & \cdot\\
}
\\
a {\in} \{\lbeta,\\\quad\cp,\lapp\}
\end{array}
\\
%%%%%%%%%%%%%%%%%%%%%%%%%%%%%%%%%%%
%%%%%%%%%%%%%%%%%%%%%%%%%%%%%%%%%%%
%%%%%%%%%%%%%%%%%%%%%%%%%%%%%%%%%%%
%%%%%%%%%%%%%%%%%%%%%%%%%%%%%%%%%%%
\end{array}
\]
\end{minipage}

\begin{minipage}[t]{.27\textwidth}
\textbf{Base cases:}
Let $s \xrightarrow{\SCtxt,\gc\vee\ucp} t$.

If $s$ is a WHNF, then $t$ is a WHNF.

If $t$ is a WHNF, then $s \xrightarrow{\sr,\llet,0\vee 1} s'
\xrightarrow{\sr,\cp,0\vee 1} s''$ where $s''$ is a WHNF.

~~\\[-1ex]
{\bfseries TRS $R_1$ for forking diagrams:}
\small$$\begin{array}{@{}r@{}c@{}lr@{}c@{}lr@{}c@{}l}
\Sug(\SR(x)) &\to& \SR(x)
\\
\Sug(\SR(x)) &\to& \SR(\Sug(x))
\\
\Sug(\SRlll(x)) &\to& \SRlll(\Sug(x))
\\
\Sug(\SR(x)) &\to& \Sug(x)
\\
\Sug(\SRlll(x)) &\to& W(k,x)
\\
W(s(k),x) &\to& \SRlll(W(k,x))
\\
\Sug(\SRlll(x)) &\to& \SRlll(\Sug(x))
\\
\Sug(\SRlll(x)) &\to& \Sug(x)
\\
W(s(k),x) &\to& \SRlll(\Sug(x))
\end{array}
$$
\end{minipage}\begin{minipage}[t]{.73\textwidth}
{\bfseries TRS $R_2$ for commuting diagrams:}\\
\small
$\begin{array}{r@{}c@{}l@{}}
\Sug(\SR(x)) &\to& \SR(\Sug(x))\\
\Sug(\SR(x)) &\to& \SR(x)\\
\Sug(\SR(x)) &\to& \SR(\SR(\Sug(x)))\\
\Sug(\SRlll(x)) &\to& \SRlll(\Sug(x))\\
\Sug(\SR(x)) &\to& \SR(\SRlll(\Sug(x)))\\
\Sug(\SR(x)) &\to& \SRlll(\SR(\Sug(x)))\\
\Sug(\SR(x)) &\to& \SRlll(\SR(\SR(\Sug(x))))\\
\Sug(\SRlll(x)) &\to& \SRlll(\SRlll(\Sug(x)))\\
\Sug(\SR(x)) &\to& \SRlll(\SR(\SRlll(\Sug(x))))\\
\Sug(\SR(x)) &\to& \SRlll(\SRlll(\SR(\Sug(x))))\\
\Sug(\SRlll(\SRlll(x))) &\to& \SRlll(\Sug(x))\\
\Sug(\SRlll(\SRlll(x))) &\to&\SRlll(\SRlll(\SRlll(\Sug(x))))\\
\Sug(\SR(x)) &\to& \SRlll(\SR(\SR(\SRlll(\Sug(x)))))\\
\Sug(\SR(x)) &\to& \SRlll(\SRlll(\SR(\SR(\Sug(x)))))\\
\Sug(\SRlll(W(x))) &\to& V(k,x)\\
\end{array}$~$\begin{array}{@{}r@{}c@{}l}
\Sug(\SRlll(W(\SRlll(x)))) &\to& \Sug(\SRlll(W(x)))\\
\Sug(\SRlll(\SRlll(x))) &\to& \SRlll(\SRlll(\Sug(x)))\\
\Sug(\SRlll(x)) &\to& \SRlll(\SRlll(\SRlll(\Sug(x))))\\
\Sug(\SRlll(x)) &\to& \SRlll(V(k,x))\\
\Sug(\SRlll(\SRlll(x))) &\to& \Sug(\SRlll(W(x)))\\
V(s(k),x) &\to& \SRlll(V(k,x))\\
V(s(k),x) &\to& \SRlll(\SRlll(\Sug(x)))\\
V(s(k),x) &\to& \SRlll(\SR(\SR(\Sug(x))))\\
V(s(k),x) &\to& \SRlll(\SR(\Sug(x)))\\
\Sug(\SRlll(x)) &\to& V(k,x)\\
\Sug(\SR(x)) &\to& \SRlll(V(k,x))\\
\Sug(\SR(x)) &\to& V(k,x)\\
\Sug(\SRlll(W(x))) &\to& \SRlll(\SRlll(\Sug(x)))\\
\Sug(\SRlll(W(x))) &\to& \SRlll(V(k,x))\\
\Sug(\SRlll(W(x))) &\to& \SRlll(\Sug(x))\\
\end{array}$
\end{minipage}
\normalsize
\caption{Diagrams, Base Cases, and TRSs for $(gc)$ and $(ucp)$\label{fig:ucp-diagrams}\label{fig:ug-trs}}
\end{figure*}

\begin{proposition}
The transformations $\xrightarrow{gc}$ and $\xrightarrow{\ucp}$ are correct.
\end{proposition}
\begin{proof}
The proof is analogous to the proof of \cref{prop:correctness-cp}, where the diagrams and the bases cases for (ucp) and (gc) are applied. They show that 
prob-sequences are not changed. For proving termination of diagram application, we use automated termination techniques, where, however, we have to encode the transitive closure that occurs in the diagrams. In \cite{rau-sabel-schmidtschauss:12} they were encoded by integer rewrite systems and free integer variables to guess any number. We use a similar TRS where free variables occur on the right-hand sides and are interpreted as any constructor term. We encode numbers using the Peano-encoding. An adapted version of AProVE and the certifier CeTA can handle those termination problems. For the forking diagrams, AProVE shows innermost termination of the TRS $R_1$ %with free variables shown 
in \cref{fig:ug-trs}, where $\SR$ represents any $\xleftarrow{\sr,a}$-reduction that is not an $\xleftarrow{\sr,lll}$-step, $\SRlll$ represents $\xleftarrow{\sr,lll}$-reductions $\Sug$ represents the union of $\xrightarrow{\SCtxt,ucp}$ and $\xrightarrow{\SCtxt,gc}$,
$k$ and $x$ are variables, $s$ represents the successor of Peano-numbers,
$W$ is used to generate the transformations from the guessed number. CeTA certifies the proof.

For the commuting diagrams the TRS  $R_2$ shown in \cref{fig:ug-trs} is shown to be innermost terminating by AProVE and the proof is certified by CeTA.  In the TRS $R_2$, $\SR$ represents any $\xrightarrow{\sr,a}$-reduction that is not an $\xrightarrow{\sr,lll}$-step, $\SRlll$ represents $\xrightarrow{\sr,lll}$-reductions, $\Sug$ represents the union of $\xrightarrow{\SCtxt,ucp}$ and $\xrightarrow{\SCtxt,gc}$,
$k$ and $x$ are variables, $s$ is the successor-constructor of Peano-numbers,
$V,W$ are used to generate the transformations from the guessed number.
\end{proof}

\ignore{
\subsection{On the Correctness of the prob-Transformations}
% The $\prob{}{}$-operator is not associative, since applying it, the corresponding transformation changes the probabilities, i.e.:
% \begin{proposition}
% The transformation $\probassoc$ is not correct.
% \end{proposition}
% \begin{proof}
% Consider three semantically different WHNFs $w_1,w_2,w_3$ (for instance, $w_i = \lambda x_1.\lambda x_2.\lambda x_3.x_i$).
% Then  $\prob{(\prob{w_1}{w_2})}{w_3}\downarrow_{0.5} w_1$ but 
% $\prob{w_1}{(\prob{w_2}{w_3})}\downarrow_{0.25} w_1$, which is incompatible with the definition of contextual equivalence.
% \end{proof}

\manfredcomment{ Jetzt ist die Korrektheit der probs  weiter oben in anderer Form}

\davidcomment{Mit dem extended context lemma bewiesen:}
\davidins{
\begin{theorem}
The transformations  $(\probid)$, $(\probcomm)$, $(\probprob)$   are correct
\end{theorem}
\begin{proof}
We use the extended context lemma to argue for correctness. 
\begin{enumerate}
 \item 
Let $s \xrightarrow{\RCtxt,\probid} t$. This means $s = R[\prob{s'}{s'}] \xrightarrow{\sr,\problabl} R[s'] =t$  as well as $R[\prob{s'}{s'}] \xrightarrow{\sr,\problabr} R[s']$. 
Then $\EC(R[\prob{s'}{s'}],0) = 0$ and for all $k > 0$:
 $$\begin{array}{@{}l@{~}c@{~}l@{}}
 \EC(R[\prob{s'}{s'}],k) 
 &=& 0.5\EC(R[\prob{s'}{s'}],k-1)\\
&&+0.5\EC(R[\prob{s'}{s'}],k-1) \\
&=& \EC(R[s'],k-1)
\end{array}$$
Since $\EC(R[s'],k-1) \leq \EC(R[s'],k)$ this shows
for all $k \geq 0$: $\EC(R[\prob{s'}{s'},k]) \leq \EC(R[s'],k)$.

For the other direction the equation
$\EC(R[\prob{s'}{s'}],k) = \EC(R[s'],k-1)$ for all $ k > 0$
implies also
$\EC(R[\prob{s'}{s'}],k+1) \geq \EC(R[s'],k)$ for all $ k \geq 0$.

Hence using the extended context-lemma \cref{thm:context-lemma-ext} in both directions we conclude that $\probid$ is correct.
\item
Let $s \xrightarrow{\RCtxt,\probcomm} t$. This means $s=R[\prob{s'}{t'}]$ and $t=R[\prob{t'}{s'}]$.
Then $s \xrightarrow{\sr,\problabl} R[s']$,
$s \xrightarrow{\sr,\problabr}R[t']$,
$t \xrightarrow{\sr,\problabr} R[s']$,
$t' \xrightarrow{\sr,\problabl}R[t']$, which shows
for all $k \geq > 0$
$\EC(R[s],k) = 0.5\EC(R[s'],k-1) + 0.5\EC(R[t'],k-1) = \EC(R[t],k)$. Since $\EC(R[s],0) = 0 = \EC(R[t],0)$ we can apply the context lemma (\cref{thm:context-lemma-ext}) (in both directions) which shows correctness of $\probcomm$.
\item 
Let $s \xrightarrow{\RCtxt,\probprob} t$. This means $s=R[\prob{r'}{(\prob{s'}{t'})}]$ and 
$t=R[\prob{(\prob{r'}{s'})}{(\prob{r'}{t'})}]$.
Then
$$
\xymatrix@R=-2mm@C=18mm{
                         \\&R[r']
\\ 
s \ar[ur]^{\sr,\problabl}   
      \ar[dr]_{\sr,\problabr} && R[s']\\
                           &R[\prob{s'}{t'}]
                            \ar[ur]^{\sr,\problabl}
                            \ar[dr]_{\sr,\problabr} 
                            \\
                            && R[t']
}
$$
$$
\xymatrix@R=-2mm@C=18mm{
&&R[r']
\\&R[\prob{r'}{s'}]
                            \ar[ur]^{\sr,\problabl}
                            \ar[dr]_{\sr,\problabr} 
\\ 
&&R[s']
\\
t \ar[uur]^{\sr,\problabl}   
      \ar[ddr]_{\sr,\problabr} && \\
      &&R[r']\\
                           &R[\prob{r'}{t'}]
                            \ar[ur]^{\sr,\problabl}
                            \ar[dr]_{\sr,\problabr} 
                            \\
                            && R[t']
}
$$
This shows for all $k \geq 2$:
$$\begin{array}{@{}l@{~}c@{}l@{}}
\EC(s,k) &=& 0.5\EC(R[r'],k-1)\\
&&+0.25\EC(R[s'],k-2)\\
&&+0.25\EC(R[t'],k-2)
\\
\EC(t,k) &=& 0.5\EC(R[r'],k-2)\\
&&+0.25\EC(R[s'],k-2)\\
&&+0.25\EC(R[t'],k-2)
\end{array}
$$
and 
$\EC(s,1) = 0.5(\EC(R[r'],0)$,
$\EC(t,1)=0$
and $\EC(s,0)=0=\EC(t,0)$

Thus for all $k \geq 0$: $\EC(t,k) \leq \EC(s,k)$
and $\EC(s,k) \leq \EC(t,k+1)$.
The context lemma (\cref{thm:context-lemma-ext}) applied for both inequations shows the correctness of $\probprob$.
\qedhere\end{enumerate}

\end{proof}
 %enddavidins
 }
 \endignore}
\ignore{
\manfredcomment{erlaub mir mal, das folgende zu versuchen}

\begin{theorem}
The transformations  $(\probid)$, $(\probcomm)$, $(\probprob)$   are correct
\end{theorem}
\begin{proof}
We use the context lemma with reduction contexts to argue for correctness. 
\begin{enumerate}
 \item 
 \davidcomment{Kontexlemma-Voraussetzung stimmt aber trotzdem nicht für k=0 und von rechts nach links 
 ($\EC(t,0)=\EC(R[s'],0)$ aber $\EC(R[s],0) = 0$ (da jede Evaluation prob-length $> 0$ hat.}
Let $s \xrightarrow{\RCtxt,\probid} t$. This means $s = R[\prob{s'}{s'}] \xrightarrow{\sr,\problabl} R[s'] =t$  as well as $R[\prob{s'}{s'}] \xrightarrow{\sr,\problabr} R[s']$. 
Then  the sets of evaluations  $\Eval(R[\prob{s'}{s'}])$ and $\Eval(R[s'])$ are related: 
$$\begin{array}{@{}l@{~}c@{~}l@{}}
\Eval({R[\prob{s'}{s'}]})& =& \{(1,s) \evaluation_{(sr,\problabl),L} (0.5 p,s'')
\mid (1,t)\evaluation_{L} (p,s'') \in \Eval({R[s']})\} \\
&\cup& 
  \{(1,s) \evaluation_{(sr,\problabr),L} (0.5 p,s'')
\mid (1,t) \evaluation_{L} (p,s'') \in \Eval({R[s']})\}.
  \end{array}
$$
We see that the expected convergence is the same for $R[\prob{s'}{s'}]$ and $R[s']$. Hence using the context-lemma \cref{thm:context-lemma} in both directions we conclude that $\probid$ is correct. 
\item
Let $s \xrightarrow{\RCtxt,\probcomm} t$. This means $s=R[\prob{s'}{t'}]$ and $t=R[\prob{t'}{s'}]$.
Then $s \xrightarrow{\sr,\problabl} R[s']$,
$s \rightarrow{\sr,\problabr}R[t']$,
$t \xrightarrow{\sr,\problabr} R[s']$,
$t' \rightarrow{\sr,\problabl}R[t']$, which shows:
$$\begin{array}{@{}l@{~}c@{~}l@{}}
\Eval(s) &=& 
\{(1,s) \evaluation_{(sr,\problabl),L} (0.5p,s'')
\mid (1,s')\evaluation_{L} (p,s'') \in \Eval(s')\}\\
&\cup&
\{(1,s) \evaluation_{(sr,\problabr),L} (0.5p,t'')
\mid (1,t')\evaluation_{L} (p,t'') \in \Eval(t')\} 
\\
\Eval(t) &=& 
\{(1,t) \evaluation_{(sr,\problabr),L} (0.5p,s'')
\mid (1,s')\evaluation_{L} (p,s'') \in \Eval(s')\}\\
&\cup&
\{(1,t) \evaluation_{(sr,\problabl),L} (0.5p,t'')
\mid (1,t')\evaluation_{L} (p,t'') \in \Eval(t')\} 
\end{array}
$$
Again, we see that the expected convergence is the same for $s$ and $t$. Hence using the context-lemma \cref{thm:context-lemma} in both directions we conclude that $\probid$ is correct. 
\item 
 \davidcomment{Kontexlemma-Voraussetzung stimmt hier auch nicht für k=1}
Let $s \xrightarrow{\RCtxt,\probprob} t$. This means $s=R[\prob{r'}{(\prob{s'}{t'})}]$ and 
$t=R[\prob{(\prob{r'}{s'})}{(\prob{r'}{t'})}]$.
Then
$$
\xymatrix@R=2mm@C=18mm{
                         \\&(0.5p,R[r'])
\\ 
(p,s) \ar[ur]^{wsr,\problabl}   
      \ar[dr]_{wsr,\problabr} && (0.25p,R[s'])\\
                           &(0.5p,R[\prob{s'}{t'}])
                            \ar[ur]^{wsr,\problabl}
                            \ar[dr]_{wsr,\problabr} 
                            \\
                            && (0.25p,R[t'])
}
$$
$$
\xymatrix@R=0mm@C=18mm{
&&(0.25p,R[r'])
\\&(0.5p,R[\prob{r'}{s'}])
                            \ar[ur]^{wsr,\problabl}
                            \ar[dr]_{wsr,\problabr} 
\\ 
&&(0.25p,R[s'])
\\
(p,t) \ar[uur]^{wsr,\problabl}   
      \ar[ddr]_{wsr,\problabr} && \\
      &&(0.25p,R[r'])\\
                           &(0.5p,R[\prob{r'}{t'}])
                            \ar[ur]^{wsr,\problabl}
                            \ar[dr]_{wsr,\problabr} 
                            \\
                            && (0.25p,R[t'])
}
$$

$$\begin{array}{@{}l@{~}c@{~}l@{}}
\Eval(s) &=& 
\{(1,s) \evaluation_{(sr,\problabl),L} (0.5p,s'')
\mid (1,s')\evaluation_{L} (p,s'') \in \Eval(s')\}\\
&\cup&
\{(1,s) \evaluation_{(sr,\problabr),L} (0.5p,t'')
\mid (1,t')\evaluation_{L} (p,t'') \in \Eval(t')\} 
\\
\Eval(t) &=& 
\{(1,t) \evaluation_{(sr,\problabr),L} (0.5p,s'')
\mid (1,s')\evaluation_{L} (p,s'') \in \Eval(s')\}\\
&\cup&
\{(1,t) \evaluation_{(sr,\problabl),L} (0.5p,t'')
\mid (1,t')\evaluation_{L} (p,t'') \in \Eval(t')\} 
\end{array}
$$
Again, we see that the expected convergence is the same for $s$ and $t$. Hence using the context-lemma \cref{thm:context-lemma} in both directions we conclude that $\probid$ is correct. 

\end{enumerate}

\end{proof}

%\todo[inline]{die weiteren noch beweisen}
\endignore}

\ignore{
\begin{proposition}
Let $s,t$ be   expressions with $s \sim_c t$. Let $W_s$ be the set of WHNFs $w$ such that $s$ evaluates to $w$ with positive probability. The same for $t$ and $W_t$. Then $W_s = W_t$
\davidcomment{Welche Gleichheit ist hier gemeint auf Ebene der Mengenelemente, kontextuelle Gleichheit?}

and for every $w \in W_s$: $s \downarrow_q w$ there is some $w' \sim w$ such that $W_t \downarrow_q w'$, and vice versa.
\end{proposition}
\davidcomment{Bin mir eher unsicher, dass die Behauptung stimmt. }
ich nehms mal zurueck und versuche eine erweiterte Variante von 3.4
}

\begin{theorem}\label{theo:all-correct}
Except for transformation (\probassoc) all transformations shown in \cref{trans-rules} are correct program transformations.
\end{theorem}

\section{\texorpdfstring{The Extended Calculus $\LRPROB$}{The Extended Calculus}\label{sec:extensions}}
In this section, we discuss extensions of the calculus $\LNEEDPROB$ to make it
a more realistic core language of probabilistic functional programming.
The extended calculus will be called $\LRPROB$.
Compared with core languages of the language Haskell,
we should add data like Boolean values, lists, etc., and a $\tseq$-operator 
to enforce strict evaluation of (sub-)expressions. 
\begin{figure*}[tpb]
$\begin{array}{@{}l@{~}l}
(sr,\casec) &R[\tcase_T~c_{T,i}~s_1\ldots s_n~\tof~\{\ldots~c_{T,i}~x_1 \ldots x_n\texttt{\,->\,}t;\ldots\}]
\to R[\tletrec~\{x_i=s_i\}_{i=1}^n~\tin~t]
\\
(sr,\casein) &
\tletr~\{x_i{=}x_{i+1}\}_{i=1}^{m-1},x_{m} =c_{T,i}~s_1 \ldots s_n,\env~\tin~A[\tcase_T~x_1~\tof~\{\ldots;c_{T,i}~y_1 \ldots y_n\texttt{\,->\,}t;\ldots\}]
\\&
\to
\tletr~\{x_i{=}x_{i+1}\}_{i=1}^{m-1},x_{m} =c_{T,i}~z_1 \ldots z_n,\{z_i=s_i\}_{i=1}^n,\env~\tin~A[\tletr~\{y_i=z_i\}_{i=1}^n~\tin~t]
\\
(sr,\casee) &
 \tletr~\{x_i=A_i[x_{i+1}]\}_{i=1}^{m-1},x_m = A_m[\tcase_T~z_1~\tof~\{\ldots;c_{T,i}~y_1\ldots y_n\texttt{\,->\,}t;\ldots\}], \{z_j =z_{j+1}\}_{j=1}^{k-1}, z_k = c_{T,i}~s_1~\ldots~s_n,\env~\tin~A[x_1]\\
 &\to
 \tletr~\{x_i=A_i[x_{i+1}]\}_{i=1}^{m-1},x_m = A_m[\tletrec~\{y_i=w_i\}_{i=1}^n~\tin~t], \{z_j =z_{j+1}\}_{j=1}^{k-1}, z_k = c_{T,i}~w_1\ldots w_n,\{w_i=s_1\}_{i=1}^n,\env~\tin~A[x_1]
\\
(sr,\lcase) & R[\tcase~(\tletrec~\env~\tin~s)~\tof~\ialts_T]\to R[\tletrec~\env~\tin~\tcase_T~s~\tof~\ialts_T]
\\
(sr,\seqc) &R[\tseq~v~t]\to R[t]~\text{if $v$ is an abstraction or a constructor application}
\\
(sr,\seqin) &
\tletr~\{x_i{=}x_{i+1}\}_{i=1}^{m-1},x_{m} =c_{T,i}~s_1 \ldots s_n,\env~\tin~A[\tseq~x_1~t]
\\&
\to
\tletr~\{x_i{=}x_{i+1}\}_{i=1}^{m-1},x_{m} =c_{T,i}~s_1 \ldots s_n,\env~\tin~A[t]
\\
(sr,\seqe) &
 \tletr~\{x_i=A_i[x_{i+1}]\}_{i=1}^{m-1},x_m = A_m[\tseq~z_1~t], 
 \{z_j =z_{j+1}\}_{j=1}^{k-1}, z_k = c_{T,i}~s_1~\ldots~s_n,\env~\tin~A[x_1]\\
 &\to
 \tletr~\{x_i=A_i[x_{i+1}]\}_{i=1}^{m-1},x_m = A_m[t]
 \{z_j =z_{j+1}\}_{j=1}^{k-1}, z_k = c_{T,i}~s_1~\ldots~s_n,\env~\tin~A[x_1]\\
(sr,\lseq) & R[\tseq~(\tletrec~\env~\tin~s)~t]\to R[\tletrec~\env~\tin~\tseq~s~t]
\end{array}
$
% 
% ((\lambda x.s)~t)] \to R[\tletr~x=t~\tin~s]
% \\                 
% (sr,\problabl)&R[\prob{s}{t}] \to R[s]
% \\                 
% (sr,\problabr)&R[\prob{s}{t}] \to R[t]
% \\                 
% (sr,\lapp) 
% &R[((\tletr~\env~\tin~s)~t)] \to R[\tletr~\env~\tin~(s~t)]
% \\
% (sr,\cpin)&
% \tletr~\{x_i{=}x_{i+1}\}_{i=1}^{n-1},x_{n} =\lambda y.s, \env~\tin~A[x_1]
%    \to  
%                      \tletr~\{x_i = x_{i+1}\}_{i=1}^{n-1},x_n=\lambda y.s, \env~\tin~A[\lambda y.s]
%          
% \\                 
% (sr,\cpe)&\begin{array}[t]{@{}l@{}}
% \tletr~\{x_i=A_i[x_{i+1}]\}_{i=1}^{n-1},x_n = A_n[y_1], \{y_j =y_{j+1}\}_{j=1}^{m-1}, y_m = \lambda z.s,\env~\tin~A[x_1]
% \end{array}
% \\
% &
% \to  \begin{array}[t]{@{}l@{}}
% \tletr~\{x_i=A_i[x_{i+1}]\}_{i=1}^{n-1},x_n = A_n[\lambda z.s], \{y_j =y_{j+1}\}_{j=1}^{m-1}, y_m = \lambda z.s,\env~\tin~A[x_1]
%      \end{array}
%  \text{ where $A_n \not= [\cdot],n {\geq} 1, m {\geq} 1$}
% \\
% (sr,\lletin)&
% \tletr~\env_1~\tin~\tletr~\env_2~\tin~s
%    \to
%    \tletr~\env_1,\env_2~\tin~s
% \\                 
% (sr,\llete)&
% \tletr~\{x_i = A_i[x_{i+1}]\}_{i=1}^{n-1},x_n = (\tletr~\env_1~\tin~s), \env_2~\tin~A[x_1]
% \to
% \tletr~\{x_i = A_i[x_{i+1}]\}_{i=1}^{n-1},x_n = s,\env_1,\env_2~\tin~A[x_1]
% \\
% \end{array}
% $
\caption{Standard Reduction-Rules for $\tcase$ and $\tseq$\label{fig-sr-rules-ext}}
\begin{minipage}{.7\textwidth}
$\begin{array}{@{}l@{~}l}
(\seqc) &(\tseq~v~t) \to t~\text{if $v$ is an abstraction or a constructor application}
\\
(\seqin) &
\tletr~x=c_{T,i}~s_1 \ldots s_n,\env~\tin~C[\tseq~x~t]
\to
\tletr~x=c_{T,i}~s_1 \ldots s_n,\env~\tin~C[t]
\\
(\seqe) &
 \tletr~x = C[\tseq~y~t], y = c_{T,i}~s_1~\ldots~s_n,\env~\tin~s
 \\&
 \to
 \tletr~x = C[t], y = c_{T,i}~s_1~\ldots~s_n,\env~\tin~s
\\                 
(\casec) &\tcase_T~c_{T,i}~s_1\ldots s_n~\tof~\{\ldots~c_{T,i}~x_1 \ldots x_n\texttt{\,->\,}t;\ldots\}\to \tletrec~\{x_i=s_i\}_{i=1}^n~\tin~t
\\
(\casein) &
\tletr~x =c_{T,i}~s_1 \ldots s_n,\env~\tin~C[\tcase_T~x~\tof~\{\ldots;c_{T,i}~y_1 \ldots y_n\texttt{\,->\,}t;\ldots\}]
\\&
\to
\tletr~x =c_{T,i}~z_1 \ldots z_n,\{z_i=s_i\}_{i=1}^n,\env~\tin~C[\tletr~\{y_i=z_i\}_{i=1}^n~\tin~t]
\\
(\casee) &
 \tletr~x = C[\tcase_T~z~\tof~\{\ldots;c_{T,i}~y_1\ldots y_n\texttt{\,->\,}t;\ldots\}], z = c_{T,i}~s_1~\ldots~s_n,\env~\tin~s\\
 &\to
 \tletr~\{x =C[\tletrec~\{y_i=w_i\}_{i=1}^n~\tin~t], z = c_{T,i}~w_1\ldots w_n,\{w_i=s_1\}_{i=1}^n,\env~\tin~s
\\
(\lcase) 
&\tcase_T~(\tletr~\env~\tin~s)~\tof~\ialts_T \to \tletr~\env~\tin~(\tcase~s~\tof~\ialts_T)
\\
(\lseq) 
&\tseq~(\tletr~\env~\tin~s)~t \to \tletrec~\env~\tin~\tseq~s~t 
\\
(\cpcxin)&
\tletr~x =c_{T,i}~s_1 \ldots s_n,\env~\tin~C[x]
% \\
% &
\to
\tletr~x =c_{T,i}~y_1 \ldots y_n,\{y_i=s_i\}_{i=1}^n,\env~\tin~C[c~y_1\ldots~y_n]
\\
(\cpcxe)&
\tletr~x =c_{T,i}~s_1 \ldots s_n,z=C[x],\env~\tin~s
% \\
% &
\to
\tletr~x =c_{T,i}~y_1 \ldots y_n,\{y_i=s_i\}_{i=1}^n,z=C[c~y_1\ldots~y_n],\env~\tin~s
\\
(\abs)&
\tletr~x =c_{T,i}~s_1 \ldots s_n~\env~\tin~s \to
\tletr~x =c_{T,i}~y_1 \ldots y_n,\{y_i=s_i\}_{i=1}^n~\env~\tin~s
\text{ where $n >0$}
\end{array}$
\end{minipage}\begin{minipage}{.3\textwidth}
\textbf{Unions:}\\
$\begin{array}{l@{~}l@{}}
(\cpcx) &  (\cpcxe) \cup (\cpcxin) 
 \\
(\caserule) &  (\casec) \cup (\casein) \cup (\casee)
\\
(\seq)  &  (\seqc) \cup (\seqin) \cup (\seqe)
\\
(\lacs) &  (\lapp) \cup (\lcase) \cup (\lseq)
\\
(\lllrule) &  (\llet) \cup (\lacs)
\end{array}
$
\vspace*{30mm}
\end{minipage}
\caption{Further Transformations for $\LRPROB$\label{trans-rules-lr}}
\end{figure*}

We assume a set of data types,
e.g.~\{Bool,Pair,List\}, a set of data constructors $c_{T,i}$ indexed by a type $T$ and a number $i$,
i.e.~$c_{T,i}$ is the $i$th constructor of type $T$, and every $c_{T,i}$ has a fixed arity $\ari(c_{T,i}) \in \mathbb{N}_0$. 
For instance, one may introduce abbreviations
$False = c_{Bool,1}, True = c_{Bool,2}, Nil = c_{List,1}, Cons = c_{List,2}, Pair=c_{Pair,1}$
where the arity of the former 3 constructors is 0 and for the latter two, it is 2.
We assume that there is a $\tcase_T$-operator for every type $T$, and that type $T$ has constructors $c_{T,1},\ldots,c_{T,n_T}$
The syntax of $\LRPROB$ extends the syntax of $\LNEEDPROB$ as follows:
\[
\begin{array}{@{}r@{~}c@{~}l@{}}
s,t,r \in \Expr &::=& \ldots~|~\tseq~s~t~|~c_{T,i}~s_1 \ldots s_{\ari(c_{T,i})}~|~\tcase_T~s~\tof~\ialts_T\\
\ialts_T    &::=&\{\ialt_{T,1};\ldots;\ialt_{T,n_T}\}\\
\ialt_{T,i} &::=& c_{T,i}~x_1\ldots x_{\ari(c_{T,i})} \texttt{~->~} s
\end{array}
\]
Contexts and surface contexts are extended to the new syntax: in contexts $\Ctxt$ the hole can appear at any expression-position, and in surface contexts $\SCtxt$, the hole is not inside the body of an abstraction. The $\ACtxt$-contexts are extended as:
$$
\begin{array}{rcl}
A \in \ACtxt &::=& \ldots ~|~ \tseq~A~s ~|~ \tcase_T~A~\tof~\ialts_Ts
\end{array}
$$
Reduction contexts are defined as before using the extended $\ACtxt$-contexts. WHNFs are extended such that
% to cover data constructors as values, now 
also $c_{T,i}~s_1~\ldots~s_n$,
$\tletr~\env~\tin~c_{T,i}~s_1~\ldots~s_n$, and 
$\tletr~\{x_i = x_{i+1}\}_{i=1}^{m-1},x_m=c_{T,i}~s_1\ldots s_n,\env~\tin~x_1$ are WHNFs. Standard reduction $\xrightarrow{\sr}$ is defined by the rules already introduced in \cref{fig-sr-rules} (where $\RCtxt$-contexts stem from the extended definition) and by the rules defined in \cref{fig-sr-rules-ext} allowing to evaluate $\tcase$- and $\tseq$-expressions. The rules $(sr,\lseq)$ and $(sr,\lcase)$ adjust $\tletr$-enviroments w.r.t.~\tseq- and $\tcase$-expressions.

%%%%%%%%%

Weighted standard reduction, contextual preorder, contextual equivalence, and correctness of program transformations is defined analogously for the calculus $\LNEEDPROB$ but now instantiated with the extended syntax and  standard reduction. The context lemma also holds for the extended calculus: this can be verified by checking all cases for the extended syntax. 
The transformations in \cref{trans-rules} are also transformations in $\LNEEDPROB$ (again using the extended syntax for all meta-expressions), and in \cref{trans-rules-lr} additional transformations and unions are shown where we also extend the union  $(\lllrule)$.

Correctness of the transformations $(\lbeta)$, $(\seqc)$, $(\casec)$ follows from the context lemma, (the arguments are analogous to \cref{prop-lbeta-correct}). Correctness of  $(\casein), (\casee), (\seqin), (\seqe)$
can be obtained by combining other transformations (i.e.~$(\casec)$ (or $(\seqc)$, resp.) $(\cpcx), (\abs), (\gc)$ and $(\cpx)$). Correctness of $(\abs)$ follows form the correctness of $(\ucp)$, since $(\abs)$ can be reversed by $(\ucp)$. 
Correctness of $(\probid), (\probcomm)$, $(\probreorder)$, $(\probprob)$ can be proved analogously as the proofs in  the calculus $\LNEEDPROB$.
For the remaining transformations,
%(\ie~those already shown to be correct for $\LNEEDPROB$ and $(\cpcx)$) 
the diagram-based method can be used to show correctness, where the diagrams have to be re-computed since there are more rules and an extended syntax. The diagrams computed are shown in the appendix ( \cref{fig-diagr-lrprob-1,fig-diagr-lrprob-2}). In our automated tool, we had to restrict the $\tcase$-expressions and types and data constructors, to Booleans, lists, and pairs (since the more general syntax using $c_{T,i}$ for constructors of type $T$ is not supported). However, the diagrams for the full syntax could be obtained by manually extending the cases of $\tcase$-expressions and constructors (and inspecting them). This changed the diagrams, for instance, a sequence of two $(\cpx)$-steps has to be replaced by a sequence of arbitrary many $(\cpx)$-steps in the diagrams for $(\SCtxt,\cpcx)$, since the number of steps depends on the arity of some constructor. Note also, that some diagrams of transformations are required, even if we can prove correctness of the transformation without diagrams: this is the case if the transformations occur in other diagrams.
For all diagrams, termination of the induction proofs can be shown by transforming them into term rewrite systems (with free variables on the right-hand sides to encode transitive closures) and proving innermost termination using AProVE. The diagrams show that prob-sequences are preserved and thus the correctness of the transformations holds.
The TRSs and the automated proofs are available from
\url{https://p9471.gitlab.io/prob-lneed/}.

%\davidcomment{Noch eine Webseite machen und zitieren, welche die Terminierungsbeweise enthaelt}
%\manfredcomment{Uff, Du hast ja sehr viel geschafft.}

\section{Conclusion\label{sec:conclusion}.}
% \manfredcomment{Im Praesens schreiben?}
We have introduced a call-by-need lambda-calculus with a binary operator for probabilistic computations. 
A small-step evaluation that keeps track of the probabilities results in a semantics of programs that observes expected convergence in all program contexts. Based on this new notion of contextual equivalence we have developed techniques and tools to show equivalences and correctness of program transformations. We have applied them to prove the correctness of several transformations. We have discussed extensions of the calculus to make them a more realistic model of a probabilistic programming language and have sketched how to transfer our techniques and results to the extended language. Future work may take into account extensions with (polymorphic) typing and notions of equivalence, that restrict observations to data values (or even numbers) only. A goal may be to show that the program-calculus with its contextual semantics is a fully-abstract model w.r.t.~usual probabilistic models of mathematics.

% \clearpage

\bibliography{bibshort.bib}

\ignore{

%% Hier die cp Kopie  (nur fuer ne kurze Zeit)

% \subsection{Correctness of cp}
We use the same method as in the previous  \cref{subsec:let-correct}.

\begin{lemma}\label{lemma:diagram-forking-cpS}[cpS forking]
Forking transformation part correctness of $\cpS$. 
\end{lemma}
\begin{proof} We use the same method as in \cref{lemma:lll-forking}.\\
We  consider a representation of an intermediate reduction sequence $e_0\xleftarrow{\sr,*} \cdot \xleftarrow{\sr,a} e \xrightarrow{\SCtxt,\cpS} e'$, and show that it can be transformed in a finite number of steps into 
 $e_0\xrightarrow{\SCtxt,\cpS,0 \vee 1} e''\xleftarrow{\sr,*} e'$ using the diagrams in \cref{subsec:diagrams-cp}. 
The intermediate situation is as follows:  
$e_0 \xleftarrow{\sr,*}  \underbrace{\xleftarrow{\sr,a} \cdot \xrightarrow{\SCtxt,\cpS}}_{focus}\cdot \xleftarrow{\sr,*} e'$. The focused part can be transformed using the forking diagrams and the cancel diagram into  $\xrightarrow{\SCtxt,\cpS,0\vee 1} \cdot \xleftarrow{\sr,a}$ in a number of steps. After iterating this, the result is a reduction sequence  $e_0\xleftarrow{\sr,*}   e'' \xrightarrow{\SCtxt,\cpS,0\vee 1} e'$.
The  base case is that $e''$ is a WHNF, or can be reduced to a WHNF using only $\xleftarrow{\sr}$-reduction steps.   The constructed sr-reduction has the same prob-length as the start sr-reduction.
\end{proof}

\begin{lemma}\label{lemma:diagram-commuting-cpS}[cpS commuting]
 Commuting transformation part of $\cpS$. 
\end{lemma}
\begin{proof}
We use the same method as in \cref{lemma:lll-commuting}.\\
The intermediate situation during the transformation is   $e_0 \xrightarrow{\sr,*} \cdot \underbrace{\xrightarrow{\SCtxt,cpS,0\vee 1} \cdot \xrightarrow{\sr,a} }_{focus} \cdot \xrightarrow{\sr,*} e$.  

The commuting diagrams in \cref{subsec:diagrams-cp} permit to shift the single $\xrightarrow{\sr,a}$ to the right in several steps and reach $e_0 \xrightarrow{\sr,*}  e'' \xrightarrow{\SCtxt,cpS,0\vee 1}  e$. Now the base case sows that $e''$ either is a WHNF or can be reduced to a WHNF using only $\xrightarrow{\sr,cp,*}$.
The constructed sr-reduction has the same prob-length as the start sr-reduction.
\end{proof}

These two lemmas together with the con^text lemma implies: 
\begin{lemma}
The transformation $\cpS$ is correct.
\end{lemma}

\begin{lemma}\label{lemma:diagram-forking-cpS}[cpd forking]
Forking transformation part correctness of $\cpd$. 
\end{lemma}
\begin{proof} We use the same method as in \cref{lemma:lll-forking}.\\
We  consider a representation of an intermediate reduction sequence $e_0\xleftarrow{\sr,*} \cdot \xleftarrow{\sr,a} e \xrightarrow{\SCtxt,\cpd} e'$, and show that it can be transformed in a finite number of steps into 
 $e_0\xrightarrow{\SCtxt,\cpd\vee \cpS,*}e''\xleftarrow{\sr,*} e'$ using the diagrams in \cref{subsec:diagrams-cp}. 
The intermediate situation is as follows:  
$e_0 \xleftarrow{\sr,*}  \underbrace{(\xleftarrow{\sr,a} \cdot \xrightarrow{\SCtxt,\cpd\vee\cpS,*}), *}_{focus}\cdot \xleftarrow{\sr,*} e'$. The focused part can be transformed using the forking diagrams and the cancel diagram into  $\xrightarrow{\SCtxt,\cpS,0\vee 1} \cdot \xleftarrow{\sr,a}$ in a number of steps. After iterating this, the result is a reduction sequence  $e_0\xleftarrow{\sr,*}   e'' \xrightarrow{\SCtxt,\cpS,0\vee 1} e'$.
The  base case is that $e''$ is a WHNF, or can be reduced to a WHNF using only $\xleftarrow{\sr}$-reduction steps.   The constructed sr-reduction has the same prob-length as the start sr-reduction.
\end{proof}

============================\\
XXXXXXXXXXXXxxxxxxxxxxxxxxx\\
============================\\

Hier der lll forking rest. Kann geloescht werden.

\delete{
The intermediate situation is as follows:  
$s' \xleftarrow{\sr,n}  \underbrace{\xleftarrow{\sr,a} \cdot \xrightarrow{\SCtxt,\lllrule,*}}_{focus}\cdot \xleftarrow{\sr,*} t$, where $n$ is a non-negative integer, and $a$ a transformation name. There are two cases: $a$ is an ($\lllrule$)-reduction, or not.   The strategy is to first concentrate on the focus part. Iterate the following on the focus part:  
shift the rightmost $\xrightarrow{\sr}$-reduction to the right,  and thereby moving $\lllrule$-reductions to the left.  There are four possible diagrams to consider.  
\begin{enumerate}
    \item If $a \not= \lllrule$, then the iteration will end with $ \xrightarrow{\SCtxt,\lllrule,*} \cdot \xleftarrow{\sr,a}$ and now we can select the next focus and shift the next sr-reduction to the left.   
  \item If $a = \lllrule$, then we again can apply an iterated shifting.
  The first diagram eliminates the rightmost $\xrightarrow{\sr}$-reduction,
  which means that in the next step (another) rightmost $\xrightarrow{\sr}$-reduction is shifted. There are two different possibilities: 
  \begin{enumerate}
      \item  The focus part consists only of   $\xleftrightarrow{\SCtxt,\lllrule}$-reductions. 
  Then we look for the next focus. This is the old one extended to the left with a further $\xleftarrow{\sr}$, or the whole iteration finishes,
  with result $\xleftrightarrow{\SCtxt,\lllrule,*} \cdot \xleftarrow{\sr,*}$. \item The focus contains at least one  $\xleftrightarrow{\sr,\lllrule}$-reduction. Then we can go on with shifting.
   \end{enumerate}
   Now we assume that we apply one of the diagrams 2,3,4, which is a simple shift. \\
 In the general case, the focus part is a mixture of   $\xrightarrow{\SCtxt,\lllrule}$ and $\xleftarrow{\sr,\lllrule}$ reductions. The further (rightmost) $\xleftarrow{\sr,\lllrule}$-reductions are again (iteratively) shifted  to the right.
 
 \manfredcomment{Mist: man muss den (nd) Algorithmus auf dem Focus doch genauer beschreiben.  Darf ich es auch sketchy machen? Ich versuchs mal }  
%   and to obtain $\xrightarrow{\SCtxt,\llrule,*} \cdot  \xleftarrow{\sr,\lllrule,*}$ as intermediate reduction sequences.
  \cref{lemma:lll-terminates} shows that this process will end.
 The current focus will then be
   $\xrightarrow{\SCtxt,\lllrule,*} \cdot \xleftarrow{\sr,\lllrule,0\vee 1}$. \davidcomment{0 in $0\vee 1$, wenn das erste Diagramm angewendet wird}
   \manfredcomment{Muss irgendwie anders sein, nachdenken.}
  
\end{enumerate}
Finally we obtain a sequence  %%overall the sequence
$s'\xrightarrow{\SCtxt,\lllrule,*}t' \xleftarrow{\sr,*} t$, and now can apply the base case, which shows that the \davidins{prob-sequence}
%prob-length 
of the generated sr-reduction sequence has not changed.
%The focused part can be transformed using the forking diagrams
% and the cancel diagram 
%into  $\xrightarrow{\SCtxt,\lllrule,*} \cdot \xleftarrow{\sr,a}$ in a number of steps. Note that the length of $\xrightarrow{\SCtxt,\lllrule,*}$ may
%be increased. 
%In any case the transformations can be continued until the form %$e_0\xrightarrow{\SCtxt,\lllrule,*} e'' \xleftarrow{\sr,*} e'$ has been constructed. Due to the start of the transformation, $e_0$ is a WHNF. The base cases in the (lll)-diagrams show that also $e''$ is a WHNF. 
%
% The probability as well as the prob-length of the generated sr-reduction sequence is the same.
%Due to the form of the diagrams the above rewriting of the sequence ends with a sequence 
%$s' \xrightarrow{\SCtxt,lll*} t'\xleftarrow{\sr,*} t$. 
Now we apply the base case for each $\xrightarrow{\SCtxt,lll,*}$ step
which shows that there is an evaluation of $t'$ to a WHNF (that does not change \davidins{the prob-sequence}).
\delete{Finally, consider the weighted reductions: the diagrams and the base case show that the number of $\xrightarrow{\sr,\problabl}$- and $\xrightarrow{\sr,\problabr}$-reductions is always preserved, and thus
if $(1,s) \xrightarrow{wsr,*} (q,s')$ where $s'$ is a WHNF, then there also exists an evaluation $(1,t)\xrightarrow{wsr,*} (q,t')$ where $t'$ is a WHNF.}
}

Ein zweites Save, kann geloescht werden

The intermediate situation is more complex as in the previous lemma and is as follows, where there are two cases:
\begin{enumerate}
    \item $s \xrightarrow{\sr,*} \cdot \underbrace{\xrightarrow{\SCtxt,\lllrule,*} \cdot \xrightarrow{\sr,a} \cdot \xrightarrow{\SCtxt,\lllrule,*}}_{focus} \cdot \xrightarrow{\sr,*} t''$, where $a$ is not an $\lllrule$-reduction.  In this case the commuting diagrams permit to shift the single $\xrightarrow{\sr,a}$ to the left in several steps and reach $\xrightarrow{\sr,a} \cdot \xrightarrow{\SCtxt,\lllrule,*}$.
    \item  $s \xrightarrow{\sr,*} \cdot \underbrace{(\xrightarrow{\SCtxt,\lllrule,*} \cdot \xrightarrow{\sr,\lllrule})^*}_{focus} \cdot \xrightarrow{\sr,*} t''$.\\ 
    We iterate the following two steps: 
    \begin{enumerate}
        \item We allow for changes of $\xrightarrow{\SCtxt,\lllrule}$ into $\xrightarrow{\sr,\lllrule}$. 
        \item If the leftmost reduction is an  $\xrightarrow{\sr,\lllrule}$, then the focused part is smaller and starts with   $\xrightarrow{\SCtxt,\lllrule}$. 
       \item If the leftmost reduction in the focus is not an sr-reduction then the leftmost $\xrightarrow{\sr,\lllrule})$ is shifted to left using a commuting diagram.  
    \end{enumerate}
    \Cref{lemma:lll-terminates} shows that this process of focus-modification  terminates, since there are no infinite (lll)-reductions.
\end{enumerate}

In every case, we will reach a reduction sequence of the form 
$s \xrightarrow{\sr,*} s'  \xrightarrow{\SCtxt,\lllrule,*} t''$. The base case of the diagrams shows that $s'  \xrightarrow{\sr,\lllrule,*} s''$,
where $s''$ is a WHNF. This shows that we obtain a standard reduction sequence with the same probability and prob-sequence as the original sequence.  

\delete{Finally, we consider weighted expressions. Since the prob-length is preserved in the above proof, we also have: if $(1,t)\xrightarrow{wsr,*} (q,t')$ where $t'$ is a WHNF, then there exists an evaluation $(1,s) \xrightarrow{wsr,*}(q,s')$ where $s'$ is a WHNF. }
%\end{proof}

%\manfredcomment{conceptual fertig..., noch checken}.

\endignore
}

\clearpage
\appendix
\section{Larger Examples}
\begin{example}\label{ex:eval:app}
We show the two evaluations for the expression
$$
    \tletrec\,z{=}\prob{\mathit{K}}{\mathit{K2}}\,\tin\,(z\,(z\,a\,b)\,(z\,c\,d))
$$
One evaluation is:
$$
\begin{array}[t]{@{}l@{}l@{}}
                           &\tletrec\,z{=}\prob{\mathit{K}}{\mathit{K2}}\,\tin\,(z\,(z\,a\,b)\,(z\,c\,d))
                           \\
\xrightarrow{\sr,\problabl} &\tletrec\,z{=}\mathit{K}\,\tin\,(z\,(z\,a\,b)\,(z\,c\,d))

\\
\xrightarrow{\sr,\cpin}     &\tletrec\,z{=}\mathit{K}\,\tin\,(\mathit{K}\,(z\,a\,b)\,(z\,c\,d))
\\
\xrightarrow{\sr,\lbeta}    &\tletrec\,z{=}\mathit{K}\,\tin\,((\tletrec\,x{=}(z\,a\,b)\,\tin\,\lambda y.x)\,(z\,c\,d))
\\
\xrightarrow{\sr,\lapp}    &\tletrec\,z{=}\mathit{K}\,\tin\,\tletrec\,x{=}(z\,a\,b)\,\tin\,((\lambda y.x)\,(z\,c\,d))
\\
\xrightarrow{\sr,\lletin}    &\tletrec\,z{=}\mathit{K},\,x{=}(z\,a\,b)\,\tin\,((\lambda y.x)\,(z\,c\,d))
\\
\xrightarrow{\sr,\lbeta}    &\tletrec\,z{=}\mathit{K},\,x{=}(z\,a\,b)\,\tin\,\tletrec\,y{=}(z\,c\,d)\,\tin\,x
\\
\xrightarrow{\sr,\lletin}   
&\tletrec\,z{=}\mathit{K},\,x{=}(z\,a\,b){,}y{=}(z\,c\,d)\,\tin\,x
\\
\xrightarrow{\sr,\cpe}  
&\tletrec\,z{=}\mathit{K},\,x{=}(\mathit{K}\,a\,b){,}y{=}(z\,c\,d)\,\tin\,x
\\
\xrightarrow{\sr,\lbeta}  
&\tletrec\,z{=}\mathit{K},\,x{=}((\tletrec\,x_1{=}a\,\tin\,\lambda y_1.x_1)\,b){,}y{=}(z\,c\,d)\,\tin\,x
\\
\xrightarrow{\sr,\lapp}  
&\tletrec\,z{=}\mathit{K},\,x{=}(\tletrec\,x_1{=}a\,\tin\,((\lambda y_1.x_1)\,b)){,}y{=}(z\,c\,d)\,\tin\,x
\\
\xrightarrow{\sr,\llete}  
&\tletrec\,z{=}\mathit{K},\,x{=}((\lambda y_1.x_1)\,b),x_1{=}a{,}y{=}(z\,c\,d)\,\tin\,x
\\
\xrightarrow{\sr,\lbeta}  
&\tletrec\,z{=}\mathit{K},\,x{=}\tletrec\,y_1{=}b\,\tin\,x_1,x_1{=}a{,}y{=}(z\,c\,d)\,\tin\,x
\\
\xrightarrow{\sr,\llete}  
&\tletrec\,z{=}\mathit{K},\,x{=}x_1,y_1{=}b,x_1{=}a{,}y{=}(z\,c\,d)\,\tin\,x
\\
\xrightarrow{\sr,\cpin}  
&\tletrec\,z{=}\mathit{K},\,x{=}x_1,y_1{=}b,x_1{=}a{,}y{=}(z\,c\,d)\,\tin\,a
\end{array}
$$
Another evaluation is:
$$
\begin{array}[t]{@{}l@{}l@{}}
                           &\tletrec\,z{=}\prob{\mathit{K}}{\mathit{K2}}\,\tin\,(z\,(z\,a\,b)\,(z\,c\,d))
                           \\
\xrightarrow{\sr,\problabr} &\tletrec\,z{=}\mathit{K2}\,\tin\,(z\,(z\,a\,b)\,(z\,c\,d))

\\
\xrightarrow{\sr,\cpin}     &\tletrec\,z{=}\mathit{K2}\,\tin\,(\mathit{K2}\,(z\,a\,b)\,(z\,c\,d))
\\
\xrightarrow{\sr,\lbeta}    &\tletrec\,z{=}\mathit{K2}\,\tin\,((\tletrec\,x{=}(z\,a\,b)\,\tin\,\lambda y.y)\,(z\,c\,d))
\\
\xrightarrow{\sr,\lapp}    &\tletrec\,z{=}\mathit{K2}\,\tin\,\tletrec\,x{=}(z\,a\,b)\,\tin\,((\lambda y.y)\,(z\,c\,d))
\\
\xrightarrow{\sr,\lletin}    &\tletrec\,z{=}\mathit{K2},\,x{=}(z\,a\,b)\,\tin\,((\lambda y.y)\,(z\,c\,d))
\\
\xrightarrow{\sr,\lbeta}    &\tletrec\,z{=}\mathit{K2},\,x{=}(z\,a\,b)\,\tin\,\tletrec\,y{=}(z\,c\,d)\,\tin\,y
\\
\xrightarrow{\sr,\lletin}   
&\tletrec\,z{=}\mathit{K2},\,x{=}(z\,a\,b){,}y{=}(z\,c\,d)\,\tin\,y
\\
\xrightarrow{\sr,\cpe} 
&\tletrec\,z{=}\mathit{K2},\,x{=}(z\,a\,b){,}y{=}(\mathit{K2}\,c\,d)\,\tin\,y
\\
\xrightarrow{\sr,\lbeta}  
&\tletrec\,z{=}\mathit{K2},\,x{=}(z\,a\,b){,}y{=}((\tletrec\,x_1{=}c\,\tin\,\lambda y_1.y_1)\,d)\,\tin\,y
\\
\xrightarrow{\sr,\lapp}  
&\tletrec\,z{=}\mathit{K2},\,x{=}(z\,a\,b){,}y{=}\tletrec\,x_1{=}c\,\tin\,((\lambda y_1.y_1)\,d)\,\tin\,y
\\
\xrightarrow{\sr,\llete}  
&\tletrec\,z{=}\mathit{K2},\,x{=}(z\,a\,b){,}y{=}((\lambda y_1.y_1)\,d),x_1{=}c\,\tin\,y
\\
\xrightarrow{\sr,\lbeta}  
&\tletrec\,z{=}\mathit{K2},\,x{=}(z\,a\,b){,}y{=}\tletrec\,y_1{=}d\,\tin\,y_1,x_1{=}c\,\tin\,y
\\
\xrightarrow{\sr,\llete}  
&\tletrec\,z{=}\mathit{K2},\,x{=}(z\,a\,b){,}y{=}y_1,y_1{=}d,x_1{=}c\,\tin\,y
\\
\xrightarrow{\sr,\cpin}  
&\tletrec\,z{=}\mathit{K2},\,x{=}(z\,a\,b){,}y{=}y_1,y_1{=}d,x_1{=}c\,\tin\,d
\end{array}
$$
The four evaluations of the expression
 $$\tletrec\,z{=}\lambda x.\lambda y.\prob{x}{y}\,\tin\,z\,(z\,a\,b)\,(z\,c\,d)$$
are as follows: all of them start with

\[
\begin{array}[t]{@{}l@{}l@{}}
&
\tletrec\,z{=}\lambda x.\lambda y.\prob{x}{y}\,\tin\,z\,(z\,a\,b)\,(z\,c\,d)
\\
\xrightarrow{\sr,\cpin}
&
\tletrec\,z{=}\lambda x.\lambda y.\prob{x}{y}\,\tin\,(\lambda x.\lambda y.\prob{x}{y})\,(z\,a\,b)\,(z\,c\,d)
\\
\xrightarrow{\sr,\lbeta}
&
\tletrec\,z{=}\lambda x.\lambda y.\prob{x}{y}\,\tin\,((\tletrec\,x{=}(z\,a\,b)\,\tin\,\lambda y.\prob{x}{y})\,(z\,c\,d))
\\
\xrightarrow{\sr,\lapp}
&
\tletrec\,z{=}\lambda x.\lambda y.\prob{x}{y}\,\tin\,\tletrec\,x{=}(z\,a\,b)\,\tin\,((\lambda y.\prob{x}{y})\,(z\,c\,d))
\\
\xrightarrow{\sr,\lletin}
&
\tletrec\,z{=}\lambda x.\lambda y.\prob{x}{y},x{=}(z\,a\,b)\,\tin\,((\lambda y.\prob{x}{y})\,(z\,c\,d))
\\
\xrightarrow{\sr,\lbeta}
&
\tletrec\,z{=}\lambda x.\lambda y.\prob{x}{y},x{=}(z\,a\,b)\,\tin\,\tletrec\,y{=}(z\,c\,d)\,\tin\,\prob{x}{y}
\\
\xrightarrow{\sr,\lletin}
&
\tletrec\,z{=}\lambda x.\lambda y.\prob{x}{y},x{=}(z\,a\,b),y{=}(z\,c\,d)\,\tin\,\prob{x}{y}
\end{array}
\]
Then there are the following continuations:
\begin{enumerate}
 \item 
\[
\begin{array}[t]{@{}l@{}l@{}}
&\tletrec\,z{=}\lambda x.\lambda y.\prob{x}{y},x{=}(z\,a\,b),y{=}(z\,c\,d)\,\tin\,\prob{x}{y}
\\
\xrightarrow{\sr,\problabl}
&\tletrec\,z{=}\lambda x.\lambda y.\prob{x}{y},x{=}(z\,a\,b),y{=}(z\,c\,d)\,\tin\,x
\\
\xrightarrow{\sr,\cpe}
&\tletrec\,z{=}\lambda x.\lambda y.\prob{x}{y},x{=}((\lambda x_1.\lambda y_1.\prob{x_1}{y_1})\,a\,b),y{=}(z\,c\,d)\,\tin\,x
\\
\xrightarrow{\sr,\lbeta}
&\tletrec\,z{=}\lambda x.\lambda y.\prob{x}{y},x{=}((\tletrec\,x_1{=}a\,\tin\,(\lambda y_1.\prob{x_1}{y_1}))\,b),\\
&\phantom{\tletrec\,}y{=}(z\,c\,d)\,\tin\,x
\\
\xrightarrow{\sr,\lapp}
&\tletrec\,z{=}\lambda x.\lambda y.\prob{x}{y},x{=}\tletrec\,x_1{=}a\,\tin\,(\lambda y_1.\prob{x_1}{y_1})\,b),\\
&\phantom{\tletrec\,}y{=}(z\,c\,d)\,\tin\,x
\\
\xrightarrow{\sr,\llete}
&\tletrec\,z{=}\lambda x.\lambda y.\prob{x}{y},x_1{=}a,x{=}(\lambda y_1.\prob{x_1}{y_1})\,b,y{=}(z\,c\,d)\,\tin\,x
\\
\xrightarrow{\sr,\lbeta}
&\tletrec\,z{=}\lambda x.\lambda y.\prob{x}{y},x_1{=}a,x{=}(\tletrec\,y_1{=}b\,\tin\,\prob{x_1}{y_1}),\\
&\phantom{\tletrec\,}y{=}(z\,c\,d)\,\tin\,x
\\
\xrightarrow{\sr,\llete}
&\tletrec\,z{=}\lambda x.\lambda y.\prob{x}{y},x_1{=}a,y_1{=}b,x{=}\prob{x_1}{y_1},y{=}(z\,c\,d)\,\tin\,x
\end{array}
\]
Now again{,}two continuations exist:
\begin{enumerate}
 \item 
\[
\begin{array}[t]{@{}l@{}l@{}}
& \tletrec\,z{=}\lambda x.\lambda y.\prob{x}{y},x_1{=}a,y_1{=}b,x{=}\prob{x_1}{y_1},y{=}(z\,c\,d)\,\tin\,x
\\
\xrightarrow{\sr,\problabl} 
&\tletrec\,z{=}\lambda x.\lambda y.\prob{x}{y},x_1{=}a,y_1{=}b,x{=}x_1,y{=}(z\,c\,d)\,\tin\,x
\\
\xrightarrow{\sr,\cpin} 
&\tletrec\,z{=}\lambda x.\lambda y.\prob{x}{y},x_1{=}a,y_1{=}b,x{=}x_1,y{=}(z\,c\,d)\,\tin\,a
\end{array}
\]
 \item
\[
\begin{array}[t]{@{}l@{}l@{}}
& \tletrec\,z{=}\lambda x.\lambda y.\prob{x}{y},x_1{=}a,y_1{=}b,x{=}\prob{x_1}{y_1},y{=}(z\,c\,d)\,\tin\,x
\\
\xrightarrow{\sr,\problabr} 
&\tletrec\,z{=}\lambda x.\lambda y.\prob{x}{y},x_1{=}a,y_1{=}b,x{=}y_1,y{=}(z\,c\,d)\,\tin\,x
\\
\xrightarrow{\sr,\cpin} 
&\tletrec\,z{=}\lambda x.\lambda y.\prob{x}{y},x_1{=}a,y_1{=}b,x{=}y_1,y{=}(z\,c\,d)\,\tin\,b
\end{array}
\]
\end{enumerate}
\item 
\[
\begin{array}[t]{@{}l@{}l@{}}
&\tletrec\,z{=}\lambda x.\lambda y.\prob{x}{y},x{=}(z\,a\,b),y{=}(z\,c\,d)\,\tin\,\prob{x}{y}
\\
\xrightarrow{\sr,\problabr}&\tletrec\,z{=}\lambda x.\lambda y.\prob{x}{y},x{=}(z\,a\,b),y{=}(z\,c\,d)\,\tin\,y
\\
\xrightarrow{\sr,\cpe}&\tletrec\,z{=}\lambda x.\lambda y.\prob{x}{y},x{=}(z\,a\,b),\\
&\phantom{\tletrec\,}y{=}((\lambda x_1.\lambda y_1.\prob{x_1}{y_1})\,c\,d)\,\tin\,y
\\
\xrightarrow{\sr,\lbeta}&\tletrec\,z{=}\lambda x.\lambda y.\prob{x}{y},x{=}(z\,a\,b),\\
&\phantom{\tletrec\,}y{=}((\tletrec\,x_1{=}c\,\tin\,\lambda y_1.\prob{x_1}{y_1})\,d)\,\tin\,y
\\
\xrightarrow{\sr,\lapp}&\tletrec\,z{=}\lambda x.\lambda y.\prob{x}{y},x{=}(z\,a\,b),\\
&\phantom{\tletrec\,}y{=}\tletrec\,x_1{=}c\,\tin\,((\lambda y_1.\prob{x_1}{y_1})\,d)\,\tin\,y
\\
\xrightarrow{\sr,\llete}&\tletrec\,z{=}\lambda x.\lambda y.\prob{x}{y},x{=}(z\,a\,b),x_1{=}c,y{=}((\lambda y_1.\prob{x_1}{y_1})\,d)\,\tin\,y
\\
\xrightarrow{\sr,\lbeta}&\tletrec\,z{=}\lambda x.\lambda y.\prob{x}{y},x{=}(z\,a\,b),x_1{=}c,\\
&\phantom{\tletrec\,}y{=}(\tletrec\,y_1{=}d\,\tin\,\prob{x_1}{y_1})\,\tin\,y
\\
\xrightarrow{\sr,\llete}&\tletrec\,z{=}\lambda x.\lambda y.\prob{x}{y},x{=}(z\,a\,b),x_1{=}c,y_1{=}d,y{=}\prob{x_1}{y_1}\,\tin\,y
\end{array}
\] 
Now again{,}two continuations exist:
\begin{enumerate}
 \item 
\[
\begin{array}[t]{@{}l@{}l@{}}
&\tletrec\,z{=}\lambda x.\lambda y.\prob{x}{y},x{=}(z\,a\,b),x_1{=}c,y_1{=}d,y{=}\prob{x_1}{y_1}\,\tin\,y
\\
\xrightarrow{\sr,\problabl}&\tletrec\,z{=}\lambda x.\lambda y.\prob{x}{y},x{=}(z\,a\,b),x_1{=}c,y_1{=}d,y{=}x_1\,\tin\,y
\\
\xrightarrow{\sr,\cpin}&
\tletrec\,z{=}\lambda x.\lambda y.\prob{x}{y},x{=}(z\,a\,b),x_1{=}c,y_1{=}d,y{=}x_1\,\tin\,c
\end{array}
\]
 \item 
\[
\begin{array}[t]{@{}l@{}l@{}}
&\tletrec\,z{=}\lambda x.\lambda y.\prob{x}{y},x{=}(z\,a\,b),x_1{=}c,y_1{=}d,y{=}\prob{x_1}{y_1}\,\tin\,y
\\
\xrightarrow{\sr,\problabl}&\tletrec\,z{=}\lambda x.\lambda y.\prob{x}{y},x{=}(z\,a\,b),x_1{=}c,y_1{=}d,y{=}y_1\,\tin\,y
\\
\xrightarrow{\sr,\cpin}&
\tletrec\,z{=}\lambda x.\lambda y.\prob{x}{y},x{=}(z\,a\,b),x_1{=}c,y_1{=}d,y{=}y_1\,\tin\,d
\end{array}
\]
\end{enumerate}
\end{enumerate}
\end{example}

\section{Proofs}

{\renewcommand{\thetheorem}{\ref{prop-lbeta-correct}}
\begin{proposition}
The transformation $\xrightarrow{lbeta}$  is correct.
\end{proposition}
\begin{proof}
We use the context lemma (\cref{thm:context-lemma-ext})
for $n=1$ (and where $d$ is always 0). Let $s \xrightarrow{lbeta} t$ and $R$ be a reduction context. 
The structure of reduction contexts implies $R[s] \xrightarrow{\sr,lbeta} R[t]$.
Thus there is
%is an evaluation
$(1,R[s])\evaluation_L(p,s')\in\Eval(R[s])$
if{,}and only if there is 
%an evaluation
$(1,R[t])\evaluation_L(p,s')\in\Eval(R[t])$.
This implies for all $k\geq0:$ $\EC(R[s],k) = \EC(R[t],k)$. 
%The reduction rule $s \xrightarrow{\sr,lbeta} t$ also implies $\EC(R[t]) = \EC(R[s]$,
%since this is the only standard reduction of $R[s]$.
With the context lemma we obtain
 $\EC(C[s]) = \EC(C[t])$ for all contexts $C${,}and hence $s \sim_c t$.
\end{proof}
\addtocounter{theorem}{-1}
}
\subsection{A Manual Proof for (lll)\label{appendix-lll}}
We first show that applying $(\lllrule)$-steps in arbitrary contexts terminates:
\begin{lemma}\label{lemma:lll-terminates}
Every sequence of $\xrightarrow{(\Ctxt,\lllrule)}$-steps is finite.
\end{lemma}
\begin{proof}
Let the measure $\LMP(s)$ be defined as the pair $(\#(\tletr,s){,}\LM(s))${,}ordered lexicographically{,}where $\#(\tletr,s)$ is the number of occurrences of {\tletr} in $s${,}and $\LM$ is a polynomial measure as
$\LM(x) %= \LM(c) 
= 1$;
$\LM(\lambda x.s) = 1+\LM(s)$; $\LM(s~t) =  2\LM(s) + \LM(t)$;
$\LM(\tletr~\env~\tin ~s) =  2\LM(\env) + \LM(s)$; $\LM(x_1 = s_1;\ldots; x_n = s_n)$ $=$ $\LM(s_1) + \ldots + \LM(s_n)$. 

The definitions show that $\xrightarrow{\Ctxt,\llete}$- and $\xrightarrow{\Ctxt,\lletin}$-steps strictly reduce the number of let-expressions and thus{,}the first component of the measure is strictly decreased. For the $\xrightarrow{\Ctxt,\lapp}$-steps,
the first component is not changed{,}but the measure $\LM(\cdot)$ is strictly decreased.
Hence there are no infinite reduction sequences consisting only of $\xrightarrow{\Ctxt,\lllrule}$-steps.
\end{proof}

\begin{lemma}\label{lemma-lll-fork}
Let $s,t$ be expressions such that $s \xrightarrow{\SCtxt,\lllrule} t$. We show that then for every evaluation $s \evaluation_L s'\in\Eval(s)${,}there exists an evaluation $t \evaluation_L t' \in \Eval(t)$.
\end{lemma}
\begin{proof}
The base case is covered by the base cases for $(\lllrule)$.
In the general case{,}we  consider a representation of a reduction sequence{,}i.e. $s'\xleftarrow{\sr,*} \cdot \xleftarrow{\sr,a} s \xrightarrow{\SCtxt,\lllrule} t$ where $s'$ is a WHNF and to show by specifying a rewrite strategy (covering all possible instances){,}that it can be transformed in a finite number of steps into 
$s'\xrightarrow{\SCtxt,\lllrule,*}t'\xleftarrow{\sr,*} t$.
Then as a base case we have that $t'$ is a WHNF{,}or can be reduced to a WHNF not changing the  prob-sequence.  

the following strategy is used to transform the reduction sequence:
\begin{enumerate}
    \item\label{focus-lll-start} First let $llls$ be the smallest substring
    containing all  $\xrightarrow{\SCtxt,\lllrule}$ reductions. 
    \item If $llls$ is empty{,}then we finish.
     \item If the reduction sequence is of the form  
       $llls ~\cdot \xleftarrow{\sr,*} t${,}then shifting is finished. 
      \item Otherwise{,}it is of the form   $\xleftarrow{\sr,+}  llls ~\cdot \xleftarrow{\sr,*} t${,}and we select the focus as  $\xleftarrow{\sr} \cdot ~ llls${,}such that the current sequence is 
     $\xleftarrow{\sr,*} \cdot ~  focus ~ \cdot \xleftarrow{\sr,*}$.
     Now the transformations take only place on the focus part:
   \begin{enumerate}
       \item  Repeat the following step until the rightmost element of the focus is a $\xleftarrow{\sr}$-reduction{,}or there are no more 
       $\xleftarrow{\sr}$ in the focus. 
       \item Use one of the forking diagrams for $(\lllrule)$ to shift the rightmost $\xleftarrow{\sr}$-reduction in the focus to the right{,}   thereby moving $\xrightarrow{\SCtxt,\lllrule}$-steps
       to the left.  
  \end{enumerate}   
   \item jump to \cref{focus-lll-start}
\end{enumerate}

Now we check the property of the non-deterministic transformation algorithm:
There are two cases:
\begin{itemize}
    \item The start focus is $\xleftarrow{\sr,a} \cdot  \xrightarrow{\SCtxt,\lllrule,*}$ and $a \not= \lllrule$. The diagrams show that the iterated shift ends with $\xrightarrow{\SCtxt,\lllrule,*} \cdot \xleftarrow{\sr,a}$. This shows that prob-reductions are moved without change. 
\item The start focus is $\xleftarrow{\sr,\lllrule} \cdot  \xrightarrow{\SCtxt,\lllrule,*}$. The the situation is similar{,}however{,}the focus part may be a mix of $\xleftarrow{\sr,\lllrule}$ and $\xrightarrow{\SCtxt,\lllrule,*}$ reductions. It is easy to see that the effect now is that more and more $\xleftarrow{\sr,\lllrule}$ are generated to the right of the focus. This stops after generating a finite number due to \cref{lemma:lll-terminates}. 
\end{itemize}
As a summary{,}we see that our strategy will produce a final situation
$s'\xrightarrow{\SCtxt,\lllrule,*}t' \xleftarrow{\sr,*} t${,}and we can apply the base case of the diagrams.
\end{proof}

\begin{lemma}\label{lemma:lll-commuting}
Let $s,t$ be expressions such that $s \xrightarrow{\SCtxt,\lllrule} t$. 
Then for every evaluation $t \evaluation_L t'\in\Eval(t)$,
there exists an evaluation $s \evaluation_L s' \in \Eval(s)$.
\end{lemma}
\begin{proof}
%
%\davidins{We first ignore the weights.}
Let $s,t$ be expressions with $s \xrightarrow{\SCtxt,\lllrule} t$.
We show that an evaluation of $t$ can be transformed into an evaluation of $s$ where the set of commuting diagrams is used as transformations (on the reduction sequence){,}and where we have to take into account the base cases. 

We consider a representation of a  sequence $s \xrightarrow{\SCtxt,\lllrule} t \xrightarrow{\sr,a}   \cdot \xrightarrow{\sr,*}t'$ where $t'$ is a WHNF. We show that it can be transformed in a finite number of steps into 
% $e_0\xrightarrow{\SCtxt,\lllrule,*}\cdot \xleftarrow{\sr,*} e'$.
%\davidins{}
$s\xrightarrow{\sr,*} s' \xrightarrow{\SCtxt,\lllrule,*} t'$.
The base case is that $s'$ is a WHNF{,}or can be reduced to a WHNF without changing the prob-sequence of the evaluation. This case holds by inspecting the base case given in \cref{fig:lll-diagrams}.
Now{,}in the general case{,}the following strategy is used to transform the  sequence:
\begin{enumerate}
    \item\label{focus-lll-commuting-start} First let $llls$ be the smallest substring containing all  $\xrightarrow{\SCtxt,\lllrule}$ reductions. 
    \item If $llls$ is empty{,}then we finish.
     \item If the reduction sequence is of the form  
       $s \xrightarrow{\sr,*} ~\cdot~llls~t'${,}then shifting is finished. 
     \item Otherwise{,}it is of the form   $s~\xrightarrow{\sr,*}   \cdot~ llls ~\cdot \xrightarrow{\sr,*}${,}and we select the focus as  $llls \cdot~\xrightarrow{\sr}${,}such that the current sequence is 
     $\xrightarrow{\sr,*} \cdot ~  focus ~ \cdot \xrightarrow{\sr,*}$.
     Now the transformations take only place on the focus part:
   \begin{enumerate}
       \item  Repeat the following step until the leftmost element of the focus is a $\xrightarrow{\sr}$-reduction{,}or there are no more 
       $\xrightarrow{\sr}$ in the focus. 
       \item Use one of the commuting diagrams for $\lllrule$ to shift the rightmost $\xrightarrow{\sr}$-reduction in the focus to the left{,}   thereby moving $\xrightarrow{\SCtxt,\lllrule}$-steps
        to the right.  
   \end{enumerate}   
   \item jump to \cref{focus-lll-commuting-start}
\end{enumerate}

This transformation terminates{,}since similar as for the forking case{,}only a focus that completely consists of $\xrightarrow{\SCtxt,\lllrule}$-steps may potentially generate longer $\xrightarrow{\sr}$-reduction sequences. However,
\cref{lemma:lll-terminates} shows the termination of this generation. 
It is also the case that the diagrams are complete in the sense that the transformation never gets stuck.

The diagrams show that the freshly constructed evaluation of $s$ has the same prob-sequence as the initial evaluation of $t$.
\end{proof}

\subsection{Correctness of (cpx) and (xch)}
The diagrams and base cases for (cpx) are shown in \cref{fig:cpx-diagrams}.
\begin{figure}[t]
\begin{minipage}{.45\textwidth}
\textbf{Forking diagrams:}
\[
\begin{array}[t]{@{}c@{}}
\xymatrix@R=5mm@C=5mm{
 \cdot \ar[d]_{\sr,a}\ar[r]^{\SCtxt,cpx}  & \cdot \ar@{-->}[d]^{\sr,a}\\
 \cdot \ar@{-->}[r]_{\SCtxt,cpx}         & \cdot \\
}
\end{array}
%%%%%%%%%%%%%%%%%%%%%%%%%%%%%%%%%%%
\begin{array}[t]{@{}c@{}}
\xymatrix@R=5mm@C=5mm{
 \cdot \ar[d]_{\sr,a}\ar[r]^{\SCtxt,cpx}  & \cdot \ar@{-->}[dl]^{\sr,a}\\
 \cdot \\
}
\\
a \in \{\cp,\problabl,\problabr\}
\end{array}
%%%%%%%%%%%%%%%%%%%%%%%%%%%%%%%%%%%
\begin{array}[t]{@{}c@{}}
\xymatrix@R=5mm@C=5mm{
 \cdot \ar[d]_{\sr,\cp}\ar[rr]^{\SCtxt,cpx}&                            & \cdot \ar@{-->}[d]^{\sr,\cp}\\
 \cdot \ar@{-->}[r]_{\SCtxt,cpx}         & \cdot \ar@{-->}[r]_{\SCtxt,cpx} & \cdot\\
}
\end{array}
\]
\end{minipage}
\begin{minipage}{.45\textwidth}
\textbf{Commuting diagrams:}
\[
\begin{array}[t]{@{}c@{}}
\xymatrix@R=5mm@C=5mm{
 \cdot \ar@{-->}[d]_{\sr,a}\ar[r]^{\SCtxt,cpx}  & \cdot \ar[d]^{\sr,a}\\
 \cdot \ar@{-->}[r]_{\SCtxt,cpx}         & \cdot \\
}
\end{array}
%%%%%%%%%%%%%%%%%%%%%%%%%%%%%%%%%%%
\begin{array}[t]{@{}c@{}}
\xymatrix@R=5mm@C=5mm{
 \cdot \ar@{-->}[dr]_{\sr,a}\ar[r]^{\SCtxt,cpx}  & \cdot \ar[d]^{\sr,a}\\
 &\cdot \\
}
\\
a \in \{\cp,\problabl,\problabr\}
\end{array}
%%%%%%%%%%%%%%%%%%%%%%%%%%%%%%%%%%%
\begin{array}[t]{@{}c@{}}
\xymatrix@R=5mm@C=5mm{
 \cdot \ar@{-->}[d]_{\sr,\cp}\ar[rr]^{\SCtxt,cpx}&                            & \cdot \ar[d]^{\sr,\cp}\\
 \cdot \ar@{-->}[r]_{\SCtxt,cpx}         & \cdot \ar@{-->}[r]_{\SCtxt,cpx} & \cdot\\
}
\end{array}
\]
\end{minipage}
\begin{minipage}{.45\textwidth}
\textbf{Base cases:} 
If $s \xrightarrow{\SCtxt,cpx} s'${,}then $s$ is a WHNF iff $s'$ is a WHNF.\\[1.1ex]
\textbf{TRS $R$ for forking and commuting diagrams:}
\small$$\begin{array}{r@{}c@{}lr@{}c@{}lr@{}c@{}l}
\Scpx(\SR(x)) &\to& \SR(x)
&\Scpx(\SR(x)) &\to& \SR(\Scpx(x))
\\
\Scpx(\SR(x)) &\to& \SR(\Scpx(\Scpx(x)))
\end{array}$$\normalsize
\end{minipage}
\caption{Diagrams{,}Base Cases and TRS for $(cpx)$ \label{fig:cpx-diagrams}}
\end{figure}

\begin{proposition}%[Correctness of (cpx)]
The transformation $(cpx)$ is correct.
\end{proposition}
\begin{proof}
The proof is analogous to the proof of \cref{prop:correctness-cp}{,}where the diagrams and the bases cases for (cpx) are applied. One has to verify that prob-sequences are not changed by the diagrams and the base cases (which holds).
We also have to verify that diagram application terminates.
In both cases (forking diagrams and commuting diagrams){,}termination can be
shown by proving (innermost) termination of the TRS $R$ 
shown in \cref{fig:cpx-diagrams}.
In the case of forking diagrams{,}the symbol  $\SR$ is interpreted as an $\xleftarrow{\SR}$-step{,}while in the case of commuting diagrams it represents an $\xrightarrow{\SR}$-step. 
$R$ is shown to be innermost terminating using TTT2.
% the Knuth-Bendix-order
%   with weight-function $w(\SR) = 1${,}$w(\Scp) = 0$ and precedence
%    $\Scp > \SR$.
\end{proof}

\begin{figure}[t]
\begin{minipage}{.25\textwidth}
\textbf{Forking diagrams:}
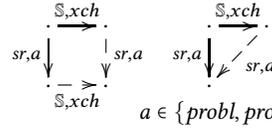
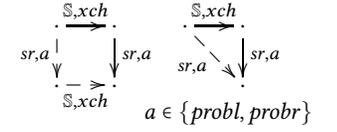
\[
\begin{array}[t]{@{}c@{}}
\xymatrix@R=5mm@C=5mm{
 \cdot \ar[d]_{\sr,a}\ar[r]^{\SCtxt,xch}  & \cdot \ar@{-->}[d]^{\sr,a}\\
 \cdot \ar@{-->}[r]_{\SCtxt,xch}         & \cdot \\
}
\end{array}\!\!\!
%%%%%%%%%%%%%%%%%%%%%%%%%%%%%%%%%%%
\begin{array}[t]{@{}c@{}}
\xymatrix@R=5mm@C=5mm{
 \cdot \ar[d]_{\sr,a}\ar[r]^{\SCtxt,xch}  & \cdot \ar@{-->}[dl]^{\sr,a}\\
 \cdot \\
}
\\
a \in \{\problabl,\problabr\}
\end{array} 
\]
\end{minipage}\begin{minipage}{.25\textwidth}
\textbf{Commuting diagrams:}
\[
\begin{array}[t]{@{}c@{}}
\xymatrix@R=5mm@C=5mm{
 \cdot \ar@{-->}[d]_{\sr,a}\ar[r]^{\SCtxt,xch}  & \cdot \ar[d]^{\sr,a}\\
 \cdot \ar@{-->}[r]_{\SCtxt,xch}         & \cdot \\
}
\end{array}\!\!\!\!
%%%%%%%%%%%%%%%%%%%%%%%%%%%%%%%%%%%
\begin{array}[t]{@{}c@{}}
\xymatrix@R=5mm@C=5mm{
 \cdot \ar@{-->}[dr]_{\sr,a}\ar[r]^{\SCtxt,xch}  & \cdot \ar[d]^{\sr,a}\\
 &\cdot \\
}
\\
a \in \{\problabl,\problabr\}
\end{array}
%%%%%%%%%%%%%%%%%%%%%%%%%%%%%%%%%%%
\]
\end{minipage}
\textbf{Base cases:}
Let $s \xrightarrow{\SCtxt,xch} s'$. Then $s$ is a WHNF iff  $s'$ is a WHNF.
\caption{Diagrams and Base Cases for $(xch)$\label{fig:xch-diagrams}}
\end{figure}

% \subsection{Correctness of (\xch)}
The diagrams and base cases for (\xch) are shown in \cref{fig:xch-diagrams}.

\begin{proposition}
%[Correctness of (\xch)]
The transformation $(xch)$ is correct.
\end{proposition}
\begin{proof}
The proof is analogous to the proof of \cref{prop:correctness-cp}{,}where the diagrams and the bases cases for (xch) are applied.
Termination of the diagram application is obvious{,}and also that prob-sequences are preserved.
\end{proof}

{\renewcommand{\thetheorem}{\ref{prop-cpx-xch-correct}}
\begin{proposition}
The transformations $\xrightarrow{\cpx}$ and $\xrightarrow{\xch}$
are correct.
\end{proposition}
\addtocounter{theorem}{-1}
}

\section{Diagrams}
\begin{figure*}
\begin{subfigure}{\textwidth}
\begin{minipage}{.45\textwidth}
\textbf{Forking diagrams:}
$$
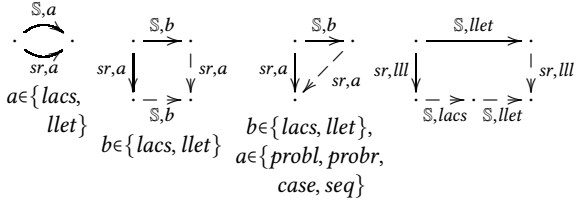
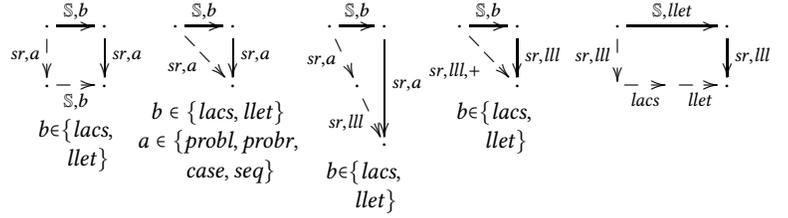
\begin{array}[t]{@{}c@{}}
\xymatrix@R=5mm@C=5mm{
 \cdot \ar@/^-.5pc/[r]_{\sr,a}\ar@/^.5pc/[r]^{\SCtxt,a}  & \cdot
}
\\
a{\in}\{\lacs,\\
\qquad\llet\}
\end{array}
\begin{array}[t]{@{}c@{}}
\xymatrix@R=5mm@C=5mm{
\cdot\ar[r]^{\SCtxt,b}       
     \ar[d]_{\sr,a}
                                &\cdot\ar@{-->}[d]^{\sr,a}
\\
\cdot\ar@{-->}[r]_{\SCtxt,b}       &\cdot
}
\\
b {\in} \{\lacs,\llet\}
\end{array}
\begin{array}[t]{@{}c@{}}
\xymatrix@R=5mm@C=5mm{
\cdot\ar[r]^{\SCtxt,b}       
     \ar[d]_{\sr,a}
                                &\cdot\ar@{-->}[dl]^{\sr,a}
\\
\cdot
}
\\
b {\in} \{\lacs,\llet\},\\
a {\in} \{\problabl,\problabr,\\
\quad\caserule,\seq\}
\end{array}
\begin{array}[t]{@{\!\!\!\!\!}c@{}}
\xymatrix@R=5mm@C=5mm{
\cdot\ar[rr]^{\SCtxt,\llet}       
     \ar[d]_{\sr,\lllrule}
                       &         &\cdot\ar@{-->}[d]^{\sr,\lllrule}
\\
\cdot\ar@{-->}[r]_{\SCtxt,\lacs}       &\cdot\ar@{-->}[r]_{\SCtxt,\llet} &\cdot
}
\end{array}
$$
\end{minipage}\begin{minipage}{.55\textwidth}
\textbf{Commuting diagrams:}
$$
\begin{array}[t]{@{\!\!}c@{}}
\xymatrix@R=5mm@C=5mm{
\cdot\ar[r]^{\SCtxt,b}       
     \ar@{-->}[d]_{\sr,a}
                                &\cdot\ar[d]^{\sr,a}
\\
\cdot\ar@{-->}[r]_{\SCtxt,b}       &\cdot
}
\\
b {\in} \{\lacs,\\
\quad\llet\}
\end{array}
\begin{array}[t]{@{\!\!\!}c@{}}
\xymatrix@R=5mm@C=5mm{
\cdot\ar[r]^{\SCtxt,b}       
     \ar@{-->}[dr]_{\sr,a}
                                &\cdot\ar[d]^{\sr,a}
\\
                                &\cdot
}
\\
b \in \{\lacs,\llet\}\\
a \in \{\problabl,\problabr,\\
\quad\caserule,\seq\}
\end{array}
\begin{array}[t]{@{}c@{}}
\xymatrix@R=5mm@C=1mm{
\cdot\ar[rr]^{\SCtxt,b}       
     \ar@{-->}[dr]_{\sr,a} &&\cdot\ar[dd]^{\sr,a}
\\
                        &\cdot\ar@{-->}[dr]_{\sr,\lllrule}
\\                                
                        &                             &\cdot
}
\\
b{\in}\{\lacs,\\
\quad\llet\}
\end{array}
\begin{array}[t]{@{\!\!}c@{}}
\xymatrix@R=5mm@C=5mm{
\cdot\ar[r]^{\SCtxt,b}       
     \ar@{-->}[dr]_{\sr,\lllrule,+}
                                &\cdot\ar[d]^{\sr,\lllrule}
\\
                                &\cdot
}
\\
b{\in}\{\lacs,\\
\quad\llet\}
\end{array}
\begin{array}[t]{@{}c@{}}
\xymatrix@R=5mm@C=5mm{
\cdot\ar[rr]^{\SCtxt,\llet}       
     \ar@{-->}[d]_{\sr,\lllrule}
                                &&\cdot\ar[d]^{\sr,\lllrule}
\\
  \ar@{-->}[r]_{\lacs}&\ar@{-->}[r]_{\llet}                          &\cdot
}
\end{array}
$$
\end{minipage}\\\begin{minipage}{\textwidth}
\textbf{Base cases:}
If $s \xrightarrow{\SCtxt,\lacs} s'$ then $s'$ is a WHNF iff $s$ is a WHNF.\\
\phantom{Base cases:\quad\!\!}If $s \xrightarrow{\SCtxt,\llet} s'$ then: if $s$ is a WHFN{,}then $s'$ is a WHNF{,}and if $s'$ is a WHNF{,}then $s$ is a WHNF or $s\xrightarrow{\sr,\lllrule} s''$ where $s''$ is a WHNF.
\end{minipage}
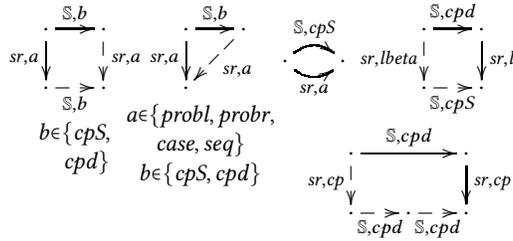
\caption{Diagrams for $(\lllrule)$}
\end{subfigure}

\begin{subfigure}{\textwidth}
\begin{minipage}[t]{.41\textwidth}
\textbf{Forking diagrams:}
\[
\begin{array}[t]{@{\!\!}c@{}}
\xymatrix@R=5mm@C=5mm{
 \cdot \ar[d]_{\sr,a}\ar[r]^{\SCtxt,b}  & \cdot \ar@{-->}[d]^{\sr,a}\\
 \cdot \ar@{-->}[r]_{\SCtxt,b}         & \cdot \\
}
\\
b {\in}\{\cpS,\\\quad\cpd\}
\end{array}
%%%%%%%%%%%%%%%%%%%%%%%%%%%%%%%%%%%
\begin{array}[t]{@{\!\!\!\!\!\!}c@{}}
\xymatrix@R=5mm@C=5mm{
 \cdot \ar[d]_{\sr,a}\ar[r]^{\SCtxt,b}  & \cdot \ar@{-->}[dl]^{\sr,a}\\
 \cdot \\
}
\\
a{\in}\{\problabl,\problabr,\\
\caserule,\seq\}\\
b {\in}\{\cpS,\cpd\}
\end{array}
% %%%%%%%%%%%%%%%%%%%%%%%%%%%%%%%%%%%
% \begin{array}[t]{@{}c@{}}
% \xymatrix@R=5mm@C=2mm{
%  \cdot \ar@{-->}[dr]_{\sr,a}\ar[rr]^{\SCtxt,cpS}  && \cdot \ar[dd]^{\sr,a}\\
%  &\cdot\ar@{-->}[dr]_{\sr,\cp} \\
%  &&\cdot
% }
% \\
% a\in\{\lbeta,\seq\}
% \end{array}
% %%%%%%%%%%%%%%%%%%%%%%%%%%%%%%%%%%%
% \begin{array}[t]{@{}c@{}}
% \xymatrix@R=5mm@C=5mm{
%  \cdot \ar@{-->}[d]_{\sr,a}\ar[r]^{\SCtxt,cpd}  & \cdot \ar[d]^{\sr,a}\\
%  \cdot \ar@{-->}[r]_{\SCtxt,cpd}         & \cdot \\
% }
% \end{array}
%%%%%%%%%%%%%%%%%%%%%%%%%%%%%%%%%%%
\begin{array}[t]{@{}c@{}}
\begin{array}[t]{@{}c@{}}
\\[-3ex]
\xymatrix@R=5mm@C=5mm{
 \cdot \ar@/^-.5pc/[r]_{\sr,a}\ar@/^.5pc/[r]^{\SCtxt,\cpS}  & \cdot
}
\end{array}
\begin{array}[t]{@{}c@{}}
\xymatrix@R=5mm@C=5mm{
 \cdot \ar@{-->}[d]_{\sr,\lbeta}\ar[r]^{\SCtxt,cpd}  & \cdot \ar[d]^{\sr,\lbeta}\\
 \cdot \ar@{-->}[r]_{\SCtxt,cpS}         & \cdot \\
}
\end{array}
\\
% %%%%%%%%%%%%%%%%%%%%%%%%%%%%%%%%%%%
% \begin{array}[t]{@{}c@{}}
% \xymatrix@R=5mm@C=5mm{
%  \cdot \ar@{-->}[dr]_{\sr,a}\ar[r]^{\SCtxt,cpd}  & \cdot \ar[d]^{\sr,a}\\
%                                      & \cdot \\
% }
% \\
% a \in \{\problabl,\problabr,\\
% \caserule,\seq\}
% \end{array}
% %%%%%%%%%%%%%%%%%%%%%%%%%%%%%%%%%%%
\begin{array}[t]{@{\!\!\!}c@{}}
\xymatrix@R=5mm@C=5mm{
 \cdot \ar@{-->}[d]_{\sr,\cp}\ar[rr]^{\SCtxt,cpd}&                            & \cdot \ar[d]^{\sr,\cp}\\
 \cdot \ar@{-->}[r]_{\SCtxt,cpd}         & \cdot \ar@{-->}[r]_{\SCtxt,cpd} & \cdot\\
}
\end{array}
\end{array}
\]
% \begin{verbatim}
% Set of forking diagrams for -cpT->
% ====================
% Forking diagram: <-SR,a- . -cpT-> ~~> -cpT-> . <-SR,a-
%  where a in {lbeta,probl,probr,cp,lll,case,seq}
% Forking diagram: <-SR,a- . -cpT-> ~~> <-SR,a-
%  where a in {probl,probr,case,seq}
% Forking diagram: <-SR,cp- . -cpT-> ~~> 
% Forking diagram: <-ANSWER- . -cpT-> ~~> <-ANSWER-
% 
% Set of forking diagrams for -cpd->
% Forking diagram: <-SR,a- . -cpd-> ~~> -cpd-> . <-SR,a-
%  where a in {lbeta,probl,probr,cp,lll,case,seq}
% Forking diagram: <-SR,a- . -cpd-> ~~> <-SR,a-
%  where a in {probl,probr,case,seq}
% Forking diagram: <-SR,lbeta- . -cpd-> ~~> -cpT-> . <-SR,lbeta-
% Forking diagram: <-SR,cp- . -cpd-> ~~> -cpd-> . -cpd-> . <-SR,cp-
% Forking diagram: <-ANSWER- . -cpd-> ~~> <-ANSWER-
% \end{verbatim}
\end{minipage}\begin{minipage}[t]{.59\textwidth}
\textbf{Commuting diagrams:}
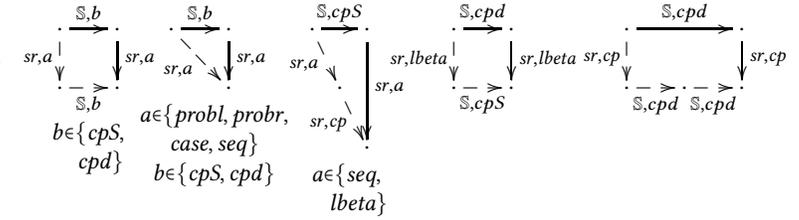
\[
\begin{array}[t]{@{}c@{}}
\xymatrix@R=5mm@C=5mm{
 \cdot \ar@{-->}[d]_{\sr,a}\ar[r]^{\SCtxt,b}  & \cdot \ar[d]^{\sr,a}\\
 \cdot \ar@{-->}[r]_{\SCtxt,b}         & \cdot \\
}
\\
b{\in}\{\cpS,\\\quad\cpd\}
\end{array}
%%%%%%%%%%%%%%%%%%%%%%%%%%%%%%%%%%%
\begin{array}[t]{@{\!\!\!\!\!\!}c@{}}
\xymatrix@R=5mm@C=5mm{
 \cdot \ar@{-->}[dr]_{\sr,a}\ar[r]^{\SCtxt,b}  & \cdot \ar[d]^{\sr,a}\\
 &\cdot \\
}
\\
a{\in}\{\problabl,\problabr,\\
\caserule,\seq\}
\\
b{\in}\{\cpS,\cpd\}
\end{array}
%%%%%%%%%%%%%%%%%%%%%%%%%%%%%%%%%%%
\begin{array}[t]{@{\!\!}c@{}}
\xymatrix@R=5mm@C=1mm{
 \cdot \ar@{-->}[dr]_{\sr,a}\ar[rr]^{\SCtxt,cpS}  && \cdot \ar[dd]^{\sr,a}\\
 &\cdot\ar@{-->}[dr]_{\sr,\cp} \\
 &&\cdot
}
\\
a{\in}\{\seq,\\\quad\lbeta\}
\end{array}
% %%%%%%%%%%%%%%%%%%%%%%%%%%%%%%%%%%%
% \begin{array}[t]{@{}c@{}}
% \xymatrix@R=5mm@C=5mm{
%  \cdot \ar@{-->}[d]_{\sr,a}\ar[r]^{\SCtxt,cpd}  & \cdot \ar[d]^{\sr,a}\\
%  \cdot \ar@{-->}[r]_{\SCtxt,cpd}         & \cdot \\
% }
% \end{array}
%%%%%%%%%%%%%%%%%%%%%%%%%%%%%%%%%%%
\begin{array}[t]{@{\!\!\!\!\!\!\!\!}c@{}}
\xymatrix@R=5mm@C=5mm{
 \cdot \ar@{-->}[d]_{\sr,\lbeta}\ar[r]^{\SCtxt,cpd}  & \cdot \ar[d]^{\sr,\lbeta}\\
 \cdot \ar@{-->}[r]_{\SCtxt,cpS}         & \cdot \\
}
\end{array}
%%%%%%%%%%%%%%%%%%%%%%%%%%%%%%%%%%%
% \begin{array}[t]{@{}c@{}}
% \xymatrix@R=5mm@C=5mm{
%  \cdot \ar@{-->}[dr]_{\sr,a}\ar[r]^{\SCtxt,cpd}  & \cdot \ar[d]^{\sr,a}\\
%                                      & \cdot \\
% }
% \\
% a \in \{\problabl,\problabr,\\
% \caserule,\seq\}
% \end{array}
%%%%%%%%%%%%%%%%%%%%%%%%%%%%%%%%%%%
\begin{array}[t]{@{\!\!}c@{}}
\xymatrix@R=5mm@C=5mm{
 \cdot \ar@{-->}[d]_{\sr,\cp}\ar[rr]^{\SCtxt,cpd}&                            & \cdot \ar[d]^{\sr,\cp}\\
 \cdot \ar@{-->}[r]_{\SCtxt,cpd}         & \cdot \ar@{-->}[r]_{\SCtxt,cpd} & \cdot\\
}
\end{array}
\]
\end{minipage}\\
\begin{minipage}[t]{\textwidth}
\textbf{Base cases}: If $s \xrightarrow{\SCtxt,\cp} t${,}then if $s$ is a WHNF{,}then $t$ is a WHNF{,}and if $t$ is a WHNF 
then either $s$ is a WHNF or $s \xrightarrow{\SR,\cp} s'$ and $s'$ is a WHNF.
\end{minipage}
\caption{Diagrams for $(\cp)$}
\end{subfigure}
\begin{subfigure}{.38\textwidth}
\begin{minipage}{.46\textwidth}
\textbf{Forking diagrams:}
$$
\begin{array}[t]{@{\!\!\!}c@{\!\!}}
\xymatrix@R=5mm@C=5mm{
\cdot\ar[r]^{\SCtxt,\xch}\ar[d]_{\sr,a} &\cdot\ar@{-->}[d]^{\sr,a}
\\
\cdot\ar@{-->}[r]_{\SCtxt,\xch}         &\cdot
}
\end{array}\begin{array}[t]{@{\!\!\!\!}c@{}}
\xymatrix@R=5mm@C=5mm{
\cdot\ar[r]^{\SCtxt,\xch}\ar[d]_{\sr,a} &\cdot\ar@{-->}[dl]^{\sr,a}
\\
\cdot
}
\\
a{\in}\{\problabl,\problabr,\\
\quad\caserule,\seq\}
\end{array}
$$
\end{minipage}\begin{minipage}{.5\textwidth}
\textbf{Commuting diagrams:}
$$
\begin{array}[t]{@{}c@{}}
\xymatrix@R=5mm@C=5mm{
\cdot\ar[r]^{\SCtxt,\xch}       
     \ar@{-->}[d]_{\sr,a}             &\cdot\ar[d]^{\sr,a}
\\
\cdot\ar@{-->}[r]_{\SCtxt,\xch}       &\cdot
}
\end{array}
\begin{array}[t]{@{\!\!\!\!\!\!\!\!}c@{}}
\xymatrix@R=5mm@C=5mm{
\cdot\ar[r]^{\SCtxt,\xch}       
     \ar@{-->}[dr]_{\sr,a}
                                &\cdot\ar[d]^{\sr,a}
\\
&\cdot
}
\\
a{\in}\{\problabl,\problabr,\\\quad\caserule,\seq\}
\end{array}
$$
\end{minipage}\\\begin{minipage}{\textwidth}
\textbf{Base case:}
\\
For $s \xrightarrow{\SCtxt{,}\xch} t$,
$s$ is a WHNF iff $t$ is a WHNF.

\end{minipage}
\caption{Diagrams for $(\xch)$}
\end{subfigure}\begin{subfigure}{.62\textwidth}
\begin{minipage}{.5\textwidth}
\textbf{Forking diagrams:}
$$
\begin{array}[t]{@{\!\!}c@{}}
\xymatrix@R=5mm@C=5mm{
\cdot\ar[r]^{\SCtxt,\cpx}       
     \ar[d]_{\sr,a}
                                &\cdot\ar@{-->}[d]^{\sr,a}
\\
\cdot\ar@{-->}[r]_{\SCtxt,\cpx}       &\cdot
}
\end{array}
\begin{array}[t]{@{\!\!\!\!\!\!}c@{}}
\xymatrix@R=5mm@C=5mm{
\cdot\ar[r]^{\SCtxt,\cpx}       
     \ar[d]_{\sr,a}
                                &\cdot\ar@{-->}[dl]^{\sr,a}
\\
\cdot
}
\\
a{\in}\{\problabl,\problabr,\\\caserule,\seq\}
\end{array}
\begin{array}[t]{@{\!\!\!\!\!}c@{}}
\xymatrix@R=5mm@C=4mm{
\cdot\ar[rr]^{\SCtxt,\cpx}       
     \ar[d]_{\sr,\cp}
                                &&\cdot\ar@{-->}[d]_{\sr,\cp}
\\
\cdot\ar@{-->}[r]_{\SCtxt,\cpx} &\cdot\ar@{-->}[r]_{\SCtxt,\cpx}        &\cdot
}
\\
\end{array}
$$
\end{minipage}
% Set of forking diagrams for -cpx->
% ====================
% Forking diagram: <-SR,a- . -cpx-> ~~> -cpx-> . <-SR,a-
%  where a in {lbeta,probl,probr,cp,lll,case,seq}
% Forking diagram: <-SR,a- . -cpx-> ~~> <-SR,a-
%  where a in {probl,probr,case,seq,cp}
% Forking diagram: <-SR,cp- . -cpx-> ~~> -cpx-> . -cpx-> . <-SR,cp-
% Forking diagram: <-ANSWER- . -cpx-> ~~> <-ANSWER-
\begin{minipage}{.5\textwidth}
\textbf{Commuting diagrams:}
$$
\begin{array}[t]{@{}c@{}}
\xymatrix@R=5mm@C=5mm{
\cdot\ar[r]^{\SCtxt,\cpx}       
     \ar@{-->}[d]_{\sr,a}
                                &\cdot\ar[d]^{\sr,a}
\\
\cdot\ar@{-->}[r]_{\SCtxt,\cpx}       &\cdot
}
\end{array}
\begin{array}[t]{@{\!\!\!}c@{}}
\xymatrix@R=5mm@C=5mm{
\cdot\ar[r]^{\SCtxt,\cpx}       
     \ar@{-->}[dr]_{\sr,a}
                                &\cdot\ar[d]^{\sr,a}
\\
                                &\cdot
}
\\
a{\in}\{\problabl,\problabr,\\
\caserule,\seq,\cp\}
\end{array}
\begin{array}[t]{@{\!\!\!\!\!\!\!}c@{}}
\xymatrix@R=5mm@C=4mm{
\cdot\ar[rr]^{\SCtxt,\cpx}       
     \ar@{-->}[d]_{\sr,\cp}
                                &&\cdot\ar[d]_{\sr,\cp}
\\
\cdot\ar@{-->}[r]_{\SCtxt,\cpx} &\cdot\ar@{-->}[r]_{\SCtxt,\cpx}        &\cdot
}
\\
\end{array}
$$
\end{minipage}
\\
\begin{minipage}{\textwidth}
\textbf{Base case:} For $s \xrightarrow{\SCtxt,\cpx} t${,}$s$ is a WHNF iff $t$ is a WHNF.
\end{minipage}
\caption{Diagrams for $(\cpx)$}
\end{subfigure}

\begin{subfigure}{\textwidth}
\begin{minipage}{\textwidth}
\textbf{Forking diagrams:}
$$
\begin{array}[t]{@{}c@{}}
\xymatrix@R=5mm@C=5mm{
\cdot\ar[r]^{\SCtxt,\cpcx}       
     \ar[d]_{\sr,a}
                                &\cdot\ar@{-->}[d]^{\sr,a}
\\
\cdot\ar@{-->}[r]_{\SCtxt,\cpcx}       &\cdot
}
\end{array}
\begin{array}[t]{@{}c@{}}
\xymatrix@R=5mm@C=5mm{
\cdot\ar[r]^{\SCtxt,\cpcx}       
     \ar[d]_{\sr,a}
                                &\cdot\ar@{-->}[dl]^{\sr,a}
\\
                    \cdot
}
\\
a \in\{\problabl,\problabr,\\
\caserule,\seq\}
\end{array}
\begin{array}[t]{@{}c@{}}
\xymatrix@R=5mm@C=5mm{
\cdot\ar[r]^{\SCtxt,\cpcx}       
     \ar[d]_{\sr,a}
                                &\cdot\ar@{-->}[d]^{\sr,a}
\\
\cdot\ar@{-->}[r]_{\SCtxt,\absrule}       &\cdot
}
\\
a \in\{\problabl,\problabr,\\\caserule,\seq\}
\end{array}
\begin{array}[t]{@{\!\!\!}c@{}}
\xymatrix@R=5mm@C=8mm{
\cdot\ar[rrr]^{\SCtxt,\cpcx}       
     \ar[d]_{\sr,\caserule}
              &                  &&\cdot\ar@{-->}[d]^{\sr,\caserule}
\\
\cdot\ar@{-->}[r]_{\SCtxt,\cpcx\vee\absrule} &\cdot\ar@{-->}[r]_{\SCtxt,\cpx.+}    &\cdot\ar@{-->}[r]_{\SCtxt,\xch,+} &\cdot
}
\end{array}
\begin{array}[t]{@{\!\!\!}c@{}}
\xymatrix@R=5mm@C=5mm{
\cdot\ar[rr]^{\SCtxt,\cpcx}       
     \ar[d]_{\sr,\cp}
                                &&\cdot\ar@{-->}[d]^{\sr,\cp}
\\
\cdot\ar@{-->}[r]_{\SCtxt,\cpcx} &\cdot\ar@{-->}[r]_{\SCtxt,\cpcx}        &\cdot
}
\end{array}
\begin{array}[t]{@{\!\!\!}c@{}}
\xymatrix@R=5mm@C=6mm{
\cdot\ar[rrr]^{\SCtxt,\cpcx}       
     \ar[d]_{\sr,\cp}
              &                  &&\cdot\ar@{-->}[d]^{\sr,\cp}
\\
\cdot\ar@{-->}[r]_{\SCtxt,\cpcx,+} &\cdot\ar@{-->}[r]_{\SCtxt,\cpx.+}    &\cdot\ar@{-->}[r]_{\SCtxt,\gc} &\cdot
}
\\
\end{array}
$$
% 
% Set of forking diagrams for -cpcx->
% ====================
% Dependencies are diagrams for: -abs->,-cpx->,-xch->,-gc->
% ====================
% Forking diagram: <-SR,a- . -cpcx-> ~~> -cpcx-> . <-SR,a-
%  where a in {lbeta,probl,probr,cp,lll,case,seq}
% Forking diagram: <-SR,a- . -cpcx-> ~~> <-SR,a-
%  where a in {probl,probr,case,seq}
% Forking diagram: <-ANSWER- . -cpcx-> ~~> <-ANSWER-
% Forking diagram: <-SR,a- . -cpcx-> ~~> -abs-> . <-SR,a-
%  where a in {probl,probr,case,seq}
% Forking diagram: <-SR,case- . -cpcx-> ~~> -cpcx-> . -cpx,+-> . -xch,+-> . <-SR,case-
% Forking diagram: <-SR,case- . -cpcx-> ~~> -abs-> . -cpx,+-> . -xch,+-> . <-SR,case-
% Forking diagram: <-SR,cp- . -cpcx-> ~~> -cpcx-> . -cpcx-> . <-SR,cp-
% Forking diagram: <-SR,cp- . -cpcx-> ~~> -cpcx-> . -cpcx-> . -cpx,+-> . -gc-> . <-SR,cp-

\textbf{Commuting diagrams:}
$$
\begin{array}[t]{@{}c@{}}
\xymatrix@R=5mm@C=5mm{
\cdot\ar[r]^{\SCtxt,\cpcx}       
     \ar@{-->}[d]_{\sr,a}
                                &\cdot\ar[d]^{\sr,a}
\\
\cdot\ar@{-->}[r]_{\SCtxt,\cpcx}       &\cdot
}
\end{array}
\begin{array}[t]{@{}c@{}}
\xymatrix@R=5mm@C=5mm{
\cdot\ar[r]^{\SCtxt,\cpcx}       
     \ar@{-->}[dr]_{\sr,a}
                                &\cdot\ar[d]^{\sr,a}
\\
                                &\cdot
}
\\
a \in\{\problabl,\problabr,\\
\caserule,\seq\}
\end{array}
\begin{array}[t]{@{}c@{}}
\xymatrix@R=5mm@C=5mm{
\cdot\ar[r]^{\SCtxt,\cpcx}       
     \ar@{-->}[d]_{\sr,a}
                                &\cdot\ar[d]^{\sr,a}
\\
\cdot\ar@{-->}[r]_{\SCtxt,\absrule}       &\cdot
}
\\
a \in\{\problabl,\problabr,\\\caserule,\seq\}
\end{array}
\begin{array}[t]{@{\!\!\!}c@{}}
\xymatrix@R=5mm@C=8mm{
\cdot\ar[rrr]^{\SCtxt,\cpcx}       
     \ar@{-->}[d]_{\sr,\caserule}
              &                  &&\cdot\ar[d]^{\sr,\caserule}
\\
\cdot\ar@{-->}[r]_{\SCtxt,\cpcx\vee\absrule} &\cdot\ar@{-->}[r]_{\SCtxt,\cpx.+}    &\cdot\ar@{-->}[r]_{\SCtxt,\xch,+} &\cdot
}
\end{array}
\begin{array}[t]{@{\!\!\!}c@{}}
\xymatrix@R=5mm@C=5mm{
\cdot\ar[rr]^{\SCtxt,\cpcx}       
     \ar@{-->}[d]_{\sr,\cp}
                                &&\cdot\ar[d]^{\sr,\cp}
\\
\cdot\ar@{-->}[r]_{\SCtxt,\cpcx} &\cdot\ar@{-->}[r]_{\SCtxt,\cpcx}        &\cdot
}
\end{array}
\begin{array}[t]{@{\!\!\!}c@{}}
\xymatrix@R=5mm@C=6mm{
\cdot\ar[rrr]^{\SCtxt,\cpcx}       
     \ar@{-->}[d]_{\sr,\cp}
              &                  &&\cdot\ar[d]^{\sr,\cp}
\\
\cdot\ar@{-->}[r]_{\SCtxt,\cpcx,+} &\cdot\ar@{-->}[r]_{\SCtxt,\cpx.+}    &\cdot\ar@{-->}[r]_{\SCtxt,\gc} &\cdot
}
\\
\end{array}
$$
\textbf{Base cases:} If $s \xrightarrow{\SCtxt,\cpcx} s'$ and $s'$ is a WHNF{,}then $s$ is a WHNF.
 
\end{minipage}
\end{subfigure}

\caption{Diagrams for $\LRPROB$\label{fig-diagr-lrprob-1}}

\end{figure*}

\cref{fig-diagr-lrprob-1,fig-diagr-lrprob-2} show the diagrams for transformations in the calculus $\LRPROB$.

\begin{figure*}

\begin{subfigure}{\textwidth}

\begin{minipage}{.5\textwidth}
\textbf{Forking diagrams:}
$$
\begin{array}[t]{@{}c@{}}
\xymatrix@R=5mm@C=5mm{
\cdot\ar[r]^{\SCtxt,\absrule}       
     \ar[d]_{\sr,a}
                                &\cdot\ar@{-->}[d]^{\sr,a}
\\
\cdot\ar@{-->}[r]_{\SCtxt,\absrule}       &\cdot
}
\end{array}
\begin{array}[t]{@{}c@{}}
\xymatrix@R=5mm@C=5mm{
\cdot\ar[r]^{\SCtxt,\absrule}       
     \ar[d]_{\sr,a}
                                &\cdot\ar@{-->}[dl]^{\sr,a}
\\
                            \cdot
}
\\
a {\in}\{\problabl,\problabr,\\\quad\caserule,\seq\}
\end{array}
\begin{array}[t]{@{}c@{}}
\xymatrix@R=5mm@C=6mm{
\cdot\ar[rrr]^{\SCtxt,\absrule}       
     \ar[d]_{\sr,\caserule}
              &                  &&\cdot\ar@{-->}[d]^{\sr,\caserule}
\\
\cdot\ar@{-->}[r]_{\SCtxt,\absrule} &\cdot\ar@{-->}[r]_{\SCtxt,\cpx.+}    &\cdot\ar@{-->}[r]_{\SCtxt,\xch,+} &\cdot
}
\\
\end{array}
$$
% Set of forking diagrams for -abs->
% ====================
% Dependencies are diagrams for: -cpx->,-xch->
% ====================
% Forking diagram: <-SR,a- . -abs-> ~~> -abs-> . <-SR,a-
%  where a in {lbeta,probl,probr,cp,lll,case,seq}
% Forking diagram: <-SR,a- . -abs-> ~~> <-SR,a-
%  where a in {probl,probr,case,seq}
% Forking diagram: <-SR,case- . -abs-> ~~> -abs-> . -cpx,+-> . -xch,+-> . <-SR,case-
% Forking diagram: <-ANSWER- . -abs-> ~~> <-ANSWER-
\end{minipage}\begin{minipage}{.5\textwidth}
\textbf{Commuting diagrams:}
$$
\begin{array}[t]{@{}c@{}}
\xymatrix@R=5mm@C=5mm{
\cdot\ar[r]^{\SCtxt,\absrule}       
     \ar@{-->}[d]_{\sr,a}
                                &\cdot\ar[d]^{\sr,a}
\\
\cdot\ar@{-->}[r]_{\SCtxt,\absrule}       &\cdot
}
\end{array}
\begin{array}[t]{@{}c@{}}
\xymatrix@R=5mm@C=5mm{
\cdot\ar[r]^{\SCtxt,\absrule}       
     \ar@{-->}[dr]_{\sr,a}
                                &\cdot\ar[d]^{\sr,a}
\\
                                &\cdot
}
\\
a{\in}\{\problabl,\problabr,\\\quad\caserule,\seq\}
\end{array}
\begin{array}[t]{@{}c@{}}
\xymatrix@R=5mm@C=6mm{
\cdot\ar[rrr]^{\SCtxt,\absrule}       
     \ar@{-->}[d]_{\sr,\caserule}
              &                  &&\cdot\ar[d]^{\sr,\caserule}
\\
\cdot\ar@{-->}[r]_{\SCtxt,\absrule} &\cdot\ar@{-->}[r]_{\SCtxt,\cpx.+}    &\cdot\ar@{-->}[r]_{\SCtxt,\xch,+} &\cdot
}
\\
\end{array}
$$
\end{minipage}
\\
\begin{minipage}{1\textwidth}
\textbf{Base cases:} If $s \xrightarrow{\SCtxt,\absrule} t${,}then $s$ is a WHNF iff $t$ is a WHNF.
\end{minipage}

\caption{Diagrams for $(\abs)$}
\end{subfigure}

\begin{subfigure}{\textwidth}
% (\gc) \davidcomment{Hier weiter}
\begin{minipage}[t]{.21\textwidth}
\textbf{Forking diagrams:}
$$\begin{array}[t]{@{}c@{}}
\xymatrix@R=5mm@C=5mm{
\cdot\ar[r]^{\SCtxt,\gc}       
     \ar[d]_{\sr,a}
                                &\cdot\ar@{-->}[d]^{\sr,a}
\\
\cdot\ar@{-->}[r]_{\SCtxt,\gc}       &\cdot
}
\end{array}
\begin{array}[t]{@{\!\!\!\!\!\!\!}c@{}}
\xymatrix@R=5mm@C=5mm{
\cdot\ar[r]^{\SCtxt,\gc}       
     \ar[d]_{\sr,a}
                                &\cdot\ar@{-->}[dl]^{\sr,a}
\\
                                &\cdot
}
\\
a{\in}\{\problabl,\problabr,\\\caserule,\seq,\lllrule\}
\end{array}
$$
% Forking diagram: <-SR,a . -gc-> ~~> -gc-> . <-SR,a-
%    where a in {lbeta,probl,probr,cp,lll,case,seq}
% Forking diagram: <-SR,a- . -gc-> ~~> <-SR,a-
%    where a in {probl,probr,case,seq}
% 
% Forking diagram: <-ANSWER- . -gc-> ~~> <-ANSWER-
\end{minipage}\begin{minipage}[t]{.47\textwidth}
\textbf{Commuting diagrams:}
$$
\begin{array}[t]{@{}c@{}}
\xymatrix@R=5mm@C=5mm{
\cdot\ar[r]^{\SCtxt,\gc}       
     \ar@{-->}[d]_{\sr,a}
                                &\cdot\ar[d]^{\sr,a}
\\
\cdot\ar@{-->}[r]_{\SCtxt,\gc}       &\cdot
}
\end{array}
\begin{array}[t]{@{}c@{}}
\xymatrix@R=5mm@C=5mm{
\cdot\ar[r]^{\SCtxt,\gc}       
     \ar@{-->}[dr]_{\sr,a}
                                &\cdot\ar[d]^{\sr,a}
\\
                                &\cdot
}
\\
a {\in}\{\problabl,\problabr,\\\caserule,\seq\}
\end{array}
\begin{array}[t]{@{\!\!\!\!}c@{}}
\xymatrix@R=5mm@C=5mm{
\cdot\ar[r]^{\SCtxt,\gc}       
     \ar@{-->}[d]_{\sr,\lllrule,+}  &\cdot\ar[dd]^{\sr,a}
     \\
   \ar@{-->}[d]_{\sr,a}           &                  &
\\
\cdot\ar@{-->}[r]_{\SCtxt,\gc}&\cdot
}
\\
\end{array}
\begin{array}[t]{@{\!\!\!\!\!\!\!}c@{}}
\xymatrix@R=5mm@C=5mm{
\cdot\ar[r]^{\SCtxt,\gc}       
     \ar@{-->}[d]_{\sr,\lllrule,+}  &\cdot\ar[ddd]^{\sr,a}
     \\
   \ar@{-->}[d]_{\sr,a}           &                  &
     \\
   \ar@{-->}[d]_{\sr,\lllrule}           &                  &
\\
\cdot\ar@{-->}[r]_{\SCtxt,\gc}&\cdot
}
\\
a{\in}\{\lbeta,\lllrule,\caserule\}
\end{array}
$$
\end{minipage}\begin{minipage}[t]{.32\textwidth}
\textbf{Base cases:} If $s \xrightarrow{\SCtxt,gc} t${,}then if $s$ is a WHNF{,}then $t$ is a WHNF; and if $t$ is a WHNF{,}then $s$ is a WHNF
or $s \xrightarrow{\sr,\lllrule} s'$ and $s'$ is a WHNF
\end{minipage}
\caption{Diagrams for $(\gc)$}
\end{subfigure}
\begin{subfigure}{\textwidth}
\begin{minipage}[t]{.3\textwidth}
\textbf{Forking diagrams:}
\[
\begin{array}{@{}c@{}}
\begin{array}[t]{@{}c@{}}
\xymatrix@R=5mm@C=5mm{
 \cdot \ar[d]_{\sr,a}\ar[r]^{\SCtxt,\ucp}  & \cdot \ar@{-->}[d]^{\sr,a}\\
 \cdot \ar@{-->}[r]_{\SCtxt,\ucp}         & \cdot \\
}
\end{array}
%%%%%%%%%%%%%%%%%%%%%%%%%%%%%%%%%%%
\begin{array}[t]{@{}c@{}}
\xymatrix@R=5mm@C=5mm{
 \cdot \ar[d]_{\sr,\cp}\ar[r]^{\SCtxt,\ucp}  & \cdot\\
 \cdot \ar@{-->}[ur]_{\SCtxt,\gc}         \\
}
\end{array}
\\
%%%%%%%%%%%%%%%%%%%%%%%%%%%%%%%%%%%
\begin{array}[t]{@{}c@{}}
\xymatrix@R=5mm@C=5mm{
 \cdot \ar[d]_{\sr,a}\ar[r]^{\SCtxt,\ucp}  & \cdot \ar@{-->}[dl]^{\sr,a}\\
 \cdot \\
}
\\
a{\in}\{\problabl,\problabr,\\
\caserule,\seq\}
\end{array}
%%%%%%%%%%%%%%%%%%%%%%%%%%%%%%%%%%%
\begin{array}[t]{@{}c@{}}
\xymatrix@R=5mm@C=5mm{
 \cdot \ar[d]_{\sr,a}\ar[r]^{\SCtxt,\ucp}  & \cdot \ar@{-->}[d]^{\sr,a}\\
 \cdot \ar@{-->}[r]_{\SCtxt,\gc}         & \cdot \\
}
\\
a{\in}\{
\caserule,\seq,\cp\\
\problabl,\problabr\}
\end{array}
\\
%%%%%%%%%%%%%%%%%%%%%%%%%%%%%%%%%%%
\begin{array}[t]{@{}c@{}}
\xymatrix@R=5mm@C=5mm{
 \cdot \ar[d]_{\sr,\lllrule}\ar[r]^{\SCtxt,\ucp}  & \cdot \ar@{-->}[d]^{\sr,\lllrule,*}\\
 \cdot \ar@{-->}[r]_{\SCtxt,\ucp}         & \cdot \\
}
\end{array}
%%%%%%%%%%%%%%%%%%%%%%%%%%%%%%%%%%%
%%%%%%%%%%%%%%%%%%%%%%%%%%%%%%%%%%%
\begin{array}[t]{@{\!\!\!}c@{}}
\xymatrix@R=5mm@C=5mm{
 \cdot \ar[d]_{\sr,\caserule}\ar[rr]^{\SCtxt,\ucp}  && \cdot \ar@{-->}[d]^{\sr,\caserule}\\
\cdot\ar@{-->}[r]_{\SCtxt,\gc}&\cdot\ar@{-->}[r]_{\SCtxt,\ucp,+} &\cdot \\
}
\end{array}
\end{array}
\]
% \begin{verbatim}
% Forking diagram: <-SR,a- . -ucp-> ~~> -ucp-> . <-SR,a-
%    where a in {lbeta,probl,probr,case,seq,lll}
% Forking diagram: <-SR,a- . -ucp-> ~~> <-SR,a-
%    where a in {probl,probr,case,seq}
%    
% Forking diagram: <-SR,a- . -ucp-> ~~> -gc-> . <-SR,a-
%    where a in {probl,probr,case,seq}
% Forking diagram: <-SR,cp- . -ucp-> ~~> -gc->
% Forking diagram: <-SR,cp- . -ucp-> ~~> -gc-> . <-SR,cp-
% Forking diagram: <-SR,lll- . -ucp-> ~~> -ucp-> . <-SR,lll,+-
% Forking diagram: <-SR,case- . -ucp-> ~~> -gc-> . -ucp,+-> . <-SR,case-
% Forking diagram: <-SR,lll- . -ucp-> ~~> -ucp->
% Forking diagram: <-ANSWER- . -ucp-> ~~> <-ANSWER-
% \end{verbatim}

\end{minipage}\begin{minipage}[t]{.7\textwidth}
\textbf{Commuting diagrams:}
\[
\begin{array}{@{}c@{}}
%%%%%%%%%%%%%%%%%%%%%
\begin{array}[t]{@{}c@{}}
\begin{array}[t]{@{}c@{}}
\xymatrix@R=5mm@C=5mm{
 \cdot \ar@{-->}[d]_{\sr,a}\ar[r]^{\SCtxt,\ucp}  & \cdot \ar[d]^{\sr,a}\\
 \cdot \ar@{-->}[r]_{\SCtxt,\ucp}         & \cdot \\
}
\end{array}
%%%%%%%%%%%%%%%%%%%%%%%%%%%%%%%%%%%
\begin{array}[t]{@{}c@{}}
\xymatrix@R=5mm@C=5mm{
 \cdot \ar@{-->}[dr]_{\sr,a}\ar[r]^{\SCtxt,\ucp}  & \cdot \ar[d]^{\sr,a}\\
 &\cdot \\
}
\\
a \in \{\problabl,\problabr,\\
\caserule,\seq\}
\end{array}
%%%%%%%%%%%%%%%%%%%%%%%%%%%%%%%%%%%
\begin{array}[t]{@{}c@{}}
\xymatrix@R=5mm@C=5mm{
 \cdot \ar@{-->}[d]_{\sr,a}\ar[r]^{\SCtxt,\ucp}  & \cdot \ar[d]^{\sr,a}\\
 \cdot \ar@{-->}[r]_{\SCtxt,\gc}         & \cdot \\
}
\\
a \in \{\problabl,\problabr,\\
\caserule,\seq\}
\end{array}
%%%%%%%%%%%%%%%%%%%%%%%%%%%%%%%%%%%
\begin{array}[t]{@{}c@{}}
\xymatrix@R=5mm@C=5mm{
 \cdot \ar@{-->}[d]_{\sr,\lbeta}\ar[r]^{\SCtxt,\ucp}  & \cdot \ar[dd]^{\sr,\lbeta}\\
 \cdot\ar@{-->}[d]_{\sr,\lllrule}\\
\cdot\ar@{-->}[r]_{\SCtxt,\ucp}&\cdot \\
}
\end{array}
\\
%%%%%%%%%%%%%%%%%%%%%%%%%%%%%%%%%%%
\begin{array}[t]{@{}c@{}}
\xymatrix@R=5mm@C=5mm{
 \cdot \ar@{-->}[d]_{\sr,\lllrule,*}\ar[r]^{\SCtxt,\ucp}  & \cdot \ar[ddd]^{\sr,a}\\
 \cdot\ar@{-->}[d]_{\sr,\cp}\\
 \cdot\ar@{-->}[d]_{\sr,a}\\
 \cdot\ar@{-->}[r]_{\SCtxt,\gc}&\cdot \\
}
\\
a{\in}\{\lbeta,\seq\}
\end{array}
%%%%%%%%%%%%%%%%%%%%%%%%%%%%%%%%%%%
\begin{array}[t]{@{\!\!\!\!}c@{}}
\xymatrix@R=5mm@C=5mm{
 \cdot \ar@{-->}[d]_{\sr,\lllrule,+}\ar[r]^{\SCtxt,\ucp}  & \cdot \ar[dd]^{\sr,a}\\
 \cdot\ar@{-->}[d]_{\sr,a}\\
 \cdot\ar@{-->}[r]_{\SCtxt,\ucp\vee\gc}&\cdot \\
}
\\
a{\in}\{\lbeta,\cp,\problabl,\\
\problabr,
\caserule,\seq\}
\end{array}
%%%%%%%%%%%%%%%%%%%%%%%%%%%%%%%%%%%
\begin{array}[t]{@{\!\!\!}c@{}}
\xymatrix@R=5mm@C=5mm{
 \cdot \ar@{-->}[d]_{\sr,\lllrule}\ar[r]^{\SCtxt,\ucp}  & \cdot \ar[ddd]^{\sr,\caserule}\\
 \cdot\ar@{-->}[d]_{\sr,\caserule}\\
 \cdot\ar@{-->}[d]_{\sr,\lllrule}\\
\cdot\ar@{-->}[r]_{\SCtxt,\ucp}&\cdot \\
}
\end{array}
%%%%%%%%%%%%%%%%%%%%%%%%%%%%%%%%%%%
\begin{array}[t]{@{\!\!\!\!\!}c@{}}
\xymatrix@R=5mm@C=5mm{
 \cdot \ar@{-->}[d]_{\sr,\lllrule}\ar[rr]^{\SCtxt,\ucp}  && \cdot \ar[ddd]^{\sr,\caserule}\\
 \cdot\ar@{-->}[d]_{\sr,\caserule}\\
 \cdot\ar@{-->}[d]_{\sr,\lllrule,0 \vee 1}\\
\cdot\ar@{-->}[r]_{\SCtxt,\gc}&\cdot\ar@{-->}[r]_{\SCtxt,\ucp,+} &\cdot \\
}
\end{array}
%%%%%%%%%%%%%%%%%%%%%%%%%%%%%%%%%%%
\end{array}
\!\!%%%%%%%%%%%%%%%%%%%%%%%%%%%%%%%%%%%
\begin{array}[t]{@{}c@{}}
%%%%%%%%%%%%%%%%%%%%%%%%%%%%%%%%%%%
\begin{array}[t]{@{}c@{}}
\xymatrix@R=5mm@C=5mm{
 \cdot \ar@{-->}[d]_{\sr,\lllrule,+}\ar[r]^{\SCtxt,\ucp}  & \cdot \ar[d]^{\sr,\lllrule,+}\\
 \cdot \ar@{-->}[r]_{\SCtxt,\ucp}         & \cdot \\
}
\end{array}
\\ 
\begin{array}[t]{@{}c@{}}
\xymatrix@R=5mm@C=5mm{
 \cdot \ar@{-->}[d]_{\sr,\lllrule}\ar[r]^{\SCtxt,\ucp}  & \cdot \ar[dddd]^{\sr,\lbeta}\\
 \cdot\ar@{-->}[d]_{\sr,\cp}\\
 \cdot\ar@{-->}[d]_{\sr,\lbeta}\\
  \cdot\ar@{-->}[d]_{\sr,\lllrule}\\
\cdot\ar@{-->}[r]_{\SCtxt,\gc}&\cdot \\
}
\end{array}
\end{array}
%%%%%%%%%%%%%%%%%%%%%%%%%%%%%%%%%%%
\end{array}
\]
\end{minipage}
\\
\begin{minipage}{\textwidth}
\textbf{Base cases:} If $s \xrightarrow{\SCtxt,\ucp} t${,}then if $s$ is a WHNF then $t$ is a WHNF{,}if $t$ is an WHNF{,}then $s \xrightarrow{\sr,\lllrule,0\vee 1}\xrightarrow{\sr,\cp,0\vee 1} s'$ where $s'$ is a WHNF.
\end{minipage}
\caption{Diagrams for $(\ucp)$}
\end{subfigure}
\caption{Diagrams for $\LRPROB$ (cont.)\label{fig-diagr-lrprob-2}}
\end{figure*}

\ignore{
\cleardoublepage
\section{Reste}
\davidcomment{Die automatische Diagrammberechnung kann das allgemeine case nicht{,}sondern nur konkret mit Bool,Pair und List. Man könnte sich entweder komplett darauf beschränken{,}oder man muss die Diagramme dann nochmal händisch prüfen. Könnte schon sein{,}das man z.B. aus 2 cpx dann cpx,* machen muss usw.}

\manfredcomment{Eine Sprache  waere gut{,}die Zahlkonstanten hat: unendlich viele{,}und nur Zahlen als Endergebnis erlaubt. und dann bloederweise ein unendlich grosses case bräuchte? 
Das wäre aber kompatibel mit dem was ich mir vorstelle zu oerationale = kontextuelle Gleichheit.
}

\davidcomment{Das passt ja irgendwie nicht zu den Programmiersprachen? Da wäre richtiger Listen von Bits und typisieren und dann nur diesen Typ in der kontextuellen Gleichheit beobachten? Vielleicht müssten diese Listen auch strikt sein (also strikte Konstruktoren) damit sie nicht unendlich werden usw). Ich seh das aber eher als Future Work,
da es den Kalkül schon stark ändern würde.}

\manfredcomment{Ja ich sehe dass das zu weit geht im Moment. 
Ich wollte das Probem umgehen{,}das wenn man eine Menge $(p_1,w_1){,}
... (p_i,w_i),...$  quasi als Ergbnisse hat{,}dass man dann für eine operationale Gleichhheit dann noch die $w_i$ weiter analysieren muss{,}in Kontexte stecken usw. Das macht es halt viel komplizierter. Endliche Listen von Bits wären ok. 

Als einfache Variante{,}die nicht soviel aendern würde könnte man endlich viele Konstanten betrachten.  OK future work
}
\section{ToDo}

\davidcomment{
als Notizen (todo) für später: 
\begin{itemize}
    \item  Notation vereinheitlichen: r,s,t für Ausdrücke statt e{,}e'
    \item Notation $(\RCtxt,T)$ statt $(R,T)$ und analog für $(\Ctxt,T)$ und $(\SCtxt,T)$
    \item Evtl noch einbauen (Reste):
    {General Remarks on the Diagrams}

The following subsections contain diagrams for the following purpose:\\
given an evaluation starting from $t${,}i.e. $t \xrightarrow{\sr,*} t_0${,}
and another reduction  $t \xrightarrow{\SCtxt,r} t_1$ (the forking case){,}or $t \xleftarrow{\SCtxt,r} t_0$ (the commuting case){,}where $r$ is a specification of the reduction{,}the diagrams cover {\bf all} possibilities of a (small-step) modification of the combined reduction. 

The goal is to iterate the modification steps given by the diagrams until an evaluation of $t_0$ is constructed. This is done in a series of Lemmas.

A convention is that the trivial cases are omitted in the diagram presentation but have to be considered in the proofs.
These are: 
\begin{enumerate}
    \item   $t \xrightarrow{\SCtxt,r} t'$ may also be an sr-reduction{,}after
the transformation application.
    \item The situation $t_0 \xleftarrow{\SCtxt,a} t \xrightarrow{\sr,a} t_1$ may have  $t_0 = t_1$ as an instance. 
\end{enumerate}

We also use variables in the diagrams (for instance in labeled arrows $\xrightarrow{\sr,a}$. The scope of variable $a$ is the whole diagram. This means that the instances of the diagrams are generated by replacing all occurrences of $a$ with the same label.

\end{itemize}
Vorschläge (nicht entschieden für mich)
\begin{itemize}
    \item Multicontexte mit $M$ statt $C$ schreiben?
    \item weighted expressions nahezu eliminieren{,}da man auch alles über $|L|$ machen kann ($L$ als prob-sequence der evaluation)
    \item \ldots
\item Evtl. noch eine (ohne irgendeine Bedeutung) Konstante $c$ dazu{,}sonst gibt es zuwenig closed expressions. (Kommentar von Manfred)
\end{itemize}

}
\endignore}

\end{document}